%% file: main.tex
\documentclass[opre,nonblindrev]{template_files/informs3_clean}

\OneAndAHalfSpacedXI 

\usepackage{endnotes}
\let\footnote=\endnote
\let\enotesize=\normalsize
\def\notesname{Endnotes}%
\def\makeenmark{\hbox to1.275em{\theenmark.\enskip\hss}}
\def\enoteformat{\rightskip0pt\leftskip0pt\parindent=1.275em
  \leavevmode\llap{\makeenmark}}
  
\usepackage{natbib}
 \bibpunct[, ]{(}{)}{,}{a}{}{,}%
 \def\bibfont{\small}%

 \usepackage{hyperref}
 \usepackage[capitalise]{cleveref}
 \usepackage{accents,algorithm,algpseudocode,enumitem,mathtools,booktabs,diagbox}
 \newcommand{\RR}{\mathbb{R}}
 \newcommand{\st}{{\rm s.t.}\ }

 \newcommand{\cX}{\mathcal{X}}
 \newcommand{\interior}{\textnormal{int}\,}
 \newcommand{\conv}{\textnormal{conv}\,}
 \newcommand{\almeve}{\textnormal{a.e.}}
 \newcommand{\ones}{\mathbf{1}}
 \newcommand{\EE}{\mathbb{E}}
 \newcommand{\ubar}[1]{\underaccent{\bar}{#1}} 
 
 \newcommand{\poly}{\textnormal{Poly}}

\TheoremsNumberedThrough     
\ECRepeatTheorems

\EquationsNumberedThrough    

\begin{document}


\RUNAUTHOR{Gao and Kroer}

\RUNTITLE{Infinite-Dimensional Fisher markets}

\TITLE{Infinite-Dimensional Fisher Markets and Tractable Fair Division}
\ARTICLEAUTHORS{%
\AUTHOR{Yuan Gao,  Christian Kroer}
\AFF{
	Department of Industrial Engineering and Operations Research, Columbia University, New York, NY 10027
\EMAIL{gao.yuan@columbia.edu}, \EMAIL{christian.kroer@columbia.edu}
} 
} 

\ABSTRACT{%
  Linear Fisher markets are a fundamental economic model with applications in fair division as well as large-scale Internet markets. In the finite-dimensional case of $n$ buyers and $m$ items, a market equilibrium can be computed using the Eisenberg-Gale convex program. 
  Motivated by large-scale Internet advertising and fair division applications, this paper considers a generalization of a linear Fisher market where there is a finite set of buyers and a continuum of items. 
  We introduce generalizations of the Eisenberg-Gale convex program and its dual to this infinite-dimensional setting, which leads to Banach-space optimization problems. 
  We establish existence of optimal solutions, strong duality, as well as necessity and sufficiency of KKT-type conditions. 
  All these properties are established via non-standard arguments, which circumvent the limitations of duality theory in optimization over infinite-dimensional Banach spaces. 
  Furthermore, we show that there exists a pure equilibrium allocation, i.e., a division of the item space. 
  When the item space is a closed interval and buyers have piecewise linear valuations, we show that the Eisenberg-Gale-type convex program over the infinite-dimensional allocations can be reformulated as a finite-dimensional convex conic program, which can be solved efficiently using off-the-shelf optimization software based on primal-dual interior-point methods. 
  Based on our convex conic reformulation, we develop the first polynomial-time cake-cutting algorithm that achieves Pareto optimality, envy-freeness, and proportionality.
  For general buyer valuations or a very large number of buyers, we propose computing market equilibrium using stochastic dual averaging, which finds approximate equilibrium prices with high probability. 
  Finally, we discuss how the above results easily extend to the case of quasilinear utilities.
}%



\maketitle

\input{text/intro.tex}

\input{text/fisher_market.tex}

\input{text/equilibrium_and_duality.tex}
\input{text/conic_reform_pwl.tex}

\input{text/sda.tex}
\input{text/ql.tex}
\input{text/conclusion.tex}
\vspace{30pt}
\bibliographystyle{template_files/informs2014} 
\bibliography{refs}

\ECSwitch


\ECHead{Technical Appendix}
\input{text/appendix.tex}
\end{document}

%% file: text/intro.tex
\section{Introduction}
Market equilibrium (ME) is a classical concept from economics, where the goal is to find an allocation of a set of items to a set of buyers, as well as corresponding prices, such that the market clears. 
One of the simplest equilibrium models is the (finite-dimensional) linear \emph{Fisher market}. A Fisher market consists of a set of $n$ buyers and $m$ divisible items, where the utility for a buyer is linear in their allocation.
Each buyer $i$ has a budget $B_i$ and valuation $v_{ij}$ for each item $j$. A ME consists of an allocation (of items to buyers) and prices (of items) such that (i) each buyer receives a bundle of items that maximizes their utility subject to their budget constraint, and (ii) the market clears (all items such that $p_j>0$ are exactly allocated).
In spite of its simplicity, this model has several applications. Perhaps one of the most celebrated examples is the \emph{competitive equilibrium from equal incomes} (CEEI), where  $m$ items are to be fairly divided among $n$ agents. By giving each agent one unit of faux currency, the allocation from the resulting ME can be used as a fair division. This approach guarantees several fairness desiderata, such as envy-freeness and proportionality. Beyond fair division, linear Fisher markets also find applications in large-scale ad markets \citep{conitzer2018multiplicative,conitzer2019pacing} and fair recommender systems \citep{kroer2019computing,kroer2019scalable}.

For the case of finite-dimensional linear Fisher markets, the Eisenberg-Gale convex program computes a market equilibrium \citep{eisenberg1959consensus,eisenberg1961aggregation}.
However, in settings like Internet ad markets and fair recommender systems, the number of items is often huge~\citep{kroer2019computing,kroer2019scalable}, if not infinite or even uncountable~\citep{balseiro2015repeated}. For example, each item can be characterized by a set of features.
In that case, a natural model for an extremely large market, such as Internet ad auctions or recommender systems, is to assume that the items are drawn from some underlying distribution over a compact set of possible feature vectors.

Motivated by settings such as the above, where the item space is most easily modeled as a continuum, we study Fisher markets and its equilibria for a continuum of items.
    While equilibrium computation for finite-dimensional linear Fisher markets is well understood, nothing is known about computation of its infinite-dimensional analogue. We rectify this issue by developing infinite-dimensional convex programs over Banach spaces that generalize the Eisenberg-Gale convex program and its dual. We show that these convex programs lead to market equilibria, and give scalable first-order methods for solving the convex programs.


A problem closely related to our infinite-dimensional Fisher-market setting is the \textit{cake-cutting} or \textit{fair division} problem. 
There, the goal is to efficiently partition a ``cake''---often modeled as a compact measurable space, or simply the unit interval $[0,1]$---among $n$ agents so that certain fairness and efficiency properties are satisfied \citep{weller1985fair,brams1996fair,cohler2011optimal,procaccia2013cake,cohler2011optimal,brams2012maxsum,chen2013truth,aziz2014cake,aziz2016discrete,deng2012algorithmic}.
Focusing on the case of finding a division of a measurable space satisfying weak Pareto optimality and envy freeness, \citet{weller1985fair} shows the existence of a fair division.
When all buyers have the same budget, our definition of a \textit{pure} ME, i.e., where the allocation consist of indicator functions of a.e.-disjoint measurable sets, is equivalent to this notion of fair division.
Thus, our convex programs yield solutions to the fair division setting of \citet{weller1985fair}.
Additionally, we also give an explicit characterization of the unique equilibrium prices based on a pure equilibrium allocation under arbitrary budgets. 
This generalizes the result of \citet{weller1985fair}, which only holds for uniform budgets.

Under piecewise \emph{constant} valuations over the $[0,1]$ interval, the equivalence of fair division and market equilibrium in certain setups has been discovered and utilized in the design of cake-cutting algorithms \citep{brams2012maxsum,aziz2014cake}. 
For example, \citet{aziz2014cake} show that the special case with piecewise constant valuations can be (easily) reduced to a finite-dimensional Fisher market and hence captured by the classical Eisenberg-Gale framework. 
Our infinite-dimensional convex optimization characterization extends this connection from piecewise constant valuations to arbitrary valuations in the $L^1$ function space: we propose Eisenberg-Gale-type convex programs that characterize \textit{all} ME. This includes pure ME which, under uniform budgets, correspond to fair divisions. 

Beyond piecewise constant valuations, piecewise linear valuations have also been considered in fair division, although with different fairness and efficiency objectives. This setting is considerably more challenging, and e.g. \citet[\S 4]{cohler2011optimal} focus on the case of two agents. Unlike the piecewise constant case, one cannot cut the unit interval \emph{a priori} based on breakpoints of the pieces of buyers' valuations. Instead, as we will see, the correct such ``cuts'' inevitably depend on the equilibrium utility prices (price per unit utility) of each buyer.
Nonetheless, we will show that in the piecewise linear case, it is possible to reformulate our general convex program as a finite-dimensional convex conic program involving second-order cones and exponential cones. 
We leverage this reformulation to give a polynomial-time procedure for computing a fair division for piecewise linear utilities in complete generality, for any number of agents.
In addition to being polynomial-time computable in theory, our conic reformulation is also highly efficient numerically: it can be written as a sparse conic program that can be solved with standard convex optimization software.
A key part of our finite-dimensional conic reformulation is to show that, given linear buyer valuations over a closed interval, the set of utilities attainable by feasible allocations of the item space can be described by a small number of linear and quadratic constraints. This result may be generalized to other classes of buyer valuations and may thus be of independent interest.




\subsection{Summary of contributions}
\paragraph{Infinite-dimensional Fisher markets and equilibria.} First, we propose the notion of a market equilibrium (ME) for an infinite-dimensional Fisher market with $n$ buyers and a continuum of items $\Theta$. 
Here, buyers' valuations are nonnegative $L^1$ functions on $\Theta$. Buyers' allocations are nonnegative $L^\infty$ functions on $\Theta$. 
A special case is \emph{pure} allocations, where buyers get a.e.-disjoint measurable sets of items. 

\paragraph{Market equilibrium and convex optimization duality.}
We then give two infinite-dimensional convex programs over Banach spaces of measurable functions on $\Theta$ (problems \eqref{eq:eg-primal} and \eqref{eq:eg-dual-beta-p}), which generalize the EG convex program and its dual for finite-dimensional Fisher markets, and establish the existence of optimal solutions.
Due to the lack of a compatible constraint qualification, general duality theory does not apply to these convex programs. 
Instead, we establish various duality properties directly through nonstandard arguments. 
Based on these duality properties and the existence of a minimizer in the ``primal'' convex program, we show that a pair of allocations and prices is a ME \textit{if and only if} they are optimal solutions to the convex programs.
Furthermore, we show that the primal convex program admits a \emph{pure} optimal solution, meaning that buyers get disjoint subsets of items.
and we conclude that there exists a pure equilibrium allocation, i.e., a \textit{division} (modulo zero-value items) of the item space.

\paragraph{Properties of a market equilibrium.} 
Based on the above results, we further show that a ME under the infinite-dimensional Fisher market satisfies (budget-weighted) proportionality, Pareto optimality and budget-weighted envy-freeness. Our results on the existence of ME and its fairness properties can be viewed as generalizations of those in \citet{weller1985fair}, in which every buyer (agent) has the same budget. 

All of the above results, except the existence of a pure equilibrium allocation, also hold when the item space $\Theta$ is discrete (finite or countably infinite).

\paragraph{Tractable reformulation under piecewise linear valuations.} 
When the item space is a closed interval (e.g., $[0,1]$) and buyers have piecewise linear valuations, we show that equilibrium allocations can be computed efficiently via solving a convex conic reformulation of the infinite-dimensional Eisenberg-Gale-type convex program. 
This gives an efficient algorithm for computing a fair division under piecewise linear valuations, the first polynomial-time algorithm for this challenging problem to the best of our knowledge.
The key in the reformulation is to show that, for linear valuations on a closed interval, the set of feasible utilities spanned by all feasible allocations can be described by a small number of linear and quadratic constraints with a few auxiliary variables.

\paragraph{Stochastic optimization for general valuations.}
For more general buyer valuations or a huge number of buyers, we propose solving a finite-dimensional convex reformulation of the dual of the infinite-dimensional EG for equilibrium utility prices using the stochastic dual averaging algorithm (SDA) and establish its convergence guarantees.

\paragraph{Extension to quasilinear utilities.} 
Most of the above results easily extend to the setting where each buyer has a quasilinear utility. 
Specifically, we show that, in this case, a different pair of infinite-dimensional convex programs exhibit optimal solutions that correspond to quasilinear market equilibria. 
The convex conic reformulation can also be modified easily to capture pure quasilinear equilibrium allocations under piecewise linear valuations. 
Finally, SDA can also be easily modified to compute equilibrium utility prices in the quasilinear setting.

\subsection{Related work}
In addition to the aforementioned works, we briefly review other recent works on market equilibrium computation and fair division.
\paragraph{Market equilibrium computation.} 
For the classical Fisher market setting with finitely many items, there is a large literature on equilibrium computation algorithms, some based on solving new equilibrium-capturing convex programs. For example, 
\citet{devanur2008market} established the first polynomial-time algorithm for exact equilibrium computation for a finite-dimensional linear Fisher market, based on a primal-dual algorithm for solving the Eisenberg-Gale convex program. 
\citet{zhang2011proportional} proposed a distributed dynamics that converges to an equilibrium, which, as later analyzed in \citet{birnbaum2011distributed}, is in fact a first-order optimization algorithm applied to a specific convex program due to \citet{shmyrev2009algorithm}.
\citet{cheung2019tatonnement} studied t\^{a}tonnement dynamics and show its equivalence to gradient descent on Eisenberg-Gale-type convex programs under the more general class of CES utilities. \citet{gao2020first} studied first-order methods based on (old and new) convex programs for Fisher market equilibria under commonly used utilities. \citet{bei2019earning} studied earning and utility limits in markets with linear and spending-constraint utilities, and proposed a polynomial-time algorithm for computing an equilibrium. 

\paragraph{Fair division.}
As stated previously, a Fisher market equilibrium on finitely many divisible items is known to be a strong fair division approach. 
There is also a literature on fair division of \emph{indivisible} items via maximizing the Nash social welfare (NSW); the discrete analogue of the Eisenberg-Gale program.
This was started by \citet{caragiannis2016unreasonable}, who showed that the maximum NSW solution provides fairness guarantees in the indivisible divisible case as well, and proposed a practical algorithm based on mixed-integer linear programming.
There are also several works on approximation algorithms for this settings, see e.g. \citet{garg2018approximating,barman2018finding}.
Interestingly, we will show that our continuum setting allows us to construct allocations via convex programming, even for the indivisible setting.


%% file: text/fisher_market.tex
\section{Infinite-Dimensional Fisher Markets} \label{sec:inf-dim-fim-setup}

\paragraph{Measure-theoretic preliminaries.} First we introduce the measure-theoretic concepts that we will need. The following paragraph can be skimmed and referred back to later. 
The items will be represented by $\Theta$, a compact subset of $\RR^d$. Denote the Lebesgue measure on $\RR^d$ as $\mu$. Since $\Theta$ is compact, it is (Borel) measurable and $\mu(\Theta) < \infty$. Note that functions that are equal a.e. on $\Theta$ form an equivalence class, which we treat as the same function. In fact, any function $f$ in this equivalence class give the same linear functional $g\mapsto \int_\Theta fg d\mu$. 
The suffix a.e. will be omitted unless the emphasis is necessary. For any set $S$ of measurable functions on $\Theta$, denote $S_+ = \{ f\in S: f\geq 0\}$. 
For $f \in L^1(\Theta)$ and $g\in L^\infty(\Theta)$, denote $\langle f, g\rangle = \int_\Theta fg d \mu$. This notation aligns with the usual notation for a bilinear form, i.e., applying a linear functional to a function. Since $L^\infty(\Theta)$ is the dual space of $L^1(\Theta)$, the integration $\int_\Theta fgd\mu$ is well-defined and is finite.
Let $\ones$ be the constant function taking value $1$ on $\Theta$. For any measurable set $A\subseteq \Theta$, $\ones_A$ denotes the $\{0,1\}$-indicator function of $A$. 
For $q \in [1, \infty]$, let $L_q(\Theta)$ be the Banach space of $L^q$ (integrable) functions on $\Theta$ with the usual $L^q$ norm, i.e, for $f \in L^q(\Theta)$,
\[ 
	\|f\| = \begin{cases}
		\int_\Theta |f|^q d\mu & {\rm if} \ q< \infty,\\[1ex]
		\inf \{ M > 0: |f| \leq M\, {\rm a.e.} \text{ on $\Theta$} \} & {\rm if} \ q =\infty.
	\end{cases} 
\]
Any $\tau \in L^1(\Theta)_+$ can also be viewed as a measure on $\Theta$ via $\mu_\tau(A):= \int_A \tau d\mu $ for any measurable set $A\subseteq \Theta$. In this work, we will denote the measure $\mu_\tau$ simply as $\tau$.
Unless otherwise stated, any measure $m$ used or constructed is absolutely continuous w.r.t. the Lebesgue measure $\mu$ and hence \textit{atomless}. In other words, for any measurable set $A\subseteq \Theta$ such that $m(A) > 0$ and any $0 < c < m(A)$, there exists a measurable subset $B\subseteq A$ such that $m(B) = c$. 
Two measurable sets $A, B\subseteq \Theta$ are said to be \emph{a.e.-disjoint} if $\mu(A\cap B) = 0$.
We use equations and inequalities involving measurable functions to denote the corresponding (measurable) \emph{preimages} in $\Theta$.
For example, $\{ f \leq 0 \} := \{ \theta\in \Theta: f(\theta) \leq 0 \}$ and $\{ f \leq g \} := \{ \theta\in \Theta: f(\theta) \leq g(\theta) \}$.

\paragraph{Fisher market.} 
Here, we formally describe the infinite-dimensional (linear) Fisher market setup that we use throughout this work.
\begin{enumerate}
	\item There are $n$ buyers and an item space $\Theta$, which is a compact subset of $\RR^d$. 
	\item Each buyer has a valuation over the item space $v_i\in L^1(\Theta)_+$ (nonnegative $L^1$ functions on $\Theta$).
	\item The items' prices $p\in L^1(\Theta)_+$ live in the same space as valuations. 
	\item An allocation of items to a buyer $i$ is denoted by $x_i \in L^\infty(\Theta)_+$. 
	We use $x = (x_1, \dots, x_n) \in (L^\infty(\Theta)_+)^n$ to denote the aggregate allocation. 
	An allocation $x$ is said to be a \textit{pure} allocation (or a pure solution, when viewed as variables of a convex program) if for all $i$, $x_i = \ones_{\Theta_i}$ for \textit{a.e.-disjoint} measurable sets $\Theta_i\subseteq \Theta$ (where leftover is allowed, i.e., $\Theta \setminus (\cup_i \Theta_i) \neq \emptyset$). 
	When $x$ is a pure allocation (solution), we also denote $x$ as $\{\Theta_i\}$. 
	An allocation is \textit{mixed} if it is not pure, or equivalently, the set $\{0< x_i < 1\} \subseteq \Theta$ has positive measure for some $i$. 
	\item Each buyer has a budget $B_i > 0$ and all items have unit supply, i.e., $x$ is supply-feasible if $\sum_i x_i \leq \ones$. Without loss of generality, we also assume that $v_i(\Theta) = \|v_i \| > 0$ for all $i$ (otherwise buyer $i$ can be removed).
\end{enumerate}
Note that the market is ``linear'' means the utility each buyer $i$ receives from an allocation $x_i$ is a linear functional $ x_i \mapsto \langle v_i, x_i \rangle$. The valuation $v_i$ itself, as an $L^1$ function on the item space $\Theta$, may not be a linear function in $\theta\in \Theta$.
\begin{example}[Valuations and allocations]
	\label{ex:pure-mixed-allo}
	Let there be $n=2$ buyers and $\Theta = [0,1]$. Then, their allocations $x_1, x_2\in L^\infty([0,1])_+$ are nonnegative measurable functions on $[0,1]$. 
	An example of a pure allocation is $x_1 = \ones_{[0,1/2]}$, $x_2 = \ones_{[1/2,1]}$. We can also denote the pure allocation as $\{\Theta_i\}$, where $\Theta_1 = [0,1/2]$, $\Theta_2 = [1/2,1]$. 
	Here, $\Theta_i$ are measurable and a.e.-disjoint. 
	An example of a mixed allocation is $x_1(\theta) = 0.5 + 0.1 \theta^2$ and $x_2(\theta) = 0.5 - 0.1 \theta^2$, $\theta\in [0,1]$. In both cases, we have $x_1 + x_2 = \ones$ a.e. (with both sides viewed as nonnegative measurable functions on $[0,1]$). Let buyer $1$'s valuation be $v_1(\theta) = \theta^2$. For the allocation $x_1 = \ones_{[0,1/2]}$, the utility buyer $1$ receives is $\langle v_1, x_1 \rangle = v_1([0,1/2]) = \int_0^1 v_1(\theta)\ones_{[0,1/2]}(\theta)d\theta = \int_0^{1/2} \theta^2 d\theta = \frac{1}{24} $; for the allocation $x_1(\theta) = 0.5 + 0.1\theta^2$, the utility is $\langle v_1, x_1 \rangle = \int_0^1 v_1(\theta)x_1(\theta) d\theta = \int_0^1 \theta^2 (0.5+0.1\theta) d\theta = \frac{23}{120}$. Note that $v_1$ denotes both the $L^1$ function $\theta\mapsto\theta^2$ and the induced measure $A\mapsto \int_A \theta^2 d\theta$.
\end{example}

Given prices $p\in L^1(\Theta)_+$, the \emph{demand set} of buyer $i$ is the set of utility-maximizing allocations subject to its budget constraint:
\[ D_i(p) = \argmax \left\{ \langle v_i, x_i\rangle: x\in L^\infty(\Theta)_+,\, \langle p, x_i\rangle \leq B_i \right\}. \]
Generalizing its finite-dimensional counterpart \citep{nisan2007algorithmic}, 
a \textit{market equilibrium} is defined as a pair $(x^*, p^*) \in (L^\infty(\Theta)_+)^n \times L^1(\Theta)_+ $ satisfying the following.
\begin{itemize}
	\item Buyer optimality: for every $i\in [n]$, $x^*_i \in D_i(p^*)$. 
	\item Market clearance (up to zero-price items): $\sum_i x^*_i \leq \ones$ and $\langle p^*, \ones - \sum_i x^*_i\rangle = 0$. 
\end{itemize}
We say that $x^*\in (L^\infty(\Theta)_+)^n$ is an equilibrium allocation if $(x^*, p^*)$ is a ME for some $p^*\in L^1(\Theta)_+$. A pair $(x^*, p^*)$ is called a pure ME if it is a ME and $x^*$ is a pure allocation.
From the definition of market equilibrium, we can assume 
the following normalizations w.l.o.g.
\begin{itemize}
	\item $v_i(\Theta) = \|v_i\| = 1$ for all $i$, since $D_i(p)$ is invariant under scaling of $v_i$. 
	\item $\|B\|_1 = 1$ since if $(x^*, p^*)$ is a ME under $B = (B_i)$, then $(x^*, p^*/\|B\|_1)$ is a ME under normalized budgets $(B_i/\|B\|_1)$. 
	\item The total supply of all items is $\mu(\Theta) = \|\ones\| = 1$. Otherwise, we can scale the item space $\Theta$ via $\theta \mapsto \alpha \theta$ for some constant $\alpha$ or scale the measure $\mu$.
\end{itemize}


%% file: text/equilibrium_and_duality.tex
\section{Equilibrium and Duality} \label{sec:equi-and-dual}


Due to intrinsic limitations of general infinite-dimensional convex optimization duality theory, we cannot start with a convex program and then derive its dual.
Instead, we directly \emph{propose} two infinite-dimensional convex programs, and then proceed to show from first principles that they exhibit optimal solutions and a strong-duality-like relationship.
First, we give a direct generalization of the (finite-dimensional) Eisenberg-Gale convex program \citep{eisenberg1961aggregation,nisan2007algorithmic}:
\begin{align}
z^* = \sup_{x\in (L^\infty(\Theta)_+)^n} \sum_i B_i \log \langle v_i, x_i \rangle\ \ {\rm s.t.}\ \sum_i x_i \leq \ones. \tag{$\mathcal P_{\rm EG}$} \label{eq:eg-primal}
\end{align}

Motivated by the dual of the finite-dimensional EG convex program \cite[Lemma 3]{cole2017convex}, we also consider the following convex program:
\begin{align}
\begin{split}
w^* = & \inf_{p\in L^1(\Theta)_+,\, \beta\in \RR^n_+} \left[ \langle p, \ones \rangle - \sum_i B_i \log \beta_i\right] \\ & \quad {\rm s.t.} \ p \geq \beta_i v_i\ \almeve, \forall\,i. 
\end{split} \tag{$\mathcal D_{\rm EG}$}
\label{eq:eg-dual-beta-p}
\end{align}

We state our central theoretical results in the following theorem. Parts of this theorem are stated in more detail in subsequent lemmas and theorems. 
Proofs of all theoretical results can be found in Appendix \ref{app:proofs}. 
\begin{theorem}
	The following results hold regarding problems \eqref{eq:eg-primal} and \eqref{eq:eg-dual-beta-p}. 
	\begin{enumerate}[label=(\alph*)]
		\item The supremum $z^*$ of \eqref{eq:eg-primal} is attained via a \emph{pure} optimal solution $x^*$, that is, $x^* = (x^*_i)$ and $x^*_i = \ones_{\Theta_i}$ for a.e.-disjoint measurable subsets $\Theta_i\subseteq \Theta$. \label{item:eg-primal-attain}
		\item The infimum $w^*$ of \eqref{eq:eg-dual-beta-p} is attained via an optimal solution $(p^*, \beta^*)$, in which $\beta^*\in \RR^n_+$ is unique and $p^* = \max_i \beta^*_i v_i$ a.e. \label{item:eg-dual-p-beta-attain}
		\item A pair of allocations and prices $(x^*, p^*) \in (L^\infty(\Theta)_+)^n \times L^1(\Theta)_+$ is a ME if and only if $x^*$ is an optimal solution of \eqref{eq:eg-primal} and $(p^*, \beta^*)$ is the (a.e.-unique) optimal solution of \eqref{eq:eg-dual-beta-p}. \label{item:eg-equi-iff-opt}
	\end{enumerate}
	\label{thm:eg-equi-opt-combined}
\end{theorem}
\begin{remark}
If we view \eqref{eq:eg-dual-beta-p} as the primal, then it can be shown that its Lagrange dual is \eqref{eq:eg-primal}, and weak duality follows (see, e.g., \cite[\S 3]{ponstein2004approaches}). 
However, we cannot conclude strong duality, or even primal or dual optimum attainment, since $L^1(\Theta)_+$ has an empty interior \cite[\S 8.8 Problem 1]{luenberger1997optimization} and hence Slater's condition does not hold.
If we choose the space for valuations and prices to be $L^\infty(\Theta)$ instead of $L^1(\Theta)$ for the space of allocations $x_i$ (i.e., the underlying Banach space of \eqref{eq:eg-primal}), then \eqref{eq:eg-dual-beta-p}, with $p \in L^\infty(\Theta)_+$, does satisfy Slater's condition \cite[\S 8.8 Problem 2]{luenberger1997optimization}. 
However, its dual is \eqref{eq:eg-primal} but with the nonnegative cone $L^\infty(\Theta)_+$ (in which each $x_i$ lies) replaced by the (much larger) cone $\{ g \in L^\infty(\Theta)^*: \langle f, g \rangle \geq 0,\, \forall\, f\in L^\infty(\Theta)_+ \}\subseteq L^\infty(\Theta)^*$. In this case, not every bounded linear functional $g\in L^\infty(\Theta)$ can be represented by a measurable function $\tilde{g}$ such that $\langle f, g \rangle = \int_\Theta \tilde{g}f d\mu$ (see, e.g., \citep{day1973normed}). 
Therefore, we still cannot conclude that \eqref{eq:eg-primal} has an optimal solution in $(L^1(\Theta)_+)^n$ satisfying strong duality. 
Similar issues occur when \eqref{eq:eg-primal} is viewed as the primal instead. \label{remark:why-define-cp-then-prove}
\end{remark}

We briefly explain the proof ideas of Theorem~\ref{thm:eg-equi-opt-combined}. 
Unlike the finite-dimensional case, the feasible region of \eqref{eq:eg-primal} here, although being closed and bounded in the Banach space $L^\infty(\Theta)$, is not compact. In fact, it is easy to construct an infinite sequence in the feasible region that does not have any convergent subsequence.
This issue can be circumvented using the following lemma.
\begin{lemma}
	Define the set of feasible utilities as
	\begin{align}
		U = U(v, \Theta) = \left\{ (u_1, \dots, u_n): u_i = \langle v_i, x_i\rangle,\, x\in (L^\infty(\Theta)_+)^n,\, \sum_i x_i \leq \ones \right\} \subseteq \RR_+^n \label{eq:def-U-U(v,Theta)}
	\end{align}
	and the set of utilities attainable via pure allocations as
	\[ U' = U'(v, \Theta) = \left\{ (v_1(\Theta_1), \dots, v_n(\Theta_n)): \Theta_i \subseteq \Theta\, \text{measurable and a.e.-disjoint} \right\} \subseteq \RR_+^n. \]
	Then, $U = U'$ and this set is convex and compact.
	\label{lemma:U-convex-compact}
\end{lemma}
Using Lemma~\ref{lemma:U-convex-compact}, for part \ref{item:eg-primal-attain}, we can show that there exists $u^* \in \RR_{++}^n$ such that $z^* = \sum_i B_i \log u^*_i$, which then ensures that $z^*$ is attained by some \emph{pure} feasible solution $\{\Theta_i\}$ of \eqref{eq:eg-primal}, that is, $\Theta_i\subseteq \Theta$ are a.e.disjoint and $v_i(\Theta_i) = u^*_i$.

Part \ref{item:eg-dual-p-beta-attain} follows by reformulating \eqref{eq:eg-dual-beta-p} into a finite-dimensional convex program in $\beta\in \RR_+^n$. 
For a  fixed $\beta>0$, setting $p = \max_i \beta_i v_i$ clearly minimizes the objective of \eqref{eq:eg-dual-beta-p}. 
Since $\beta\geq 0$ and $v_i \in L^1(\Theta)_+$, we have
\[  0 \leq \max_i \beta_i v_i \leq \|\beta\|_1 \sum_i v_i, \]
where the right-hand side is $L^1$-integrable since each $v_i$ is. Hence, $\max_i \beta_i v_i \in L^1(\Theta)_+$ as well. 
Thus we can also reformulate \eqref{eq:eg-dual-beta-p} as the following convex program:
\begin{align}
\inf_{\beta\in \RR_+^n} \left[\left \langle \max_i \beta_i v_i, \ones \right \rangle - \sum_i B_i \log \beta_i\right].  \label{eq:eg-dual-beta-1}
\end{align} 
\begin{lemma}
	The infimum of \eqref{eq:eg-dual-beta-1} is attained via a unique minimizer $\beta^* > 0$. The optimal solution $(p^*, \beta^*)$ of \eqref{eq:eg-dual-beta-p} has a unique $\beta^*$ and satisfies $p^* = \max_i \beta^*_i v_i$ a.e.
	\label{lemma:beta-dual-attain}
\end{lemma}

To show Part \ref{item:eg-equi-iff-opt}, we first establish weak duality and KKT conditions (necessary and sufficient for optimality) in the following lemma. 
As mentioned before, this is necessary due to the lack of general duality results in infinite-dimensional convex optimization.
These conditions parallel those in KKT-type first-order optimality conditions in classical nonlinear optimization over Euclidean spaces (see, e.g., \citet[\S 12.3]{nocedal2006numerical} and \citet[\S 3.3.1]{bertsekas1999nonlinear}).
\begin{lemma}
	Let $C = \|B\|_1 - \sum_i B_i \log B_i$. We have
	\begin{enumerate}[label=(\alph*)]
		\item Weak duality: $C + z^* \leq w^*$. \label{item:eg-weak-duality}
		\item KKT conditions: For $x^*$ feasible to \eqref{eq:eg-primal} and $(p^*, \beta^*)$ feasible to \eqref{eq:eg-dual-beta-p}, they are both optimal (i.e., attaining the optima $z^*$ and $w^*$ respectively) if and only if
		\begin{align}
			& \left \langle p^*, \ones - \sum_i x^*_i\right \rangle = 0, \label{eq:mkt-clear} \\
			& \langle v_i, x_i^*\rangle = u^*_i :=  \frac{B_i}{\beta^*_i}, \ \forall\, i,	\label{eq:u*=B/beta*} \\
			& \langle p^* - \beta^*_i v_i, x^*_i \rangle = 0,\ \forall\, i. \label{eq:comp-slack-dual}
		\end{align} \label{item:eg-KKT-iff}
	\end{enumerate} \label{lemma:weak-duality}
\end{lemma}
Thus, we see that in spite of the general difficulties with duality theory in infinite dimensions, we have shown that \eqref{eq:eg-primal} and \eqref{eq:eg-dual-beta-p} behave like duals of each other: strong duality holds, and KKT conditions hold if and only if a pair of feasible solutions are both optimal (see, e.g., \citet[\S 5.2]{nisan2007algorithmic} for the finite-dimensional counterparts). 
Using Lemma \ref{lemma:weak-duality}, we can show the following theorem, which is an expanded version of Part \ref{item:eg-equi-iff-opt} of Theorem~\ref{thm:eg-equi-opt-combined} regarding the equivalence of market equilibrium and optimality w.r.t. the convex programs.
\begin{theorem}
	Assume $x^*$ and $(p^*, \beta^*)$ are optimal solutions of \eqref{eq:eg-primal} and \eqref{eq:eg-dual-beta-p}, respectively. 
	Then $(x^*, p^*)$ is a ME, $\langle p^*, x^*_i \rangle = B_i$ for all $i$, and the equilibrium utility of buyer $i$ is $u^*_i = \langle v_i, x^*_i \rangle = \frac{B_i}{\beta^*_i}$. 
	Conversely, if $(x^*, p^*)$ is a ME, then $x^*$ is an optimal solution of \eqref{eq:eg-primal} and $(p^*, \beta^*)$, where $\beta^*_i := \frac{B_i}{\langle v_i, x^*_i\rangle}$, is an optimal solution of \eqref{eq:eg-dual-beta-p}.
	\label{thm:eg-gives-me}
\end{theorem}
We list some direct consequences of the results we have obtained so far. Below is a direct consequence of Theorem~\ref{thm:eg-gives-me} and Part \ref{item:eg-weak-duality} of Lemma~\ref{lemma:weak-duality} on the structural properties of a market equilibrium.
\begin{corollary}
	Let $(x^*, p^*)$ be a ME. 
	Then, $x^*$ and $(p^*, \beta^*)$, where $\beta^*_i := \frac{B_i}{\langle v_i, x^*_i\rangle}$, satisfy \eqref{eq:mkt-clear}-\eqref{eq:comp-slack-dual}. In particular, \eqref{eq:comp-slack-dual} shows that a buyer's equilibrium allocation $x^*_i$ must be zero a.e. outside its ``winning'' set of items $\{p^* = \beta^*_i v_i \}$.
	\label{cor:me-structiral-properties}
\end{corollary}
	The equilibrium $\beta^*$, or equivalently, the second part of the unique optimal solution $(p^*, \beta^*)$ of \eqref{eq:eg-dual-beta-p}), is often known as the (equilibrium) \emph{utility price}, that is, $\beta^*_i = \frac{B_i}{u^*_i}$ is the price each buyer $i$ pays for a unit of utility. The above corollary shows that, at equilibrium, each buyer $i$ only gets items where its $\beta^*_i v_i$ is the maximum among all buyers, that is, where $p^* = \beta^*_i v_i$. 
	In other words, buyer $i$ only pays for items with the lowest price per unit utility, or equivalently, the most utility per unit price.
	Since $p^* \geq \beta^*_i v_i$, under prices $p^*$, buyer $i$ must pay at least $\beta^*_i$ for each unit of utility.
Given a pure optimal solution $\{\Theta_i\}$ of \eqref{eq:eg-primal}, we can construct the (a.e.-unique) optimal solution $(p^*, \beta^*)$ of \eqref{eq:eg-dual-beta-p}. In particular, such a construction ensures feasibility of $(p^*, \beta^*)$. 
\begin{corollary}
	Let $\{\Theta_i\}$ be a pure optimal solution of \eqref{eq:eg-primal}, $u^*_i = v_i(\Theta_i)$ and $\beta^*_i = \frac{B_i}{u^*_i}$.
	\begin{enumerate}[label=(\alph*)]
		\item For each $i$, we have $\beta^*_i v_i \geq \beta^*_j v_j$ a.e. for all $j\neq i$ on $\Theta_i$. \label{item:thm-eg-pure-solution-gives-dual-feas}
		\item Let $p^* := \max_i \beta^*_i v_i$. Then, $p^*(A) = \sum_i \beta^*_i v_i(A\cap \Theta_i)$ for any measurable set $A\subseteq \Theta$.
		\label{item:thm-eg-pure-solution-gives-p*}
		\item The constructed $(p^*, \beta^*)$ is an optimal solution of \eqref{eq:eg-dual-beta-p} and satisfies \eqref{eq:mkt-clear}-\eqref{eq:comp-slack-dual}. 
		\label{item:thm-eg-pure-solution-gives-dual-opt}
	\end{enumerate}
	 \label{cor:eg-pure-solution-gives-strong-duality}
\end{corollary}
Given a pure allocation, we can also verify whether it is an equilibrium allocation using the following corollary.
\begin{corollary}
	A pure allocation $\{\Theta_i\}$ is an equilibrium allocation (with equilibrium prices $p^*$) if and only if the following conditions hold with $\beta^*_i := \frac{B_i}{v_i(\Theta_i)}$ and $p^* := \max_i \beta^*_i v_i$.
	\begin{enumerate}
		\item Prices of items in $\Theta_i$ are given by $\beta^*_i v_i$: $p^* = \beta^*_i v_i$ on each $\Theta_i$, $i\in [n]$.
		\item Prices of leftover are zero: $p^*(\Theta\setminus (\cup_i \Theta_i)) = 0$.
	\end{enumerate}
	\label{cor:check-pure-alloc-ME}
\end{corollary}

\paragraph{Fairness and efficiency properties of ME.} 
Let $x\in (L^\infty(\Theta)_+)^n$, $\sum_i x_i \leq \ones$ be an allocation. It is (strongly) \emph{Pareto optimal} if there does \emph{not} exist $\tilde{x}\in (L^\infty(\Theta)_+)^n$, $\sum_i \tilde{x}_i \leq \ones$ such that $\langle v_i, \tilde{x}_i \rangle \geq \langle v_i, x_i \rangle$ for all $i$ and the inequality is strict for at least one $i$ \citep{cohler2011optimal}. 
It is \emph{envy-free} (in a budget-weighted sense) if 
\[\frac{1}{B_i}\langle v_i, x_i \rangle \geq \frac{1}{B_j}\langle v_i, x_j\rangle\] 
for any $j\neq i$ \citep{nisan2007algorithmic,kroer2019computing}. 
When all $B_i = 1$, this is sometimes referred to as being ``equitable'' \citep{weller1985fair}. 
It is \emph{proportional} if $\langle v_i, x_i \rangle \geq \frac{B_i}{\|B\|_1} v_i(\Theta)$ for all $i$, that is, each buyer gets at least the utility of its \emph{proportional share} allocation, 
$x^{\rm PS} := \frac{B_i}{\|B\|_1} \ones$. 
Similar to the finite-dimensional case, market equilibria in infinite-dimensional Fisher markets also exhibit these properties.
\begin{theorem}
	Let $(x^*, p^*)$ be a ME. Then, $x^*$ is Pareto optimal, envy-free and proportional. \label{thm:me-is-pareto-ef-prop}
\end{theorem}

\paragraph{ME as generalized fair division.}
By Theorem \ref{thm:me-is-pareto-ef-prop}, a pure ME $\{\Theta_i\}$ under uniform budgets ($B_i = 1/n$) is a fair division in the sense of \citet{weller1985fair}, that is, a Pareto optimal and envy-free division (into a.e.-disjoint measurable subsets) of $\Theta$. 
Furthermore, \citep[\S 3]{weller1985fair} shows that, there exist equilibrium prices $p^*$ such that 
\begin{itemize}
	\item $p^*(\Theta_i) = 1/n$ for all $i$.
	\item $v_i(\Theta_i) \geq v_i(A)$ for any measurable set $A \subseteq \Theta$ such that $p^*(A) \leq 1/n$.
	\item For any measurable set $A\subseteq \Theta$, $p^*(A) = \frac{1}{n}\sum_i \frac{v_i(A\cap \Theta_i)}{v_i(\Theta_i)}$.
\end{itemize}
Utilizing our results, when $B_i = 1/n$, and $\{\Theta_i\}$ is a pure ME, the first property above is a special case of $\langle p^*, x^*_i \rangle = B_i$ in Theorem~\ref{thm:eg-gives-me} (with $x^*_i = \ones_{\Theta_i}$); the second property above follows from the ME property $x^*_i \in D_i(p^*)$; the third property is a special case of Part \ref{item:thm-eg-pure-solution-gives-p*} in Corollary \ref{cor:eg-pure-solution-gives-strong-duality}, since $\beta^*_i = \frac{B_i}{u^*_i} = \frac{1}{n} \cdot \frac{1}{v_i(\Theta)}$. Hence, ME under a continuum of items can be viewed as generalized fair division.


		

\paragraph{Bounds on equilibrium quantities.} 
Using the KKT condition $u^* = \frac{B_i}{\beta^*_i}$ (Lemma \ref{lemma:weak-duality}) and an equilibrium allocation being proportional (Theorem \ref{thm:me-is-pareto-ef-prop}), we can easily establish upper and lower bounds on equilibrium quantities. These bounds will be useful in subsequent convergence analysis of stochastic optimization in \S \ref{sec:sda}. 
Similar bounds hold in the finite-dimensional case~\citep{gao2020first}. 
Recall that we assume $v_i(\Theta) = 1$ and $\|B\|_1 = 1$ w.l.o.g.
\begin{lemma}
	For any ME $(x^*, p^*)$, we have $p^*(\Theta) = 1$. 
	Furthermore, $B_i \leq u^*_i = \langle v_i, x^*_i\rangle \leq 1$ and hence $ B_i \leq \beta^*_i := \frac{B_i}{u^*_i} \leq 1$ for all $i$.
	\label{lemma:equi-bounds}
\end{lemma}

\paragraph{Special cases of discrete item spaces $\Theta$.}
All of the theory developed so far, except the existence of a \emph{pure} equilibrium allocation, holds when $\Theta$ is discrete (finite or \emph{countably} infinite).
This is because, given a discrete $\Theta$, the set of feasible utilities $U = U(v, \Theta)$ defined in \eqref{eq:def-U-U(v,Theta)} is still closed and convex; however, a pure optimal solution of \eqref{eq:eg-primal} may not exist, as it requires $v_i$ to be atomless.
We give more details for the cases of finite and countably infinite item spaces below.
\begin{itemize}
	\item For a countably infinite $\Theta$, we can w.l.o.g. assume $\Theta = \mathbb{N}$ and $\mu(\Theta) = \sum_{\theta\in \Theta} \mu(\theta) = 1$. All subsets $A\subseteq \Theta$ are measurable with measure $\mu(A) = \sum_{\theta\in A} \mu(\theta)$. 
	In this case, a buyer's valuation is a nonnegative summable sequence $v_i \in \ell^1(\Theta)_+$, i.e., $\|v_i\| = \sum_{\theta\in \Theta} v_i(\theta) <\infty$ (we can assume $\|v_i\| = 1$ w.l.o.g., as discussed in \S \ref{sec:inf-dim-fim-setup}). A buyer's allocation is $x_i \in \ell^\infty(\Theta)_+$, i.e., $\sup_{\theta\in \Theta} x_i(\theta) < \infty$. 
	\item For a finite item space $\Theta = [m]$, we can take $\mu(A) := |A|/m$ for all $A\subseteq \Theta$. Buyers' valuations $v_i$ and allocations $x_i$ are nonnegative $m$-dimensional vectors, with $\langle v_i, x_i\rangle = \sum_{j\in [m]} v_{ij} x_{ij}$. 
	Here, to align with the normalization in \S \ref{sec:inf-dim-fim-setup}, we can set the supply of each item $j\in [m]$ to be $s_j = 1/m$ so that the total supply of all items is $\mu(\Theta) = \sum_{j\in [m]} s_j = 1$. To ensure $\|v_i\| = v_i(\Theta) = \sum_{j\in [m]} s_j v_{ij} = 1$ (where the norm is w.r.t. the Banach space with measure $\mu$), which means $v_i(\Theta) = \sum_{j\in [m]} v_{ij} s_j = \frac{1}{m}\|v_i\|_1 = 1$, i.e., $\|v_i\|_1 = m$ (where the norm is the usual Euclidean $1$-norm). 
\end{itemize}

%% file: text/conic_reform_pwl.tex
\section{Tractable Convex Optimization for Piecewise Linear Valuations}
\label{sec:convex-opt-for-pwl}
In this section we show that for the case where the item space is $\Theta = [0,1]$ and the buyers' valuations $v_i$ are piecewise linear functions on $[0,1]$, it is possible to reformulate our infinite-dimensional convex program \eqref{eq:eg-primal} as a finite-dimensional convex conic program. 
This finite-dimensional program can be solved efficiently using off-the-shelf interior-point methods.
Based on the optimal solution of this reformulation, an approximate pure equilibrium allocation can be constructed easily.
This yields a highly practical approach for solving the case of piecewise linear valuations.
Unfortunately, the current theory of interior-point methods does not allows us to immediately conclude that we have a polynomial-time algorithm.
To complement our practical convex conic program, we use the ellipsoid method to 
show that there exists a theoretical algorithm that finds an $\epsilon$-approximate pure equilibrium allocation in time polynomial in $n$, $K$ and $\log \frac{\kappa}{\epsilon}$, where $\kappa$ is the inverse of the smallest buyer budget $\min_i B_i$. 

As discussed in \S\ref{sec:equi-and-dual}, when all $B_i$ are equal we are in the CEEI case, and  a pure equilibrium allocation in which each buyer gets a union of intervals is a fair division in the sense of \citep{weller1985fair}. 
Hence, our method gives a polynomial-time algorithm for finding a fair division of the unit interval under piecewise linear valuations.

The key to our practical convex conic program is to show that when the buyers' valuations $v_i$ are piecewise linear and $\Theta = [0,1]$, the set of feasible utilities $U(v, [0,1])$ defined in \eqref{eq:def-U-U(v,Theta)} can be represented by a small number of simple linear and quadratic constraints with a small number of auxiliary variables.
The first subsection shows this result for a type of normalized linear valuations on the $[0,1]$ interval. The second subsection extends the characterization to general intervals $[a,b]$ and unnormalized linear valuations, which is necessary for handling piecewise linear valuations later. 
The next two subsections show our practical and theoretical algorithmic results, respectively, followed by numerical examples and experiments.



\subsection{Characterization of the set $U(v, [0,1])$ under linear $v_i$} \label{subseq:charact-linear-vi-[0,1]}
We first characterize the set of feasible utilities when each valuation $v_i$ is \emph{linear} over the unit interval $[0,1]$. 
We will show that it can be represented by $O(n)$ linear and quadratic constraints using $O(n)$ auxiliary variables. Before proceeding, we note that, for a nonnegative linear function $v_i: \theta \mapsto c_i \theta + d_i$ on $[0,1]$, the following operations take constant time:
\begin{itemize}
	\item Eval: given $[a,b] \subseteq [0,1]$ the utility of the interval is 
	\[ v_i([a,b]) = \int_a^b v_i(\theta)d\theta = \frac{c_i}{2} (b^2 - a^2) + d_i (b-a). \]
	\item Cut: given $a \in [0,1]$ and $0\leq u_0 \leq v_i([a, 1])$, finding $b\in [a,1]$ such that $v_i([a,b]) = u_0$ amounts to solving a simple quadratic equation:
	\[ v_i([a,b]) = \frac{c_i}{2}(b^2 - a^2) + d_i(b-a) = u_0 \ \Rightarrow\ b = \frac{-d_i + \sqrt{d_i^2 - c_i(c_i a^2 + 2d_ia+2u_0)}}{c_i}.  \]
\end{itemize}
The names ``eval'' (evaluation of the utility of a given interval) and ``cut'' (finding a cut such that the utility of the resulting interval equals a given value) are customary in the cake cutting literature \citep{procaccia2014cake,robertson1998cake}.

\paragraph{Geometry of a pure equilibrium allocation.} 
Recall that, by \cref{cor:me-structiral-properties} and \cref{eq:comp-slack-dual}, if $\{\Theta_i\}$ is a pure equilibrium allocation, then it must hold that $\Theta_i \subseteq \{p^* = \beta^*_i v_i \}$ a.e., that is, each buyer only gets a non-zero allocation in regions where $\beta^*_i v_i$ is the maximum among all buyers. 
When the valuations $v_i$ are linear functions on $[0,1]$, $p^*$ must be a piecewise linear (p.w.l.) function with at most $n$ pieces. 
We will use this to show that, in the case of linear valuations, a pure equilibrium allocation only needs to consist of $(n-1)$ cuts on $[0,1]$, resulting in $n$ intervals, one for each buyer. 
If we are able to compute the equilibrium utilities $u^*_i$ and an \emph{ordering} of the buyers specifying who gets the first interval starting at $0$, who gets the second interval, and so on, then an equilibrium allocation reduces to performing at most $n$ ``cut'' operations to find the exact endpoints of these intervals. 
Such an ordering of the buyers at equilibrium can in fact be determined \emph{a priori}. 
Intuitively, a buyer with a higher valuation on the left of the unit interval should always be assigned items on the left as well. 
This motivates us to consider an ordering based on the magnitudes of each valuation intercept $v_i(0) = d_i$. 
Since equilibrium allocations are invariant under arbitrary scaling of each $v_i$, such an ordering must be independent of the absolute magnitudes of buyers' valuations $\|v_i\|$. 
Hence, we consider normalized valuations such that for each $i$, $\|v_i\| = v_i([0,1]) = 1$, and we assume that the buyer indices are sorted by their intercepts $d_i$ in descending order. 
We also assume that the valuations $v_i$ are distinct. Due to normalization, this is equivalent to the intercepts $d_i$ being distinct. 
\begin{assumption}
	The item space is $\Theta = [0,1]$. 
	The valuation of each buyer $i$ is linear and nonnegative: 
	\[ v_i(\theta) = c_i \theta + d_i \geq 0, \, \theta\in [0,1]. \] 
	The valuations are normalized so that
	\[ v_i(\Theta) = \|v_i\| = \int_0^1 v_i(\theta) d\theta = \frac{c_i}{2} + d_i = 1. \]
	The intercepts of $v_i$ are 
	\[ 2 \geq d_1 > \dots > d_n \geq 0. \] 
	\label{assump:linear-distinct-decreasing-[0,1]}
\end{assumption}
The upper and lower bounds in Assumption \ref{assump:linear-distinct-decreasing-[0,1]} are due to nonnegativity: $v_i(0) \geq 0$ and $v_i(1) \geq 0$ imply $d_i \geq 0$ and $0 \leq c_i + d_i = 2(1-d_i) + d_i \Rightarrow d_i \leq 2$. The following lemma shows that, under the above assumption, the equilibrium prices $p^*$ are p.w.l. with exactly $n$ linear pieces, corresponding to intervals that are the pure equilibrium allocations to the buyers.
\begin{lemma}
	Under Assumption~\ref{assump:linear-distinct-decreasing-[0,1]} and budgets $B_i > 0$, the equilibrium prices $p^* = \max_i \beta^*_i v_i$ are piecewise linear with exactly $n$ linear pieces.
	The (unique) breakpoints of the linear pieces $0 = a^*_0 < a^*_1 < \dots < a^*_n = 1$ induce a pure allocation: buyer $i$ receives $\Theta_i = [a^*_{i-1}, a^*_i]$, $i\in [n]$. This allocation is the unique equilibrium allocation.
	\label{lemma:all-linear-equilibrium-geometry}
\end{lemma}	
\paragraph{Recovering a pure allocation given feasible utilities.} 
Based on Lemma~\ref{lemma:all-linear-equilibrium-geometry}, we can establish Lemma~\ref{lemma:all-linear-represent-any-u} below, which ensures that partitioning the interval into $n$ subintervals is sufficient to attain any feasible utility $u\in U(v, [0,1])$.
A key fact used in the proof of Lemma~\ref{lemma:all-linear-represent-any-u} is a variant of the well-known second welfare theorem for an exchange economy with finitely many divisible items, but for our infinite-dimensional setting.
As a technical contribution, we state and prove a general second welfare theorem for the case of a continuum of items and general $L^1$ buyer valuations, which may be of independent interest. 
It states that any Pareto optimal allocation and its corresponding utilities are equilibrium allocations and equilibrium utilities of a Fisher market for some choice of buyer budgets. One technical challenge in establishing the lemma is the allocation space being a non-Euclidean Banach space. Hence, the set of feasible allocations cannot have a ``tractable'' dual space while having a nonempty interior at the same time (see Remark~\ref{remark:why-define-cp-then-prove}). 
This essentially rules out the use of a separation theorem in the allocation space, a key step in proving the classical finite-dimensional second welfare theorem. 
Instead, the proof relies on the convexity and compactness of the set of feasible utilities and the structure of an infinite-dimensional ME.
\begin{lemma} 
	\label{lemma:pareto-opt-utility-is-equil-of-some-B}
	Let each buyer $i$ have valuation $v_i \in L^1(\Theta)_+$ where $\Theta\subseteq \RR^d$ is a compact set. 
	Assume $v_i(\Theta)>0$ for all $i$.
	Let $u^\circ\in U$ be a Pareto optimal utility vector, that is, there is a Pareto optimal allocation $\{x^\circ_i\}$ such that $u^\circ_i = \langle v_i, x^\circ_i\rangle$. 
	Then, there exists $B \in \RR^n_+$ such that $\{x^\circ_i\}$ is an equilibrium allocation and $u^\circ_i$ are the corresponding equilibrium utilities of a Fisher market with $n$ buyers, each having valuation $v_i$ and budget $B_i$.
\end{lemma}
Using the above two lemmas, when $\Theta = [0,1]$ and $v_i$ are normalized linear valuations with descending intercepts $d_1\geq \dots \geq d_n$, we can show that any feasible utility vector $u\in U(v, [0,1])$ can be attained by a pure allocation of $[0,1]$ consisting of $n$ intervals (some of which can have length zero).
These intervals are allocated from left to right to buyers in the order of $1, \dots, n$.
\begin{lemma}
	Let $v_i(\theta) = c_i \theta+d_i$ be normalized linear valuations on $[0,1]$ (i.e. $ v_i([0,1]) = \frac{c_i}{2}+d_i = 1$) such that $2\geq d_1 \geq \dots \geq d_n \geq 0$. For any $u\in  U(v, [0,1])$, there exists $a_0 = 0 \leq a_1 \leq \dots \leq a_n = 1$ such that $v_i([a_{i-1}, a_i]) \geq u_i$ for all $i$. \label{lemma:all-linear-represent-any-u}
\end{lemma}
Suppose we are given a set of feasible utilities $u \in U(v, [0,1])$ and the valuations $v_i$ are normalized and sorted in descending order of their intercept $d_i$. 
Then, we can find the breakpoints $a_0 = 0\leq a_1 \leq \dots \leq a_n=1$ in Lemma~\ref{lemma:all-linear-represent-any-u} by performing $(n-1)$ ``cut'' operations sequentially from left to right, each time computing a new $a_i\in [a_{i-1}, 1]$ such that $v_i([a_{i-1}, a_i]) = u_i$ for $i=1, \dots, n-1$ (and setting $a_n = 1$ ensures $v_n([a_{n-1}, a_n]) \geq u_n$). Later, we will present a general version of this procedure that handles unnormalized and unsorted linear valuations on an arbitrary interval $[l,h]\subseteq [0,1]$.

\begin{figure}
	\centering
	\includegraphics[scale=0.5]{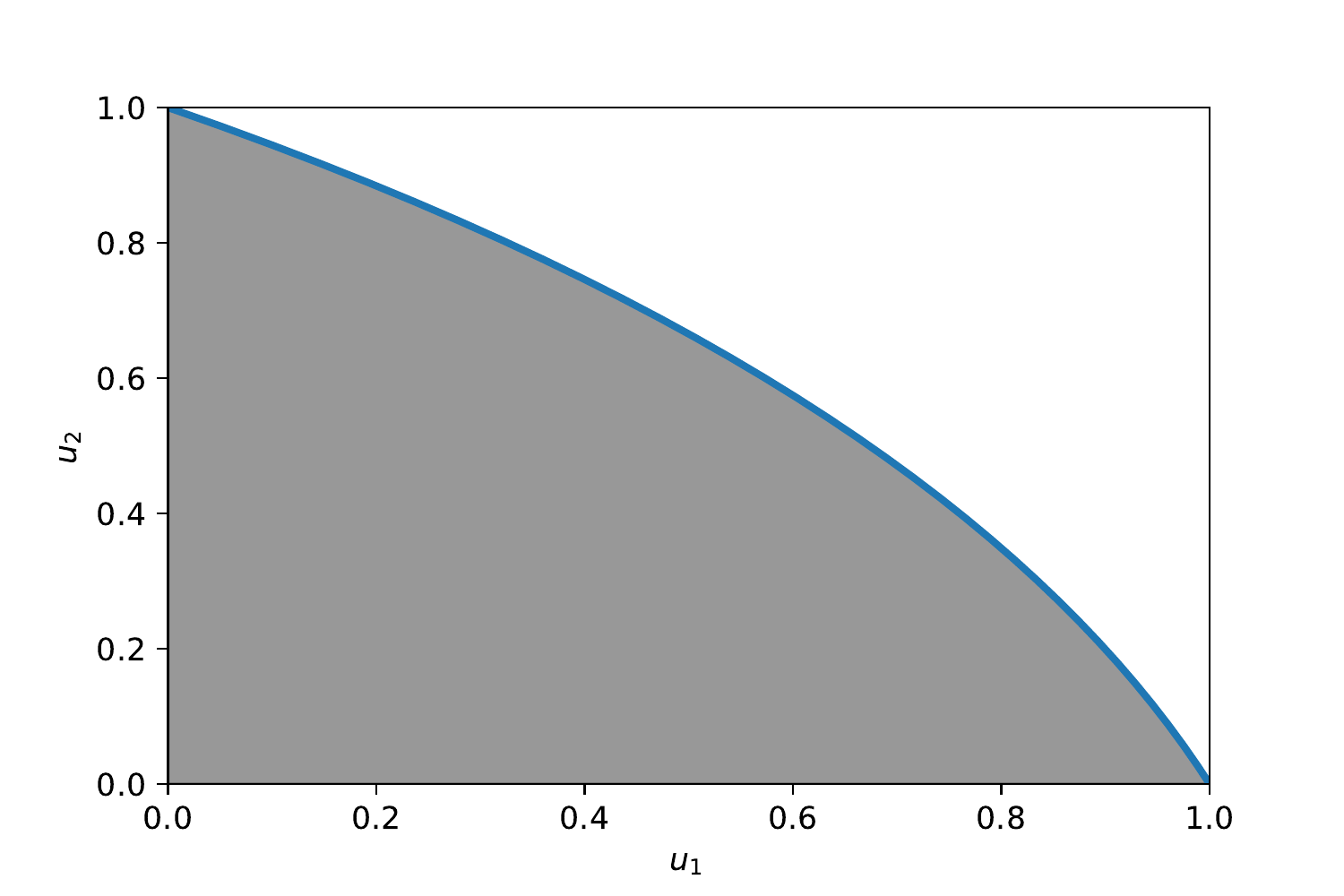}
	\caption{An illustration of the Pareto frontier $U^\circ$ (the curve connecting $(0,1)$ and $(1,0)$) and $U(v, [0,1])$ (the set of feasible utilities, i.e., the shaded region bounded by the Pareto frontier and two axes) in Example~\ref{ex:2-buyers-U}. Here, $d_1 = 1.5$, $d_2 = 0.8$ and $c_i = 2(1-d_i)$, $i=1,2$.}
	\label{fig:U-set-2}
\end{figure}

\paragraph{Convex conic representation of $U(v,[l,h])$.} 
Through the next two theorems, we show that the set $U(v, [l,h])$, where $0\leq l\leq h\leq 1$, can be represented by $O(n)$ number of linear and quadratic constraints using $O(n)$ auxiliary variables. 
We start with the case of sorted and normalized (but not necessarily distinct) $v_i$ on $[0,1]$.
\begin{theorem}
	Let $v_i(\theta) = c_i \theta + d_i$, $\theta\in [0,1]$ and $v_i([0,1]) = \frac{c_i}{2}+d_i = 1$ for all $i$. 
	Let $2\geq d_1 \geq \dots \geq d_n \geq 0$. 
	Denote by $\mathcal{C} = \{ (t_1,t_2)\in \RR^2: t_1^2 \leq t_2 \}$, and let
	\[G_i = \begin{bmatrix}
		d_i & \frac{c_i}{2}\\[1ex]
		-d_{i+1} & -\frac{c_{i+1}}{2}
 	\end{bmatrix}\in \RR^{2\times 2},\ \forall\, i\in [n-1].\] 
	Then, a vector of utilities $u$ is in $U(v, [0,1])$ if and only if it is part of a feasible solution to the following constraints together with auxiliary (real) variables $z_i, w_i, s_i, t_i$:
	\begin{align*}
		& u = (u_1, \dots, u_n) \geq 0,\\
		& u_1 \leq z_1, \\
		& u_i \leq z_i + w_{i-1},\ \forall\, i=2, \dots, n-1,\\
		& u_n \leq 1 + w_{n-1},\\
		& G_i \begin{bmatrix}
			s_i \\ t_i
		\end{bmatrix} = \begin{bmatrix}
			z_i \\ w_i
		\end{bmatrix}, \, \begin{bmatrix}
			s_i \\ t_i
		\end{bmatrix} \in \mathcal{C},\ \forall\, i \in [n-1],\\[1ex]
		& z_i \in [0,1],\, w_i \in [-1,0],\, z_i + w_i \geq 0,\ \forall\, i\in [n-1].
	\end{align*} 
	Furthermore, the above constraints imply $0\leq s_i \leq 1$, $0\leq t_i \leq 1$.
	\label{thm:U-conic-rep}
\end{theorem}
The key observation used in the proof is to express the \emph{Pareto frontier} of utilities as the image of a linear transformation of the product of quadratic curves $\{(t_1, t_2): t_1^2 \leq t_2\}$.
In short, the above theorem shows that  $U$ is the first $n$ dimensions of a convex compact set in $\mathbb{R}^{n+4(n-1)}$ represented by $O(n)$ linear and quadratic constraints. 
Next, we illustrate the derivation of the above characterization via the case of $n=2$ buyers.
\begin{example}[Representation of \text{$U(v, [0,1])$} with $n=2$ buyers] 
	\label{ex:2-buyers-U}
	Under the assumptions of Theorem~\ref{thm:U-conic-rep} (linear, normalized $v_i$ sorted by $d_i$ in descending order), let $2\geq d_1 > d_2 \geq 0$. 
	By Lemma~\ref{lemma:all-linear-represent-any-u}, for any feasible utilities $u \in U(v, [0,1])$, we can find $a\in [0,1]$ such that $u_i \leq v_i([0, a])$, $u_2 \leq v_2([a,1])$.  
	Therefore, 
		\[ U = U(v, [0,1]) = \left\{ (u_1, u_2)\geq 0: \exists\, a \in [0,1]\ \st u_1 \leq \frac{c_1}{2}a^2 + d_1 a,\, u_2 \leq \frac{c_2}{2}(1-a^2) + d_2(1-a)  \right\}. \]
	Clearly, $u\in U$ if and only if $u\geq 0$ and there exist $z_1$, $w_1$, $a$ such that
	\begin{align*}
		& u_1 \leq z_1, \ u_2 \leq 1 + w_1, \\
		& z_1 \leq \frac{c_1}{2} a^2 + d_1 a, \ w_1 \leq -\frac{c_2}{2} a^2 - d_2 a, \\ 
		& a\in[0,1].
	\end{align*}
	Since $a\in [0,1]$ and $\frac{c_i}{2} + d_i = 1$, $i=1,2$, the above inequalities imply the bounds on $(z_1, w_1)$:
	\begin{align*}
		& 0 \leq u_1 \leq z_1 \leq \frac{c_1}{2} + d_1 \leq 1, \\
		& -1 \leq -\frac{c_2}{2} - d_2 \leq w_1 \leq u_2 - 1 \leq \frac{c_2}{2} + d_2 - 1 = 0, \\
		& z_1 + w_1 = (d_1 - d_2) a + \left( (1-d_1) - (1-d_2) \right)a^2 = (d_1 - d_2) (a - a^2) \geq 0,
	\end{align*}
	where the last inequality uses $d_1 \geq d_2$ and $1 \geq a \geq a^2 \geq 0$.
	Therefore, $u\in U$ if and only if there exists $(z_1, w_1)$ and $a$ such that 
	\begin{align}
		\begin{split}
			& u_1 \leq z_1, \ u_2 \leq 1 + w_1, \\
			& z_1 \leq \frac{c_1}{2} a^2 + d_1 a, \ w_1 \leq -\frac{c_2}{2} a^2 - d_2 a, \ a\in[0,1], \\
			& 0\leq z_1 \leq 1, \ -1\leq w_1 \leq 0,\ z_1 + w_1 \geq 0.
			\label{eq:u1-u2-z1-w1-a-in-example}				
		\end{split}
	\end{align}
	Let the Pareto frontier be 
	\[ U^\circ = \left\{ \left( \frac{c_1}{2}a^2 + \frac{d_1}{2}, \frac{c_2}{2}(1-a^2) + \frac{d_2}{2}(1-a) \right): a\in [0,1] \right\}. \]
	For specific values of $d_1, d_2$, the Pareto frontier $U^\circ$ and the set $U$ are illustrated in Figure~\ref{fig:U-set-2}, where $U$ is the region bounded by the axes and $U^\circ$.
	The Pareto frontier is in fact the image under an affine transformation of a parabola:
	\[ U^\circ = (0,1) + \Gamma, \]
	where $\Gamma$ is a (linearly transformed) parabola segment:
	\begin{align*}
		& \Gamma = \left\{ \left( \frac{c_1}{2}a^2 + d_1 a, -\frac{c_2}{2} a^2 - d_2 a \right): a\in [0,1] \right \} = G_1 \{ (s_1, s_1^2): s_1\in [0,1] \}.
	\end{align*}
	The feasible region for $(z_1, w_1)$ given by \eqref{eq:u1-u2-z1-w1-a-in-example} can be described as follows:
	\[ S_1 = \conv(\Gamma) = \conv(\bar{\Gamma}) \cap T_1,\]
	where $\bar{\Gamma}$ is the \emph{entire} curve of $\Gamma$ ($a\in [0,1]$ in the parametric description replaced by $a\in \RR$) and 
	\[ T_1 = \{ (z_1, w_1)\in [0,1]\times [-1,0]: z_1 + w_1 \geq 0 \} \]
	is the triangular region given by the implied linear inequalities in \eqref{eq:u1-u2-z1-w1-a-in-example}.
	Since 
	\[ \bar{\Gamma} = G_1 \{(s_1, s_1^2): s_1 \in \RR \}, \]
	we have 
	\[ \conv(\bar{\Gamma}) = G_1 \conv( \{(s_1, s_1^2): s_1 \in \RR \} ) = G_1 \mathcal{C}. \]
	Hence, the feasible region for $(z_1, w_1)$ can be described by the linear inequalities in $T_1$ and $ (z_1, w_1) = G_1 (s_1, t_1)$, $(s_1, t_1) \in \mathcal{C}$. 
	
	We have arrived at all linear and quadratic constraints given in Theorem~\ref{thm:U-conic-rep} for the case of $n=2$. Furthermore, we can also verify that $S_1 = G_1 \conv( \{(s_1, s_1^2): s_1 \in [0,1] \} )$, where 
	\[ \conv( \{(s_1, s_1^2): s_1 \in [0,1] \} ) = \{ (s_1, t_1): 0\leq s_1 \leq 1,\, 0 \leq t_1 \leq 1,\, s_1^2 \leq t_1 \}. \]
	Therefore, we also know that $0\leq s_1, t_1 \leq 1$.

\end{example}

\subsection{The case of general linear $v_i$}
When $v_i$ are defined on a general interval $[l,h]\subseteq [0,1]$ and are neither normalized nor sorted in any particular order, we can still utilize Theorem~\ref{thm:U-conic-rep} to characterize the set of feasible utilities. 
In this case, $U(v, [l,h])$ is the image of $U(v', [0,1])$ under a simple linear transformation, where $v' = (v'_1, \dots, v'_n)$ is the corresponding set of linear valuations defined on $[0,1]$ that satisfy the assumptions of Theorem~\ref{thm:U-conic-rep}, i.e., normalized and sorted in descending order of their intercepts. 
The linear transformation is the composition of a permutation (represented by a permutation matrix) and a scaling (represented by a diagonal matrix). The following theorem describes the characterization of $U(v, [l,u])$ in detail.
\begin{theorem}
	Let $v_i(\theta) = \frac{c_i}{2}\theta+d_i \geq 0$, $\theta\in [l,h]\subseteq [0,1]$. 
	Assume 
	\[ \Lambda_i = v_i([l,h]) = \frac{c_i}{2}(h^2 - l^2) + d_i(h-l) > 0. \] 
	Denote $\hat{c}_i = (h-l)^2 c_i/ \Lambda_i$, $\hat{d}_i = (h-l) (c_i l + d_i) /\Lambda_i$. Then, for each $i$, the normalized valuation
	\[ \hat{v}_i(\theta) = \hat{c}_i(\theta)+\hat{d}_i\] 
	satisfies $\hat{v}_i([0,1]) = 1$.
	Let $\sigma$ be a permutation of $[n]$ that sorts $\hat{d}_i$ in descending order, i.e., $\hat{d}_{\sigma(1)} \geq \dots \geq \hat{d}_{\sigma(n)}$. 
	Let $P\in \{0,1\}^{n\times n}$ be the (inverse) permutation matrix of $\sigma$ with entries $P_{ij} = \mathbb{I}\{i = \sigma(j)\}$. 
	Let $D\in \RR^{n\times n}$ be a diagonal matrix with $D_{ii} = \Lambda_i := \frac{c_i}{2}(h^2-l^2) + d_i(h-l)$. Then, 
	\begin{align*}
		U(v, [l,h]) = D P \hat{U}^\sigma = \{ DP u: u\in \hat{U}^\sigma \},
	\end{align*}
	where $\hat{U}^\sigma = U(\hat{v}_{\sigma}, [0,1])$ is the set of feasible utilities given (sorted and normalized) valuations $\hat{v}_{\sigma(j)}$, $j \in [n]$ on $[0,1]$ with coefficients $\hat{c}_{\sigma(j)}, \hat{d}_{\sigma(j)}$.
	In other words, for any feasible utility vector $u\in U(v, [l,h])$, there exists $u' \in \hat{U}^\sigma$ such that 
	\[ u_{\sigma(j)} = \Lambda_{\sigma(j)} u'_j,\ \forall\,j\in [n], \]
	and vice versa. 
	\label{thm:U-conic-rep-general-vi}
\end{theorem}
In Theorem~\ref{thm:U-conic-rep-general-vi}, the set $\hat{U}^\sigma$ satisfies the assumptions in Theorem~\ref{thm:U-conic-rep} (i.e., normalized valuations sorted by intercepts) and hence can be represented by $O(n)$ linear and quadratic constraints with $O(n)$ auxiliary variables. Therefore, so is $U(v, [l,h])$, which requires $n$ additional simple linear constraints, each only involving two utility variables $u_{\sigma(j)}$ (for the true utility) and $u'_j$ (for the utility in the permuted and normalized space).

\paragraph{Recovering a pure allocation given feasible utilities (revisited).}
In \S\ref{subseq:charact-linear-vi-[0,1]}, we showed how to find the breakpoints described in Lemma~\ref{lemma:all-linear-represent-any-u} given normalized and sorted $v_i$. 
Algorithm~\ref{alg:interval-partition-given-feasible-u} solves the more general case of unnormalized, unsorted $v_i$ defined on an arbitrary interval $[l,h] \subseteq [0,1]$. 
\begin{algorithm}
	\caption{Partition $[l,h]$ according to $u\in U(v, [l,h])$}
	\textbf{Input:} coefficients $c_i, d_i$ s.t. $v_i(\theta) = c_i\theta + d_i \geq 0$, $\theta\in [l,h]$, vector of feasible utilities $u\in U(v, [l, h])$. \\
	Compute $\Lambda_i$, $\hat{c}_i$, $\hat{d}_i$, $i\in [n]$ as in Theorem~\ref{thm:U-conic-rep-general-vi}. \\
	Sort $\hat{d}_i$ in descending order and let $\sigma$ be a  permutation of $[n]$ such that $\hat{d}_{\sigma(1)} \geq \dots \geq \hat{d}_{\sigma(n)}$. \\
	Set $l_{\sigma(1)} = h_{\sigma(0)} = l$ (define $\sigma(0)  = 0$) and $h_{\sigma(n)} = h$.\\
	\textbf{For} $j = 1, \dots, n-1$:
	\begin{itemize}
		\item Set $i = \sigma(j)$ (buyer $i$ gets the $j$th interval, with left endpoint $l_i = l_{\sigma(j)} = h_{\sigma(j-1)}$).
		\item Find the right endpoint $h_i\in [l_i, h]$ such that $v_i([l_i, h_i]) = u_i$ by solving the following quadratic equation (the ``cut'' operation)
		\[ \frac{c_i}{2}(h_i^2 - l_i^2) + d_i(h_i - l_i) = u_i. \]	 
		\item Set $l_{\sigma(j+1)} = h_i$.
	\end{itemize}
	 \textbf{Return:} $[l_i, h_i]$, $i\in [n]$ s.t. $v_i([l_i, h_i])\geq u_i$, where equality holds for $i=\sigma(1), \dots, \sigma(n-1)$.
	\label{alg:interval-partition-given-feasible-u}
\end{algorithm}
Its running time is clearly $O(n\log n)$ due to the sorting step. 
The correctness of Algorithm~\ref{alg:interval-partition-given-feasible-u} is a direct consequence of Theorem~\ref{thm:U-conic-rep-general-vi} and Lemma~\ref{lemma:all-linear-represent-any-u}. 
In more detail, Lemma~\ref{lemma:all-linear-equilibrium-geometry} shows that we can partition $[0,1]$ according to the transformed valuations $\hat{v}_{\sigma(j)}$ and utility values $u_{\sigma(j)} / \Lambda_{\sigma(j)}$, and then to get an allocation of $[l,h]$ we linearly transform these intervals back to $[l,h]$ and assign the buyers in the same order. 
This can be done by directly partitioning $[l,h]$ in the order of the ordering $\sigma$ according to the scaled intercepts $\hat{d}_i$. 

Next, we give an example that illustrates the representation of $U(v, [l,h])$ in Theorem~\ref{thm:U-conic-rep-general-vi} as well as the execution of Algorithm~\ref{alg:interval-partition-given-feasible-u}. It is taken from Example~\ref{ex:pwl-numerical} in \S~\ref{subsec:numerical-examples} for recovering pure equilibrium allocations: the interval $[l,h]$ corresponds to the second predefined interval $[a_1, a_2]$ of the piecewise linear valuations $v_i$ in that example.
\begin{example}
	[Representation of $U(v,{[l,h]})$ and Algorithm~\ref{alg:interval-partition-given-feasible-u}]
	Consider $n=4$ buyers with valuations $v_i(\theta) = c_i \theta + d_i\geq 0$, $\theta\in [l, h] = [0.3741, 0.8147]$. Here, the coefficients $c_i$ and $d_i$ of the valuations are
	\begin{align*}
		c &= (1.6253, -0.2604, -1.7084, 2.5419),\\
		d & = (-0.2972, 0.4864, 1.3919, 0.6464). 
	\end{align*}
	The normalized coefficients $\hat{c}_i$, $\hat{d}_i$ and normalizing constants $\Lambda_i = v_i([l,h])$, as in Theorem~\ref{thm:U-conic-rep-general-vi}, are
	\begin{align*}
		\hat{c} &= (1.0708, -0.3460, -2.0, 0.5192),\\
		\hat{d} &= (0.4646, 1.1730, 2.0, 0.7404),\\
		\Lambda &= (0.2947, 0.1461, 0.1659, 0.9506).
	\end{align*}
	The order of descending $\hat{d}_i$ is $\sigma = (3, 2, 4, 1)$, i.e., $\hat{d}_3 \geq \hat{d}_2 \geq \hat{d}_4 \geq \hat{d}_1$. 
	Therefore, the sorted arrays of $\hat{c}_i, \hat{d}_i$ are (e.g., $\hat{c}_{\sigma(1)} = \hat{c}_3 = -2.0$):
	\begin{align*}
		\hat{c}_\sigma &= (-2.0, -0.3460, 0.5192, 1.0708),\\
		\hat{d}_\sigma &= (2.0, 1.1730, 0.7404, 0.4646). 
	\end{align*}
	Using $\hat c,\hat d$, we get the transformed valuation $\hat{v}_{\sigma(j)}(\theta) = \hat{c}_{\sigma(j)} \theta + \hat{d}_{\sigma(j)} \geq 0$ for each $j=1,\ldots,4$, with a normalized value such that $\hat{v}_{\sigma(j)}([0,1]) = 1$. The elements of the diagonal matrix $D$ are $ D_{jj} = \Lambda_{\sigma(j)}  = v_{\sigma(j)}([l,h])$, i.e.,
	\begin{align*}
		(D_{11}, D_{22}, D_{33}, D_{44}) = (\Lambda_3, \Lambda_2, \Lambda_4, \Lambda_1) = (0.1659, 0.1461, 0.9506, 0.2947).
	\end{align*}
	The statement $U(v, [l,h]) = DP \hat{U}^\sigma$ in Theorem 
	\ref{thm:U-conic-rep-general-vi} is as follows:
	\begin{align*}
		u\in U(v ,[l,h]) \ \Leftrightarrow \ \exists\, u' \in \hat{U}^\sigma\ \st
		u_3 = \Lambda_3 u'_1, \ u_2 = \Lambda_2 u'_2, \ u_4 = \Lambda_4 u'_3,\ u_1 = \Lambda_1 u'_4.
	\end{align*}
	Here, the set $\hat{U}^\sigma$ can be represented by $O(n)$ linear and quadratic constraints with $O(n)$ auxiliary variables as in Theorem~\ref{thm:U-conic-rep}, since $\hat{v}_{\sigma(j)}$ are normalized on $[0,1]$ and $\hat{d}_{\sigma(1)} \geq \dots \hat{d}_{\sigma(4)}$. Next, suppose we are given $u\in U(v, [l,h])$ and need to find a partition of $[l,h]$ into $4$ intervals that achieve $u_i$. Here, we use $u = (0.0000, 0.0732, 0.0036, 0.5646) \in U(v, [l,h])$ (which is the equilibrium utility achieved on this interval in Example~\ref{ex:pwl-numerical}). The steps of Algorithm~\ref{alg:interval-partition-given-feasible-u} are as follows. 
	\begin{itemize}
		\item The allocation order is $\sigma = (3, 2, 4, 1)$, i.e. decreasing in $\hat{d}_i$.
		\item Since $\sigma(1) = 3$, set $l_3 = l$ and find $h_3 \in [l,h]$ such that $v_3([l,h_3]) = u_3 = 0.0036$. This gives $[l_3, h_3] = [0.3741, 0.3789]$. Set $l_{\sigma(2)} = l_2 = h_3 = 0.3789$. 
		\item For $\sigma(2) = 2$ we find $h_2 \in [l_2, h]$ such that $v_2([l_2, h_2]) = u_2$, which gives $[l_2, h_2] = [0.3789, 0.5815]$.
		\item Similarly, the next buyer is $\sigma(3) = 4$ and $[l_4, h_4] = [0.5815, 0.8147]$. 
		\item The last buyer is $\sigma(4) = 1$, which gets an empty interval ($l_1 = h_1 =  h = 0.8147$), resulting in zero utility $u_1 = 0$. 
	\end{itemize}
\end{example} 

\subsection{Convex conic reformulation of \eqref{eq:eg-primal}} \label{subsec:convex-conic-reform-of-EG-pwl}
We now show how to handle the general case of piecewise linear valuations on $[0,1]$.
We first give a convex program whose variables are the utilities buyers receive from the subintervals defined by their linear pieces.
Formally, the item space is $\Theta = [0,1]$ and each $v_i$ is a piecewise linear valuation on $[0,1]$. Let the union of their breakpoints be $a_0 = 0 \leq a_1 \leq \dots \leq a_{K-1} \leq a_K = 1$. 
For each buyer $i\in [n]$ and each subinterval $k\in [K]$, $v_i$ is linear on $[a_{k-1}, a_k]$:
\[ v_i(\theta) = c_{ik} \theta + d_{ik},\ \ \theta \in [a_{k-1}, a_k]. \]
For each $k$, let the set of feasible utilities with item space $[a_{k-1},a_k]$ and valuations $v_i$ be \[ U_k := U(v, [a_{k-1}, a_k]),\] 
as defined in \eqref{eq:def-U-U(v,Theta)}. 
Consider the following convex program, whose variables denote how much utility each buyer $i$ receives from each linear segment $k$:
\begin{align}
	\begin{split}
			\sup_{(u_{ik}) \in \RR_+^{n\times K} } & \sum_{i=1}^n B_i \log \left(\sum_{k=1}^K u_{ik}\right) \\
		\st & (u_{1k}, \dots, u_{nk}) \in U_{k},\ \forall\,k \in [K].
	\end{split}
	\label{eq:convex-prog-u(ik)}
\end{align}
By Theorem~\ref{thm:U-conic-rep-general-vi}, the set $U_k$ is the image of a permutation and a scaling of another set of feasible utilities spanned by normalized valuations on $[0,1]$:
\[ U_k = D^k P^k \hat{U}_k.\]
Here, $\hat{U}_k$ is the set of feasible utilities spanned by $\hat{v}_{ik}$, where $v_{ik}$ is the restriction of $v_i$ on interval $[a_{k-1}, a_k]$ and $\hat{v}_{ik}$ (defined on $[0,1]$) are the normalized and sorted versions of $v_{ik}$ as described in Theorem~\ref{thm:U-conic-rep-general-vi}. 
We will adopt the notation of Theorem~\ref{thm:U-conic-rep-general-vi}, but since we need a set $\hat U_k$ for each piece $k$, we use an additional subscript $k$ to refer to the $k$th copy corresponding to the subinterval $[l,h] = [a_{k-1}, a_k]$.
Thus, $D^k$ is a diagonal matrix with diagonal entries $\Lambda_{ik} = v_i([a_{k-1}, a_k])$ (this corresponds to $\Lambda_i$ in Theorem~\ref{thm:U-conic-rep-general-vi} with $[l,h] = [a_{k-1}, a_k]$), $P^k$ is a permutation matrix corresponding to the permutation $\sigma^k$ that sorts $\hat{d}_{ik}$ (this corresponds to the intercepts $\hat{d}_i$ in Theorem~\ref{thm:U-conic-rep-general-vi}) in descending order. 
Both $D^k$ and $P^k$ depend on $(c_{ik}, d_{ik})$. 
The set $\hat{U}_k$ is the set of feasible utilities of normalized valuations as given in Theorem~\ref{thm:U-conic-rep}. 
We will use $(s_{ik}, t_{ik}, w_{ik}, z_{ik}, \hat{u}_{ik})$ to denote the  variables $(s_i, t_i, w_i, z_i, u_i)$ in Theorem~\ref{thm:U-conic-rep} corresponding to $\hat U_k$.

Next we will describe how the convex program~\ref{eq:convex-prog-u(ik)} can be solved efficiently in practice using industry-grade interior-point methods.
To that end, denote the $3$-dimensional second-order (quadratic) cone as $\mathcal{L} = \{ (t_1, t_2, t_3): t_1 \geq \sqrt{t_2^2+t_3^2} \}$ and the $3$-dimensional exponential cone $\mathcal{E}$ as the \textit{closure} of the following set (see, e.g., \citep{chares2009cones}, \citep{dahl2019primal}, \citep{serrano2015algorithms}):
\[ \mathcal{E}  = \left\{ (t_1, t_2, t_3): e^{t_3/t_2} \leq t_1/t_2,\, t_2 >0 \right\}. \]
Define the following standard-form convex conic program \citep{skajaa2015homogeneous,dahl2019primal,nemirovski2004interior}:
\begin{align}
	\begin{split}
		f^* = \min\ & c^\top x \\
		 \st & Ax = b,\, \\
		 	 & x\in \mathcal{K}:= \RR_+^{n_1} \times \mathcal{L}^{n_2} \times \mathcal{E}^{n_3}
	\end{split} \label{eq:conic-std-form}
\end{align}
where $A\in \RR^{m\times \bar{n}}$, $c\in \RR^{\bar{n}}$, $b\in \RR^m$, $\bar{n} = n_1 + 3n_2 + 3n_3$ (we use $\bar{n}$ to denote the dimension to distinguish it from $n$, the number of buyers). 
Problem \eqref{eq:convex-prog-u(ik)} can be reformulated into \eqref{eq:conic-std-form} via standard techniques. 
The following lemma summarizes the facts about the said reformulation. 
In additional to the small dimensions $\bar{n} = O(nK)$, $m=O(nK)$, the reformulation also ensures ${\rm nnz}(A) = O(\bar{n})$, where ${\rm nnz}(A)$ denotes the number of nonzeros in the matrix $A$. A complete convex conic reformulation can be found in the proof of the lemma.
\begin{theorem}
	The supremum of \eqref{eq:convex-prog-u(ik)} is attained and is equal to $z^*$, the supremum of \eqref{eq:eg-primal}.
	For any optimal solution $(u^*_{ik})$ of \eqref{eq:convex-prog-u(ik)}, $u^*_i := \sum_{k=1}^K u^*_{ik}$ is the equilibrium utility of buyer $i$. 
	Each interval $[a_{k-1}, a_k]$ can be divided into $n$ a.e.-disjoint subintervals $[l_{ik}, h_{ik}]$ such that $u^*_{ik} = v_i([l_{ik}, h_{ik}])$. Hence, $\Theta_i := \cup_{k=1}^K [l_{ik}, h_{ik}]$, $i\in[n]$ is an equilibrium allocation. 
	Problem \eqref{eq:convex-prog-u(ik)} can be reformulated into the standard form \eqref{eq:conic-std-form} with dimensions $n_1 = O(nK)$, $n_2 = O(nK)$, $n_3 = O(n)$, $m = O(nK)$ and ${\rm nnz}(A) = O(nK)$. The minimum of the reformulation is $f^* = -z^*$.
	An optimal solution of the reformulation contains an optimal solution of \eqref{eq:convex-prog-u(ik)}, that is, $(u_{ik})\in \RR_+^{n\times K}$ such that $(u_{1k}, \dots, u_{nK})\in U_k$ for all $k$ and $\sum_k u_{ik} = u^*_i$ for all $i$.
 	 \label{lemma:eg=>u(ik)-facts}
\end{theorem}

\paragraph{Solving \eqref{eq:conic-std-form} using an interior-point method.}
The standard-form problem \eqref{eq:conic-std-form}, in which $\mathcal{K}$ is the product of a nonnegative orthant, second-order cones, and exponential cones, can be solved via off-the-shelf optimization software based on interior-point methods even for very large instances using the \emph{Mosek} solver~\citep{mosek2010mosek,dahl2019primal}. In fact, modern optimization software usually do not require such a standard-form input and allows more general input formats.
Although the theoretical time complexity of interior-point methods for a general convex optimization problem with computable self-concordant barrier functions has been well-studied (see e.g. \citep{nesterov1994interior}, \citep[\S 5]{nesterov2018lectures}, \citep[\S 4]{nemirovski2004interior}), to the best of our knowledge, clear-cut polynomial-time complexity results are only available for the case of \emph{self-scaled} cones, that is, linear programming (LP), second-order cone programming (SOCP) and semidefinite programming (SDP).

For general convex optimization beyond self-scaled cones, there exist bounds on the number of Newton iterations that are roughly of the form ``$O\left(\sqrt{\nu}\log \frac{M}{\epsilon}\right)$''---where $\nu$ is the barrier parameter, $\epsilon$ is the tolerance level (for duality gap and infeasibility) and $M$ is an instance-dependent constant---for ``theoretical versions'' of various interior-point methods applied to strictly feasible problems (see, e.g., \cite[\S3 and \S6]{nesterov1994interior}, \cite[5.3.4]{nesterov2018lectures}, \cite[\S 4 and \S 7]{nemirovski2004interior}).
In the various forms of this type of bound, the constant $M$ depends critically on the geometry of the problem and the instance data, such as closeness of the initial solution to the boundary of the feasible region~\citep[Theorems~4.5.1 and 7.4.1]{nemirovski2004interior}. 
Whether we can extend the polynomial time complexity results for self-scaled cones to the case of exponential cones is beyond the scope of this work. 
For our purpose, it remains a challenge to bound this constant using the market data $B_i$, $c_{ik}$, $d_{ik}$ as well.  
Furthermore, we point out that mature interior-point optimization software rarely implements all components of a theoretically-convergent method. 
Instead, highly sophisticated numerical linear algebra methods and stepsizing heuristics are used to speed up and stabilize the computation of search directions (see e.g. \citep{toh2012implementation,sturm1999using}). These techniques, necessary for practically efficient implementations, often invalidate the theoretical complexity guarantees \citep{dahl2019primal,skajaa2015homogeneous}. 
In conclusion, at this time one cannot derive polynomial-time solvability of the piecewise linear problem directly from our conic reformulation in~\eqref{eq:conic-std-form}.
Nevertheless, we remark that solving the reformulation~\eqref{eq:conic-std-form} of~\eqref{eq:convex-prog-u(ik)} using industry-grade interior-point optimization software (such as Mosek) is a highly efficient and stable approach for computing a pure equilibrium allocation: Mosek easily solves problems with hundreds of variables and constraints.
After computing an optimal solution $(u^*_{ik})$ of \eqref{eq:convex-prog-u(ik)} using Mosek, a pure allocation that attains these utilities can easily be found with Algorithm~\ref{alg:interval-partition-given-feasible-u}. 


\subsection{A polynomial-time algorithm for computing equilibrium allocations} \label{subsec:compute-pure-eq-alloc-poly-time}
Due to the problems mentioned in the previous section regarding polynomial-time solvability of conic programs involving exponential cones, we next investigate polynomial-time solvability of our problem via alternative methods.
The results in this section are primarily of theoretical interest; in practice the conic program from the previous section is extremely efficient and preferable.

We are going to present an algorithm that finds an $\epsilon$-approximate pure equilibrium allocation in $\poly\left(n,k,\log \frac{1}{\epsilon}\right)$ time.
This method computes approximate equilibrium utility prices using the ellipsoid method for convex optimization and constructs a pure allocation via Algorithm~\ref{alg:interval-partition-given-feasible-u}. 

Recall our assumptions that $\|B\|_1 = 1$ and $v_i([0,1]) = 1$ for all $i$; the unit interval is divided into $K$ subintervals by breakpoints $a_0 = 0 < a_1 < \dots < a_K = 1$; for each $i$, buyer $i$'s valuation is $v_i(\theta) = c_{ik}\theta+d_{ik}$ on the $k$th subinterval $[a_{k-1}, a_k]$. 
\paragraph{The ellipsoid method for convex optimizaton.}
First, we describe the ellipsoid method for generic convex optimization \citep{shor1977cut,yudin1976informational,yudin1976evaluation,Nemirovski1977optimization}. 
We refer the readers to the survey \citep{bland1981ellipsoid} for the history of development of ellipsoid methods and further references. 
Here, we adopt the exposition in \citep{ben2019lectures}.
Consider the following generic convex program \citep[\S 4.1.4]{ben2019lectures}:
\begin{align}
	f^*:= \min_x f(x) \ \st x\in X \label{eq:generic-constrained-minimization}
\end{align}
where $f$ is convex and continuous (and hence subdifferentiable) on a compact region $X\subseteq \RR^n$.
Assume we have access to the following oracles:
\begin{itemize}
	\item The \emph{separation} oracle $\mathcal{S}$: given any $x\in \RR^n$, either report $x\in \interior X$ or return a $g\neq 0$ (representing a separating hyperplane) such that $\langle g, x\rangle \geq \langle g, y\rangle$ for any $y\in X$. 
	\item The \emph{first-order} or \emph{subgradient} oracle $\mathcal{G}$: given $x\in \interior X$ (the interior of $X$), return a subgradient $f'(x)$ of $f$ at $x$, that is, $f(y) \geq f(x) + \langle f'(x), y-x\rangle$ for any $y$.
\end{itemize} 
The time complexity of the ellipsoid method is as follows.
\begin{theorem} \cite[Theorem 4.1.2]{ben2019lectures}
	Let 
	\[V = \max_{x\in X} f(x) - f^*,\ \ R = \sup_{x\in X} \|x\|,\] and $r>0$ be the radius of a Euclidean ball contained in $X$.
	For any $\epsilon>0$, a solution $x_\epsilon\in X$ such that $f(x_\epsilon) \leq f^* + \epsilon$ can be computed using no more than $N(\epsilon)$ calls to $\mathcal{S}$ and $\mathcal{G}$, followed by no more than $O(1)n^2 N(\epsilon)$ arithmetic operations to process the outputs of the oracles, where 
	$N(\epsilon) = O(1) n^2 \log \left( 2+ \frac{V R}{\epsilon r} \right)$.
	\label{thm:4.1.2-ellipsoid-bn-notes}
\end{theorem}
Using the above theorem, we can show that problem \eqref{eq:convex-prog-u(ik)} can be solved in polynomial time. 
\begin{theorem}
	For any $0 < \epsilon < 1$, we can compute $(u_{ik})\in \RR_+^{n\times K}$ such that 
	\begin{align*}
		& (u_{1k}, \dots, u_{nk})\in U(v, [a_{k-1}, a_k]),\ \forall\, k\in [K],\\ 
		& u_i := \sum_i u_{ik} \geq u^*_i - \epsilon, \ \forall\, i \in [n]
	\end{align*}
	in $O\left( n^4 K \left( \log(nK) + \log \frac{\kappa}{\epsilon} \right) \right)$ time, where $\kappa = \frac{1}{\min_i B_i}$.
	Furthermore, a pure equilibrium allocation $\{\Theta_i\}$, where buyer $i$ receives an allocation $\Theta_i = \cup_k [l_{ik}, u_{ik}]$ of at most $K$ intervals,
	with value $v_i(\Theta_i) \geq u^*_i - \epsilon$,
	can be constructed in $O(nK)$ additional time. 
	\label{thm:ellipsoid-method-for-u(ik)-convex-program}
\end{theorem}
At a high level, in order to use Theorem \ref{thm:4.1.2-ellipsoid-bn-notes} to solve our problem \eqref{eq:convex-prog-u(ik)} in polynomial time, we need to cast it into the form \eqref{eq:generic-constrained-minimization}, construct efficient separation and first-order oracles, and bound the ratio $\frac{VR}{\epsilon r}$. To this end, all variables involved have absolute values bounded above by either absolute constants or $\kappa$. In order to ensure a nonzero radius $r>0$ of the feasible region, we can simply ``$\epsilon$-perturb'' the linear constraints and then ``$\epsilon$-discount'' the solution obtained to ensure feasibility w.r.t. the original constraints.

We remark that the constant $\kappa$ in Theorem~\ref{thm:ellipsoid-method-for-u(ik)-convex-program} can be viewed as a ``condition number'' of problem \eqref{eq:convex-prog-u(ik)}: given a fixed accuracy level $\epsilon$, the running time of the algorithm scales logarithmically (via the term $\log \kappa$) as $\kappa$ grows. 
This aligns with our intuition about a ``second-order'' method such as an interior-point method or the ellipsoid method. 
In contrast, the running time of a first-order method---such as projected gradient descent---usually scales polynomially in the problem's condition number.

\subsection{Numerical examples and experiments} \label{subsec:numerical-examples}
To round out this section, we describe two specific examples of computing a pure equilibrium over $[0,1]$ given linear and piecewise linear valuations, respectively. 
Then, we run our proposed method end-to-end on randomly generated large instances to demonstrate its scalability. 
More details can be found in Appendix~\ref{app:more-on-numerical}. 

\begin{figure}
	\centering
	\includegraphics[scale=0.5]{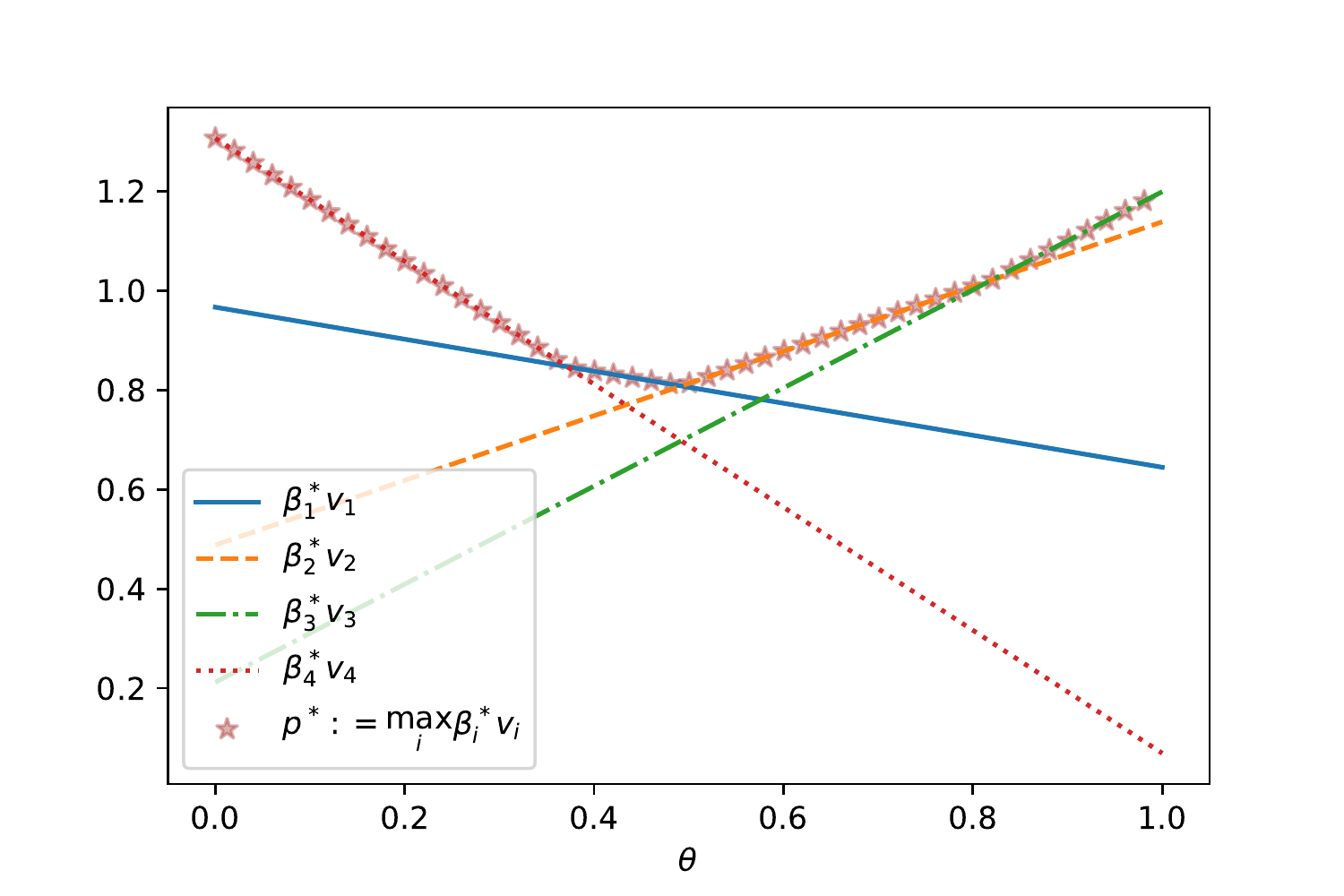}
	\caption{The equilibrium prices $p^*$ and $\beta^*_i v_i$ for $4$ buyers with linear (normalized) $v_i$ on $[0,1]$ (Example \ref{ex:linear-numerical}). 
	The stars denote $p^*$, whose linear pieces correspond to the winning segments for each buyer.
	}
	\label{fig:n-linear}
\end{figure}

\begin{example}
	[Linear $v_i$]
	\label{ex:linear-numerical}
	Let there be four buyers with budgets $B = (0.1, 0.3, 0.2, 0.4)$ and normalized linear $v_i$ with intercepts $d = (1.2, 0.6, 0.3, 1.9)$, which implies $c = (-0.4,  0.8,  1.4, -1.8)$. Ordering the buyers by decreasing intercept gives $\sigma = (4, 1, 2, 3)$. 
	Figure~\ref{fig:n-linear} illustrates the equilibrium prices $p^*$ and the scaled valuation $\beta^*_i v_i$ for each buyer. Buyer $4$ receives the leftmost interval $[0, 0.3713]$, buyer $1$ receives $[0.3713, 0.4921]$, and so on. 
	For each buyer, its allocated interval $[l_i,h_i]$ is precisely the segment where it ``wins'', i.e., $[l_i, h_i] = \{p^* = \beta^*_i v_i\}$. 
	Since all buyers have distinct $d_i$ and positive budgets, this is the unique pure allocation. The fact that this pure allocation is indeed an equilibrium allocation (with the corresponding $\beta^*$) can also be verified using Corollary~\ref{cor:check-pure-alloc-ME}.
\end{example}
%
%

\begin{figure}
	\centering
	\includegraphics[scale=0.5]{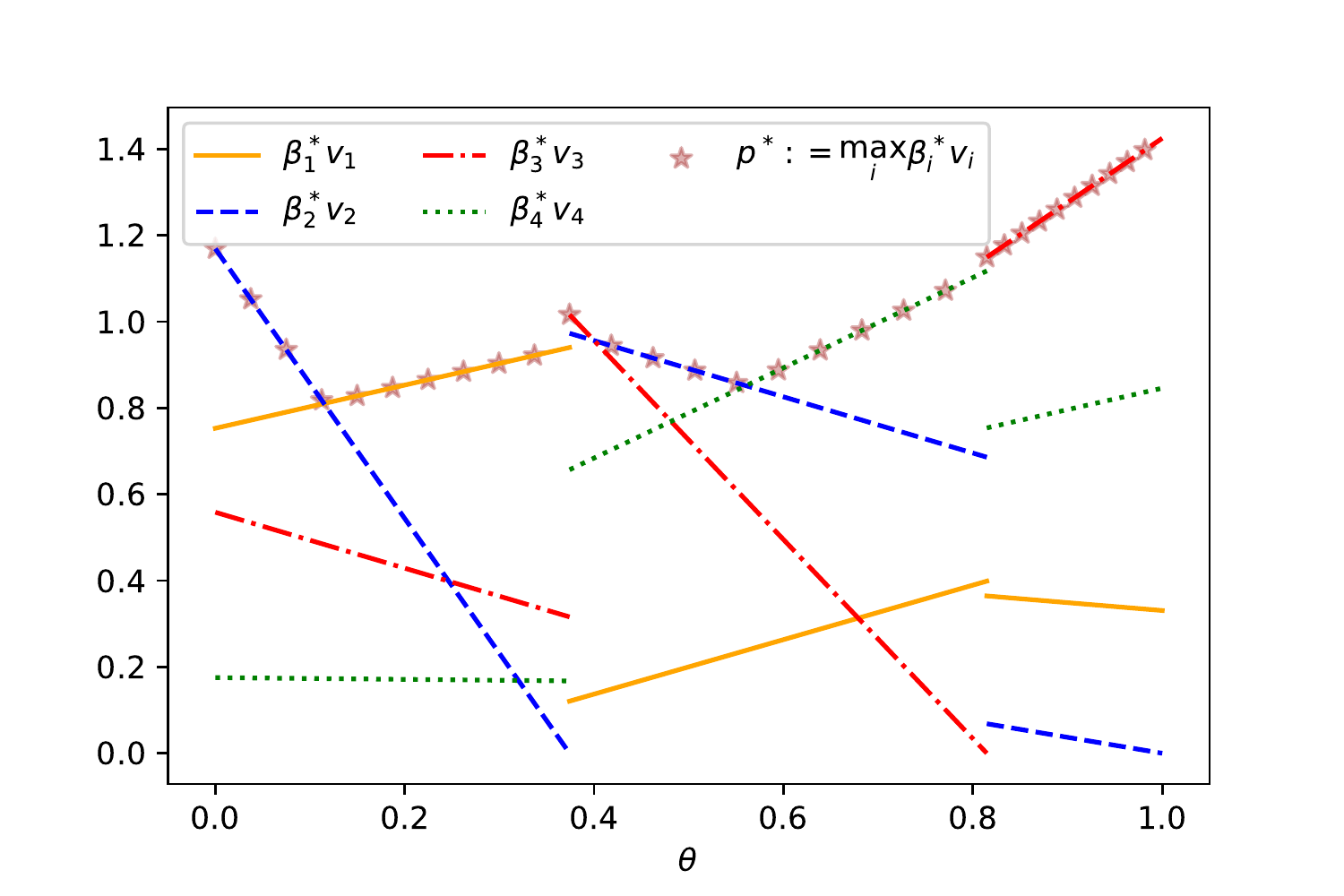}
	\caption{The equilibrium prices $p^*$ and $\beta^*_i v_i$ for $n=4$ buyers with piecewise linear $v_i$ on $[0,1]$ (Example \ref{ex:pwl-numerical}). The stars denote $p^*$, whose linear pieces correspond to the winning segments for each buyer.
	Note that each $v_i$, and hence $p^*$, are p.w.l. They are not necessarily continuous.}
	\label{fig:n-pwl}
\end{figure}

\begin{example}
	[Piecewise linear $v_i$] \label{ex:pwl-numerical}
	Let there be $n=4$ buyers, each with piecewise linear valuations $v_i$ with $K = 3$ pieces. Each buyers' linear pieces share endpoints $0 = a_0 < a_1 < a_2 < a_3 = 1$, with $a_1 = 0.3741$, $a_2 = 0.8147$. 
	Solving the convex program \eqref{eq:convex-prog-u(ik)} gives equilibrium utilities $u^*_{ik}$ for all $i,k$, where $u^*_{ik}$ is the amount of utility buyer $i$ gets from segment $[a_{k-1}, a_k]$ (which can be $0$ for some buyer-segment pairs). 
	Similarly to the linear case above, we divide each segment  $[a_{k-1}, a_k]$ among the buyers according to their $u^*_{ik}$ and their ordering according $\hat{d}_i$  on interval $k$ (c.f. Lemma~\ref{lemma:all-linear-represent-any-u}).
	In the final pure allocation, each buyer $i$ gets a union of at most $K$ intervals, one from each $[a_{k-1}, a_k]$, which is a subset of its winning set $\{p^* = \beta^*_i v_i\}$. 
	The solution is illustrated in Figure~\ref{fig:n-pwl} using $\beta^*_i v_i$ and $p^*$.
	For example, on $[a_0, a_1]$, since $p^* > \beta^*_3 v_3$ everywhere on the segment, buyer $3$ does not get allocated anything from this interval, and thus $u^*_{31} = 0$. The same is true for buyer $4$ on $[a_0, a_1]$. 
\end{example}

\paragraph{Large-scale experiments.} 
Next, we generate random instances with varying values of $n$ and $K$.
For each $(n,K)$ combination and each random seed value in $\{ 1, \dots, 8\}$, we perform the conic reformulation, solve the resulting convex program using Mosek and construct pure allocations based on the solution $(u^*_{ik})$.
For each $(n,K)$ combination, we record the mean running time and the standard error.
The results are presented in Table \ref{table:running-times-n-K-seeds}. 
As can be seen, for fixed $n$, the running time scales approximately linearly in $K$, and similarly running time scales linearly with $n$ for a fixed $K$. Thus our model should be scalable to very large instances.
We remark that, in the experiment, around $2/3$ of the running time arises from ``model building'', that is, constructing the sparse matrices and interacting with Mosek through its API, even after writing a highly vectorized implementation. 
Only around $1/3$ of the time arises from ``optimization'', that is, running the interior-point method through \texttt{Model.solve()}. The time for constructing pure allocations from $(u^*_{ik})$ is negligible compared to model building and computation. As an additional note for practitioners, we also tried the CVXPY modeling language: it was vastly slower at reformulating the model before calling Mosek.

\begin{table}
	\begin{center}
		\input{codes/large-scale.tex}
		\caption{Running times for each $(n,K)$, mean and standard error across $8$ seeds} 
		\label{table:running-times-n-K-seeds}
	\end{center}
\end{table}

%% file: codes/large-scale.tex
\begin{tabular}{|c||*{4}{c|}}\hline
\backslashbox{$n$}{$K$}
& $50$ & $100$ & $150$ & $200$\\ \hline \hline
50 & $1.53 \pm 0.05$ & $3.29 \pm 0.14$ & $5.57 \pm 0.91$ & $6.49 \pm 0.18$\\ \hline
100 & $3.34 \pm 0.14$ & $6.50 \pm 0.20$ & $10.36 \pm 0.40$ & $13.62 \pm 0.48$\\ \hline
150 & $4.84 \pm 0.12$ & $9.86 \pm 0.18$ & $16.68 \pm 1.46$ & $20.50 \pm 0.22$\\ \hline
200 & $6.44 \pm 0.07$ & $13.18 \pm 0.23$ & $20.12 \pm 0.33$ & $28.65 \pm 1.14$\\ \hline
\end{tabular}

%% file: text/sda.tex
\section{Stochastic optimization for general item spaces and valuations}\label{sec:sda}
Here, we consider the case of general valuations $v_i$ on a convex compact set $\Theta\subseteq \RR^d$ and show that we can use stochastic first-order methods to find approximate equilibrium utility prices. 

\paragraph{Efficient subgradient computation.} 
Our method will rely on oracle access to stochastic subgradients of $\phi(\beta) = \langle \max_i \beta_i v_i, \ones \rangle$, the first term in the objective of \eqref{eq:eg-dual-beta-1}.
From the proof of Lemma \ref{lemma:beta-dual-attain}, we know that the function $\phi$ is finite, convex and continuous on $\RR_+^n$. 
Hence, it is subdifferentiable on $\RR^n_{++}$ \cite[Proposition C.6.5]{ben2019lectures}. 
The next lemma shows that  $\phi$ can be viewed as the expectation of a stochastic function, and that an unbiased stochastic subgradient of $\phi$ can easily be computed. 
\begin{lemma}
	Let $\phi(\beta, \theta) = \max_i \beta_i v_i(\theta)$. For any $\theta\in \Theta \subseteq \RR^d$, a subgradient of $\phi(\cdot, \theta)$ at $\beta$ is $ g(\beta, \theta) = v_{i^*}(\theta) \mathbf{e}^{(i^*)}$, where $i^* \in \argmax_i \beta_i v_i(\theta)$ (taking the smallest index if there is a tie). Hence, a subgradient of $\phi$ at $\beta$ is $\phi'(\beta) = \int_{\Theta} g(\beta, \theta) d\theta = \mu(\Theta) \cdot \EE_\theta g(\beta, \theta) \in \partial \phi(\beta)$, where the expectation is taken over the uniform distribution $\theta\sim {\rm Unif}(\Theta)$.  
	\label{lemma:subgrad-w.r.t.-beta-fixed-theta}
\end{lemma}
Using Lemma \ref{lemma:subgrad-w.r.t.-beta-fixed-theta}, we can solve \eqref{eq:eg-dual-beta-1} using a stochastic first-order method that only requires oracle access to stochastic subgradients.
The structure of this problem is particularly suitable for the \textit{stochastic dual averaging} (SDA) algorithm \citep{xiao2010dual,nesterov2009primal}. 
It solves problems of the form:
\begin{align}
	\min_{\beta} \ \EE_\theta f(\beta, \theta) + \Psi(\beta), \label{eq:sda-reg-std-form}
\end{align}
where $\Psi$ is a convex function---often known as a \emph{regularizer} in the context of machine learning---with a closed, nonempty domain ${\rm dom}\, \Psi = \{ \beta: \Psi(\beta) < \infty \}$.
Now, assume that $\Psi$ is strongly convex, $\theta \sim \mathcal{D}$ is a random variable with distribution $\mathcal{D}$, and $f(\cdot, \theta)$ is convex and subdifferentiable on ${\rm dom}\,\Psi$ for all $\theta\in\Theta$. 
The algorithm is shown in \cref{alg:sda}.
\begin{algorithm}
	Initialize: Choose $\beta^1 \in {\rm dom}\, \Psi$ and $\bar{g}^0 = 0$\\
	\noindent\it{For} $t=1,2, \dots$:
	
		\hspace{10pt} Sample $\theta_t \sim \mathcal{D}$ and compute $g^t \in \partial_\beta f(\beta, \theta_t)$
		
		\hspace{10pt} $\bar{g}^t = \frac{t-1}{t}\bar{g}^{t-1} + \frac{1}{t}g^t$
		
		\hspace{10pt} $\beta^{t+1} = \argmin_\beta \left\{ \langle \bar{g}^t, \beta \rangle + \Psi(\beta) \right\}$ $\ (*)$
	\caption{Stochastic dual averaging (SDA)}\label{alg:sda}
\end{algorithm}

To solve \eqref{eq:eg-dual-beta-1}, we set 
\[ f(\beta, \theta) = \max_i \beta_i v_i(\theta),\ \mathcal{D} = {\rm Unif}(\Theta),\] 
where we assume $\mu(\Theta)=1$ w.l.o.g. (otherwise, we can ``shrink'' the item space $\Theta$ by a scalar and consider $\Theta' = \{\alpha\theta: \theta\in \Theta\}$ or multiply the underlying measure $\mu$ by a scalar between $0$ and $1$; valuations $v_i$ are then replaced by $v'_i(\theta') = v_i(\alpha\theta)$).

Then, $\EE_\theta[ \max_i\beta_i v_i(
\theta)] = \langle \max_i \beta_i v_i, \ones\rangle$. By Lemma \ref{lemma:subgrad-w.r.t.-beta-fixed-theta}, we can choose \[g^t = g(\beta, \theta_t) \in \partial_\beta f(\beta, \theta_t).\]
Recall the bounds on $\beta^*$ in Lemma \ref{lemma:equi-bounds}: 
\[ \ubar{\beta}_i = B_i \leq \beta^*_i \leq \bar{\beta}_i = 1. \] 
Let the regularizer be
\[ \Psi(\beta) = \begin{cases}
	-\sum_i B_i \log \beta_i & {\rm if}\ \beta\in [\ubar{\beta}, \bar{\beta}], \\
	\infty & {\rm o.w.}
\end{cases} \]
Clearly, ${\rm dom}\, \Psi = [\ubar{\beta}, \bar{\beta}]$ is closed and nonempty. 
Given these specifications, in Algorithm \ref{alg:sda}, the step $(*)$ yields a simple, explicit update: at iteration $t$, compute 
\[\beta^{t+1}_i = 
\Pi_{[\ubar{\beta}_i, \bar{\beta}_i]}  \left( \frac{B_i}{\bar{g}^t_i} \right),\ i\in [n], \] 
where $\Pi_{[a,b]}(c) = \min\{ \max\{a, c \}, b  \}$ is the projection onto a closed interval. 
This can be derived easily from its first-order optimality condition.
Using the convergence results in \citep{xiao2010dual} for strongly convex $\Psi$, we can show that the uniform average of all $\beta^t$ generated by SDA converges to $\beta^*$ both in mean square error (MSE) and with high probability, under mild additional boundedness assumptions on $v_i$.
\begin{theorem}
	Assume $v_i \in L^2(\Theta)$, that is, $\langle v_i^2, \ones\rangle = \EE_\theta[v_i(\theta)^2 ] < \infty$ for all $i$. 
	Let 
	\[ G^2 := \EE_\theta [\max_i v_i(\theta)^2]  < \infty, \ \sigma = \min_i B_i.\] 
	Let $\tilde{\beta}^t := \frac{1}{t} \sum_{\tau = 1}^t \beta^\tau$. 
	Then,
	\begin{align*}
		& \EE \| \beta^t - \beta^* \|^2 \leq \frac{6 + \log t}{t} \times \frac{G^2}{\sigma^2} , \\
		& \EE \|\tilde{\beta}^t - \beta^* \|^2 \leq \frac{6(1+\log t) + \frac{1}{2}(\log t)^2}{t} \times \frac{G^2}{\sigma^2}.
	\end{align*}
	Further assume that $v_i \leq G$ a.e. for all $i$. 
	Then, for any $\delta>0$, with probability at least $ 1 - 4\delta \log t$, we have \[\|\tilde{\beta}^t - \beta^*\|^2 \leq \frac{2M_t}{\sigma},\]
	where
	\begin{align*}
		& M_t = \frac{\Delta_t}{t} + \frac{4 G}{t}\sqrt{\frac{\Delta_t \log (1/\delta)}{\sigma}} + \max \left\{ \frac{16 G^2}{\sigma}, 6V \right\}\frac{\log (1/\delta)}{t}, \\
		& \Delta_t = \frac{G^2}{2\sigma} (6 + \log t), \\
		& V = n + \log \frac{1}{\sigma}.
	\end{align*}
	\label{thm:sda-conv}
\end{theorem}
It can be seen that the above bounds grow as the strong convexity modulus of the objective function $\sigma = \min_i B_i$ decreases.
For the CEEI case, $B_i = 1/n$ for all $i$ and $\sigma = \min_i B_i = 1/n$.
For the general case of heterogeneous buyer budgets, $\sigma \leq 1/n$.

%% file: text/ql.tex
\section{Extension to quasilinear utilities} \label{sec:ql}
We now discuss how the results in previous sections---convex optimization characterizations, a finite-dimensional reformulation under piecewise linear utilities and convergence guarantees of stochastic optimization---generalize to the case where each buyer has a quasilinear utility function. 
For the finite-dimensional case, there is a natural extension of EG to QL utilities, as shown by \citet{chen2007note,cole2017convex}. 
Furthermore, \citet{conitzer2019pacing} showed that budget management in an auction market with first-price auctions can be computed with the QL variant of EG. 

A QL utility is one such that cost is deducted from the utility, that is, $u_i(x_i) = \langle v_i - p, x_i \rangle$, where $p\in L_1(\Theta)_+$ is the vector of prices of all items. 
A market equilibrium under QL utilities (QLME) is a pair of allocations $x^* = (x_i)\in (L_1(\Theta)_+)^n$ and prices $p^*\in L_1(\Theta)_+$ such that
\begin{itemize}
    \item For each buyer $i$, their allocation is optimal: 
        \[ x^*_i \in \argmax \{ \langle v_i - p^*, x_i \rangle: \langle p^*, x_i\rangle \leq B_i,\, x_i \in L_\infty(\Theta)_+\}. \]
    \item Market clears: $\langle p^*,  \ones - \sum_i x^*_i \rangle = 0$.
\end{itemize}
In the QL case, we cannot normalize both valuations and budgets, since buyers' budgets have value outside the current market. 
Without loss of generality, we can only assume that $\|B\|_1 = 1$ and $v_i(\Theta) >0$ for all $i$ (all buyers' $v_i$ and $B_i$ must be scaled by the same constant).

Here, we consider the infinite-dimensional ``primal'' Eisenberg-Gale convex program:
\begin{align}
    \begin{split}
    \sup\, & \sum_i (B_i \log u_i - \delta_i) \\ 
    \st & u_i \leq \langle v_i, x_i \rangle + \delta_i,\, \forall\, i \in [n], \\
    & \sum_i x_i \leq \ones,\,  \\
    & u_i \geq 0,\ \delta_i \geq 0,\ x_i \in L_1(\Theta)_+,\ \forall\, i \in [n].
    \end{split}
    \label{eq:ql-eg-primal} 
    \tag{$\mathcal P_{\rm QLEG}$}
\end{align}
The ``dual'' is
\begin{align}
    \begin{split}
        \inf\, & \langle p, \ones \rangle - \sum_i B_i \log \beta_i \\
        \st & p \geq \beta_i v_i,\ \beta_i \leq 1,\ \forall\, i \in [n], \\
        & p \in L_1(\Theta)_+,\ \beta\in \RR^d_+.
    \end{split} 
    \label{eq:ql-eg-dual-p-beta}
    \tag{$\mathcal D_{\rm QLEG}$}
\end{align}
As in our earlier results, the ``primal'' and ``dual'' terminology should only be understood intuitively; the programs are not derived from each other via duality. However, the next theorem shows that they indeed behave like duals.
\begin{theorem} \label{thm:ql-me-duality-and-eq}
    The following results hold regarding \eqref{eq:ql-eg-primal} and \eqref{eq:ql-eg-dual-p-beta}.
    \begin{enumerate}[label=(\alph*)]
        \item The supremum of \eqref{eq:ql-eg-primal} is attained via an optimal solution $(x^*, \delta^*)$, in which $x^* = (x^*_i)$ is a pure allocation, that is, $x^*_i = \ones_{\Theta_i}$ for a.e.-disjoint measurable subsets $\Theta_i \subseteq \Theta$. \label{item:ql-primal-attainment}
        \item The infimum of \eqref{eq:ql-eg-dual-p-beta} is attained via an optimal solution $(p^*, \beta^*)$, in which $\beta^*\in \RR^n_+$ is unique and $p^* = \max_i \beta^*_i v_i$ a.e. \label{item:ql-dual-attainment}
        \item Given a feasible $(x^*, \delta^*)$  to \eqref{eq:ql-eg-primal} and a feasible $(p^*, \beta^*)$ to \eqref{eq:ql-eg-dual-p-beta}, they are both optimal solutions of their respective convex programs if and only if the following KKT conditions hold:
        \label{item:ql-kkt}
        \begin{align*}
            & \left \langle p^*, \ones - \sum_i x^*_i \right \rangle = 0, \\
            & u^*_i := \frac{B_i}{\beta^*_i}, \ \forall\, i, \\
            & \delta^*_i (1 - \beta^*_i) = 0,\ \forall\, i,\\
            & \langle p^* - \beta^*_i v_i, x^*_i \rangle = 0, \ \forall \, i.
        \end{align*}
        \item A pair of allocations and prices $(x^*, p^*) \in (L_\infty(\Theta)_+)^n \times L_1(\Theta)_+$ is a QLME if and only if there exists $\delta^* \in \RR_+^n$ and $\beta^*\in \RR_+^n$ such that $(x^*, \delta^*)$ and $(p^*, \beta^*)$ are optimal solutions of \eqref{eq:ql-eg-primal} and \eqref{eq:ql-eg-dual-p-beta}, respectively. 
        \label{item:ql-eq-iff-opt}
    \end{enumerate}
\end{theorem}
Note that $u^*_i$ above does \emph{not} correspond to the equilibrium utility of buyer $i$, which is $\langle v_i - p^*, x^*_i \rangle$.
Instead, by the definition of QLME and the above theorem, for each buyer $i$, there are two possibilities at equilibrium ($\Leftrightarrow$ primal and dual optimality).
\begin{itemize}
    \item If $\beta^*_i < 1$, then $\delta^*_i = 0$ and $u^*_i = \langle v_i, x^*_i \rangle$ in \eqref{eq:ql-eg-primal}. 
    Since $\langle p^* - \beta^*_i v_i, x^*_i \rangle = 0$, the equilibrium utility is 
        \[ \langle v_i - p^*, x_i \rangle = (1 - \beta^*_i) \langle v_i, x^*_i \rangle =(1-\beta^*_i) u^*_i. \]
    \item If $\beta^*_i = 1$, then $\langle p^* - \beta^*_i v_i, x^*_i \rangle = 0$ implies the equilibrium utility is 
    \[  \langle v_i - p^*, x^*_i \rangle = 0. \]
\end{itemize}

\paragraph{Tractable convex optimization under piecewise linear valuations over $[0,1]$.} Similar to \S\ref{sec:convex-opt-for-pwl}, we can reformulate \eqref{eq:ql-eg-primal} into a tractable convex program using the same characterization of the set of feasible utilities $u_i$ (which does not take prices into account) in Theorems~\ref{thm:U-conic-rep} and \eqref{thm:U-conic-rep-general-vi}. To reconstruct a pure allocation that achieves the equilibrium utilities, run Algorithm~\ref{alg:interval-partition-given-feasible-u} on the subintervals corresponding to the linear pieces of the valuations.

\paragraph{Stochastic optimization.} 
Similar to the case of linear utilities (Lemma~\ref{lemma:equi-bounds}), we can establish bounds on equilibrium quantities such as the equilibrium utility prices $\beta^*$. For a finite-dimensional Fisher market with buyers having QL utilities, \cite[Lemma~2]{gao2020first} gives bounds on equilibrium prices. Similar to their proof, we can show that 
\[u^*_i \leq v_i(\Theta) + B_i \leq 1+B_i \ \Rightarrow\ \beta^*_i = \frac{B_i}{u^*_i} \geq \frac{B_i}{ v_i(\Theta) +B_i} > 0.  \]
It follows that we can add the following bounds (together with the existing bound $\beta_i\leq 1$)
\[ \frac{B_i}{v_i(\Theta) + B_i} \leq \beta_i \leq  1, \]
to \eqref{eq:ql-eg-dual-p-beta} without affecting its (unique) optimal solution $\beta^*$. 
Hence, completely analogous to the linear case discussed in \S\ref{sec:sda}, we can use SDA to solve \eqref{eq:ql-eg-dual-p-beta}  with similar convergence guarantees.





%% file: text/conclusion.tex
\section{Summary, discussion and future research}
Motivated by applications in ad auctions and fair recommender systems, we considered a Fisher market with a continuum of items and the concept of a market equilibrium in this setting. 
By extending the finite-dimensional Eisenberg-Gale convex program and its dual, we proposed convex programs whose optimal solutions are ME, and vice versa. 
Optimality conditions for the convex programs parallel various structural properties of a market equilibrium. 
Due to the limitations of general duality theory for optimization over infinite-dimensional vector spaces, we established these properties via directly exploiting the problem structure.
In particular, we showed that, under a continuum of items, a pure market equilibrium must exist, and an equilibrium allocation is guaranteed to be Pareto optimal, envy-free and proportional. Hence, when all buyers have the same budget, a pure equilibrium allocation is a fair division.
Under piecewise linear buyer valuations over a closed interval, we showed that the infinite-dimensional Eisenberg-Gale convex program \eqref{eq:eg-primal} can be reformulated as a finite-dimensional convex program with linear and quadratic constraints, via simple characterizations of the set of feasible utilities and a sequence of reformulations. 
This yielded a highly scalable approach for computing a fair division under piecewise linear buyer valuations, and the first polynomial-time algorithm for this problem.
We also showed that, for more general valuations, the finite-dimensional convex program \eqref{eq:eg-dual-beta-1} in utility prices $\beta$ can be solved via subgradient-based stochastic optimization, for which we established mean-square convergence and high-probability convergence guarantees. 
Finally, we showed that most of the above results also extend to the case of quasilinear utilities: a class of utilities relevant to characterizing pacing equilibria in auction markets.

For future research, we would like to consider tractable convex optimization formulations for (i) a multidimensional item space $\Theta = [a_i, b_i]^d$ representing linear features of items and (ii) linear valuations $v_i(\theta) = \beta_i^\top \theta$. 
This has application in low-rank market models in ad auctions with budget constraints \citep{conitzer2019pacing}. There, a low-rank model typically assumes a distribution over the space of possible item types (which captures the extremely large space of possible impressions). Modeling it as a finite-dimensional Fisher markets require discretizing the space of items, leading to a huge number of items. Our approach can potentially provide a much more compact convex optimization reformulation, and scale much better in both the number of items and the feature dimension. Another interesting direction is to find other structured classes of valuations for which tractable reformulations similar to the piecewise linear case exist.

%% file: text/appendix.tex
\section{Proofs} \label{app:proofs}

	\subsection*{Proof of Lemma~\ref{lemma:U-convex-compact}}

	First, we state a lemma due to \citet[Theorems 1 and 4]{dvoretzky1951relations}. Here, we assume that there is an underlying $\sigma$-algebra $\mathcal{M}$ on $\Theta$. 
	A measure is a countably additive set function mapping each measurable set in $\mathcal{M}$ to a nonnegative real number. 
	A function $f$ on $\Theta$ is measurable if, for any $c\in \RR$, it holds that $\{ f<c \} = \{ \theta\in \Theta: f(\theta) < c \} \in \mathcal{M}$. 
	\begin{lemma}
		Let $\mu_i$ be finite measures on $\Theta$, $i\in m$. 
		Let 
		\begin{align*}
			& \mathcal{X} = \left\{ (x_1, \dots, x_p): \sum_j x_j = \ones, \, \almeve,\, \text{each $x_j$ is a nonnegative measurable function on $\Theta$} \right\}, \\
			& \mathcal{S} = \left\{ (\Theta_1, \dots, \Theta_p) \in \mathcal{M}^p: \text{ $\Theta_i \cap \Theta_j = \emptyset$ for all $i\neq j$, $\cup_i \Theta_i = \Theta$}\right\}.
		\end{align*}
		Define 
		\begin{align*}
			& \mathcal{U} = \left\{ u\in \RR^{m\times p}: u_{ij} = \int_{\Theta} x_j d\mu_i,\, (x_1, \dots, x_p)\in\mathcal{X} \right\}, \\
			& \mathcal{U}' = \left\{ u\in \RR^{m\times p}: u_{ij} = \mu_i(\Theta_j),\, (\Theta_1, \dots, \Theta_p) \in \mathcal{S} \right\}.
		\end{align*}
		The set $\mathcal{U}$ is convex and compact. Furthermore, if each $\mu_i$ is atomless, then, $\mathcal{U}' = \mathcal{U}$.
		\label{lemma:dvo-51}
	\end{lemma}

	Next, we prove Lemma \ref{lemma:U-convex-compact}.
	Note that $v_i$ as measures are atomless, since they are absolutely continuous w.r.t. the Lebesgue measure on $\Theta\subseteq \RR^d$.
	Consider the set 
	\[ \bar{\cX} = \{ (x, x_{n+1})\in L^\infty(\Theta)_+^{n+1}: \sum_i x_i + x_{n+1} = \ones \}.\] 
	Define
	\[ \bar{U} = \left\{ u \in \RR^{n\times (n+1)}: u_{ij} = \langle v_i, x_j\rangle,\, (i,j)\in [n]\times [n+1],\, x_i \in L^\infty(\Theta)_+\, \sum_{i=1}^{n+1} x_i = \ones \right\} \]
	and 
	\[ \bar{U}' = \left\{ u \in \RR^{n\times (n+1)}: u_{ij} = v_i(\Theta_j),\, (i,j)\in [n]\times [n+1],\, \textnormal{$\Theta_i\subseteq \Theta$ are measurable and  a.e.-disjoint}\right\}. \]
	By Lemma \ref{lemma:dvo-51} (taking $m=n$ and $p=n+1$), we have
	\[ \bar{U} = \bar{U}' \] 
	and it is a convex and compact set in $\RR^{n(n+1)}$.
	Therefore, their (Euclidean) projections on to the $n$ dimensions corresponding to $u_{ii}$, $i\in [n]$ (corresponding to the utility values $u_i$) are also equal, that is, 
	\[ U = U'.\] 
	The set $U=U'$ is convex and compact (in $\RR^n$) since projection preserves convexity and compactness in a finite-dimensional Euclidean space.

	\subsection*{Proof of Theorem~\ref{thm:eg-equi-opt-combined}}
	\paragraph{Proof Part~\ref{item:eg-primal-attain}.}
	By Lemma~\ref{lemma:U-convex-compact}, the set $U = U(v, \Theta) \subseteq \RR_+^n$ is convex and compact. Let $\rho(u) = -\sum_i B_i \log u_i$. Taking $u^0_i = \langle v_i, \ones/n \rangle = \frac{1}{n} v_i(\Theta)$ for all $i$ ensures $u^0\in \mathcal{U}$ and $\rho(u^0)$ is finite. Since
	\[u_i \leq \langle v_i, \ones \rangle = v_i(\Theta) < \infty\] 
	for all $u\in U$, we have 
	\[\sum_i B_i \log u_i \leq M:= \sum_i B_i \log v_i(\Theta) < \infty.\] 
	Hence, there exists $\epsilon>0$ sufficiently small such that, for $u\in \mathcal{U}$, if some $u_i \leq \epsilon$, then 
	\[\rho(u) \geq - B_i \log \epsilon - M > \rho(u^0). \]
	In fact, it suffices to have $- B_i \log \epsilon > \rho(u^*) + M$ for all $i$, or simply
	\[ \epsilon < e^{-\frac{\rho(u^0)+M}{\|B\|_\infty}}. \]
	Hence, removing all $u\in U$ such that $\min_i u_i < \epsilon$ does not affect the infimum of $\rho$ over $U$, that is, 
	\[ \inf_{u \in U} \rho(u) = \inf_{u\in U:\, u\geq \epsilon} \rho(u). \]
	Since $\{ u\in U: u\geq \epsilon \}$ is compact (as a closed subset of the compact set $U$) and $\rho$ is continuous on it, by the extreme value theorem, there exists a minimizer $u^*\in U$ (such that $u\geq \epsilon$). 
	By the definition of $U'$, there exists $x^*$ such that $x^*_i = \ones_{\Theta_i}$ (where $\Theta_i$ are disjoint measurable subsets of $\Theta$) and
	\[ u^*_i = \langle v_i, x^*_i \rangle = v_i(\Theta_i). \]
	Finally, if $\sup_i \Theta_i \subsetneq \Theta$, then assign $\Theta\setminus (\cup_i \Theta_i)$ to buyer $1$, that is, augment $\Theta_1$ so that $\sup_i \Theta_i = \Theta$. This does not affect $\sum_i x_i \leq \ones$, nor does it affect optimality (since $v_i(S) \leq v_i(T)$ if $S\subseteq T \subseteq \Theta$, all sets measurable). In fact, $\Theta_0 := \Theta\setminus (\cup_i \Theta_i)$ corresponds to the subset on which all buyers have valuation $0$ a.e.

	\paragraph{Proof of Part~\ref{item:eg-dual-p-beta-attain}.} See Lemma~\ref{lemma:beta-dual-attain}.

	\paragraph{Proof of Part~\ref{item:eg-equi-iff-opt}.} See Theorem~\ref{thm:eg-gives-me}.

	
	\subsection*{Proof of Lemma~\ref{lemma:beta-dual-attain}}
	Denote the objective function of \eqref{eq:eg-dual-beta-1} as $\psi(\beta)$. Notice that 
	\[ \max_i \beta_i v_i \leq \sum_i \beta_i v_i \]
	and therefore 
	\[0\leq \left\langle \max_i \beta_i v_i, \ones \right \rangle\leq \sum_i \beta_i \langle v_i, \ones \rangle,\ \forall\, \beta>0. \] 
	For any $\lambda \in [0,1]$, $\beta, \beta' \in \RR_+^n$, $\theta\in \Theta$, 
	\[ \max_i \left(\lambda\beta_i + (1-\lambda)\beta'_i\right)v_i(\theta) \leq \lambda \max_i \beta_i v_i(\theta) + (1-\lambda)\max_i \beta'_i v_i(\theta). \]
	Therefore, 
	\[ \left \langle \max_i \left(\lambda\beta_i + (1-\lambda)\beta'_i\right)v_i, \ones \right \rangle \leq  \lambda \max_i \langle \beta_i v_i, \ones\rangle + (1-\lambda) \langle \max_i \beta'_i v_i, \ones \rangle. \]
	In other words, the function
	\[\beta \mapsto \left\langle \max_i \beta_i v_i, \ones \right \rangle\]
	is convex. Since $\beta\mapsto -\sum_i B_i \log \beta_i$ is strictly convex on $\RR_{++}^n$, we know that $\psi$ is real-valued, strictly convex and hence continuous on $\RR^d_{++}$. 
	Furthermore, for any $i$, when $\beta_i\rightarrow 0$ or $\beta_i \rightarrow \infty$, we have $\psi(\beta) \rightarrow \infty$.
	Hence, for $\beta^0 = (1, \dots, 1)>0$, there exists $0 < \ubar{\beta} < \bar{\beta} < \infty$ such that 
	\[ \beta\notin [\ubar{\beta}, \bar{\beta}] \Rightarrow \psi(\beta) > \psi(\beta^0). \]
	Therefore, we can restrict $\beta$ inside a closed interval without affecting the infimum:
	\[ \inf_{\beta\in \RR_+^n } \psi(\beta) = \inf_{\beta\in [\ubar{\beta}, \bar{\beta}] }\psi(\beta). \]
	The right-hand side is the infimum of a continuous function on a compact set. Therefore, the infimum is attained at some $\beta^* \in [\ubar{\beta}, \bar{\beta}]$. Clearly, $\beta^* > 0$. It is unique since $\psi$ is strictly convex on $[\ubar{\beta}, \bar{\beta}]$.
	
	Finally, when solving \eqref{eq:eg-dual-beta-p}, for any fixed $\beta$, the objective is clearly minimized at $p = \max_i \beta_i v_i \in L^1(\Theta)_+$. Therefore, we can eliminate $p$ in this way and obtain \eqref{eq:eg-dual-beta-1}. In other words, for the optimal solution $\beta^*$ of \eqref{eq:eg-dual-beta-1}, setting $p^* := \max_i \beta^*_i v_i$ gives an optimal solution $(p^*, \beta^*)$ of \eqref{eq:eg-dual-beta-p}, which is (a.e.) unique. 

	Tighter bounds for the optimal solution $\beta^*$ are given in Lemma~\ref{lemma:equi-bounds}.
	
	\subsection*{Proof of Lemma~\ref{lemma:weak-duality}}
	\paragraph{Proof of Part~\ref{item:eg-weak-duality} (weak duality).}
	First, we introduce new variables $u = (u_i) \in \RR_+^n$ and rewrite \eqref{eq:eg-primal} into 
	\begin{align}
		\begin{split}
			z^* = \sup_{x\in (L^\infty(\Theta)_+)^n,\, u\in \RR_+^n}& \sum_i B_i \log u_i \\
			\st \ & \ u_i \leq \langle v_i, x_i \rangle,\\
			& \sum_i x_i \leq \ones. 
		\end{split}
		\label{eq:eg-primal-with-u}
	\end{align}
	Let $(x,u)$ be a feasible to \eqref{eq:eg-primal-with-u}
	and $(p, \beta)$ be a feasible solution of \eqref{eq:eg-dual-beta-p}. Using the feasibility assumptions, we have $\beta_i (u_i - \langle v_i, x_i \rangle) \leq 0$ and $\left\langle p, \sum_i x_i - \ones \right\rangle \leq 0$. 
	Hence, 
	\begin{align}
		\sum_i B_i \log u_i  
		& \leq \sum_i B_i \log u_i - \sum_i \beta_i (u_i - \langle v_i, x_i\rangle) - \left\langle p, \sum_i x_i - \ones \right\rangle \nonumber \\
		& = \sum_i (B_i \log u_i - \beta_i u_i) - \sum_i \langle p - \beta_i v_i, x_i \rangle + \langle p, \ones \rangle \nonumber \\
		& \leq \sum_i \left(B_i \log \frac{B_i}{\beta_i} - \beta_i \frac{B_i}{\beta_i}\right) - \sum_i \langle p - \beta_i v_i, x_i \rangle + \langle p, \ones \rangle \nonumber \\
		& \leq \sum_i (B_i \log B_i - B_i) + \langle p, \ones\rangle - \sum_i B_i \log \beta_i \nonumber \\
		& = \langle p, \ones\rangle - \sum_i B_i \log \beta_i - C,
		\label{eq:eg-weak-duality-ineq-chain}
	\end{align}
	where the second inequality is because $u_i = \frac{B_i}{\beta_i}$ maximizes the concave function 
	\[ u_i \mapsto B_i \log u_i - \beta_i u_i\] 
	and the third inequality is because $p \geq \beta_i v_i$ a.e. for all $i$. 
	Taking supremum on the left and infimum on the right yields 
	\[ C + z^* \leq w^*. \]

	\paragraph{Proof of Part~\ref{item:eg-KKT-iff}.} 
	Suppose $x^*$ is feasible to \eqref{eq:eg-primal} and attains $z^*$; $(p^*, \beta^*)$ is feasible to \eqref{eq:eg-dual-beta-p} and attains $w^*$. 
	Then, $(x^*, u^*)$, where $u^*_i := \langle v_i, x^*_i \rangle$, is feasible to \eqref{eq:eg-primal-with-u}. Note that $C + z^* = w^*$ if and only if all inequalities in \eqref{eq:eg-weak-duality-ineq-chain} are tight (with $x = x^*$, $u = u^*$, $p = p^*$, $\beta = \beta^*$). The first inequality being tight implies \eqref{eq:mkt-clear}, i.e., 
	\[ \left\langle p^*, \ones - \sum_i x^*_i \right \rangle = 0. \]
	The second inequality being tight implies \eqref{eq:u*=B/beta*}, i.e.,
	\[ u^* = \frac{B_i}{\beta^*_i}. \]
	The third inequality being tight implies \eqref{eq:comp-slack-dual}, i.e.,
	\[ \langle p^* - \beta^*_i v_i, x^*_i \rangle = 0,\ \forall\, i. \]
	Conversely, let $x^*$ and $(p^*, \beta^*)$ be feasible to \eqref{eq:eg-primal} and \eqref{eq:eg-dual-beta-p}, respectively. Then, $(x^*, u^*)$, where $u^*_i := \langle v_i, x^*_i \rangle$ is feasible to \eqref{eq:eg-primal-with-u}. 
	If they satisfy \eqref{eq:mkt-clear}-\eqref{eq:comp-slack-dual}, then all inequalities in \eqref{eq:eg-weak-duality-ineq-chain} are tight. 
	Hence, both $x^*$ and $(p^*, \beta^*)$ must be optimal.

	\subsection*{Proof of Theorem~\ref{thm:eg-gives-me}}
	\paragraph{Optimal solutions $\Rightarrow$ ME.}
	We first show the forward direction. Let $x^*$ and $(p^*, \beta^*)$ be optimal solutions of \eqref{eq:eg-primal} and \eqref{eq:ql-eg-dual-p-beta}, respectively. By Lemma~\ref{lemma:weak-duality}, they satisfy \eqref{eq:mkt-clear}-\eqref{eq:comp-slack-dual}.
	Here, \eqref{eq:mkt-clear} gives market clearance. 
	It remains to verify buyer optimality and budget depletion, i.e., $x^*_i \in D_i(p^*)$ and $\langle p^*, x^*_i \rangle = B_i$. 
	For each $i$, by \eqref{eq:u*=B/beta*} and \eqref{eq:comp-slack-dual},
	\begin{align}
		\langle p^*, x^*_i\rangle  = \beta^*_i \langle v_i, x^*_i\rangle = B_i. \label{eq:budget-deplete}
	\end{align}
	In words, $x^*_i$ depletes buyer $i$'s budget $B_i$ and the utility buyer $i$ receives is
	\[ \langle v_i, x^*_i\rangle = \frac{B_i}{\beta^*_i}. \]
	Consider any $x_i\in L^\infty(\Theta)_+$ such that $\langle p^*, x_i \rangle \leq B_i$. Feasibility of $(p^*, \beta^*)$ implies $p^* \geq \beta^*_i v_i$. Then,
	\[ \langle v_i, x_i\rangle \leq \frac{1}{\beta^*_i} \langle p^*, x_i \rangle \leq \frac{B_i}{\beta^*_i} = \langle v_i, x^*_i \rangle. \]
	Therefore, 
	\[x^*_i \in D_i(p^*). \]
	Hence, $(x^*, p^*)$ is a ME, where buyer $i$'s equilibrium utility is clearly $u^*_i := \langle v_i, x^*_i \rangle = \frac{B_i}{\beta^*_i}$.
	
	\paragraph{ME $\Rightarrow$ Optimal solutions.}
	Conversely, let $(x^*, p^*)$ be a ME and $\beta^*_i := \frac{B_i}{u^*_i}$, where $u^*_i := \langle v_i, x^*_i \rangle$. We first check that $(p^*, \beta^*)$ is feasible to \eqref{eq:eg-dual-beta-p}. For any $i$, suppose there exists a measurable set $A\subseteq \Theta$ such that $p^*(A) < \beta^*_i v_i(A)$. Then, consider the allocation $x_i = \frac{B_i}{p^*(A)} \cdot \ones_A$. 
	We have 
	\[\langle p^*, x_i \rangle = B_i \]
	and 
	\[ \langle v_i, x_i\rangle = B_i \cdot \frac{v_i(A)}{p^*(A)} > \frac{B_i}{\beta^*_i} = u^*_i = \langle v_i, x^*_i \rangle, \]
	which contradicts to buyer optimality $x^*_i \in D_i(p^*)$. 
	Therefore, we must have
	\[ p^* \geq \beta^*_i v_i \ \almeve,\ \forall\, i. \]
	Thus, $(p^*, \beta^*)$ is feasible to \eqref{eq:eg-dual-beta-p}. 
	We know that $x^*$ is feasible to \eqref{eq:eg-primal}, since ME already requires $\sum_i x^*_i \leq \ones$.
	Furthermore, $x^*$ and $(p^*, \beta^*)$ satisfy \eqref{eq:mkt-clear}-\eqref{eq:comp-slack-dual}. 
	Therefore, by Lemma~\ref{lemma:weak-duality}, they must be optimal to \eqref{eq:eg-primal} and \eqref{eq:eg-dual-beta-p}, respectively. 

\subsection*{Proof of Corollary~\ref{cor:me-structiral-properties} }
	Since $(x^*, p^*)$ is a ME, by Theorem~\ref{thm:eg-gives-me}, $x^*$ and $(p^*, \beta^*)$ (where $\beta^*_i = \frac{B_i}{u^*_i}$, $u^*_i = \langle v_i, x^*_i\rangle$) are optimal solutions of \eqref{eq:eg-primal} and \eqref{eq:eg-dual-beta-p}, respectively. By Lemma~\ref{lemma:weak-duality}, the KKT conditions \eqref{eq:mkt-clear}-\eqref{eq:comp-slack-dual} holds. 

\subsection*{Proof of Corollary~\ref{cor:eg-pure-solution-gives-strong-duality}}
	Let $(\tilde{p}^*, \tilde{\beta}^*)$ be the (a.e.-unique) optimal solution of \eqref{eq:eg-dual-beta-p}.
	Since $u^*_i$ are the equilibrium utilities, by Lemma~\ref{lemma:weak-duality}, $\tilde{\beta}^*_i = \frac{B_i}{u^*_i}$ for all $i$. Hence, $\tilde{\beta}^* = \beta^*$ and therefore $\tilde{p} = \max_i \beta^*_i v_i$ a.e. Lemma~\ref{lemma:weak-duality} ensures that they satisfy \eqref{eq:mkt-clear}-\eqref{eq:comp-slack-dual}. 
	Then, since $\{\Theta_i\}$ is a pure allocation, we have $\Theta_i = \{ p^* = \beta^*_i v_i \}$ a.e. (i.e., the symmetric difference of $\Theta_i$ and $\{ p^* = \beta^*_i v_i \}$ has measure zero). 
	Therefore, on each $\Theta_i$, we must have $p^* = \beta^*_i v_i \geq \beta^*_j v_j$ a.e. for any $j\neq i$. 
	Using this fact, we have
	\[ p^*(A) = \sum_i p^*(A\cap \Theta_i) = \sum_i \beta^*_i v_i(A\cap \Theta_i) \]
	for any measurable set $A\subseteq \Theta$. 

	\subsection*{Proof of Corollary~\ref{cor:check-pure-alloc-ME}}
	If $\{\Theta_i\}$ is a (pure) equilibrium allocation, then, by Theorem~\ref{cor:eg-pure-solution-gives-strong-duality},
	$p^* = \max_i \beta^*_i v_i$ is the equilibrium prices. 
	By Corollary~\ref{cor:me-structiral-properties}, $\langle p^* -\beta^*_i v_i, x^*_i\rangle = 0$, where $x^*_i = \ones_{\Theta_i}$. In other words, $p^* = \beta^*_i v_i$ on $\Theta_i$. Corollary~\ref{cor:me-structiral-properties} also implies \eqref{eq:mkt-clear}, i.e., $\left \langle p^*, \ones - \sum_i x^*_i \right\rangle = 0$. Since $\ones - \sum_i x^*_i = \ones_{\Theta_0}$, where $\Theta\setminus \left(\cup_i \Theta_i\right)$, we have $p^*(\Theta_0) = 0$. 

	Conversely, if $\{\Theta_i\}$ and $\beta^*$, $p^*$ satisfy the said conditions, we can also verify similarly that \eqref{eq:mkt-clear}-\eqref{eq:comp-slack-dual} holds. Since $\{\Theta_i\}$ is feasible to \eqref{eq:eg-primal} and $(p^*, \beta^*)$, by the construction, is feasible to \eqref{eq:eg-dual-beta-p}, by Part~\ref{item:eg-KKT-iff} Lemma~\ref{lemma:weak-duality}, they are both optimal to \eqref{eq:eg-primal} and \eqref{eq:eg-dual-beta-p}, respectively. Hence, by Theorem~\ref{thm:eg-gives-me}, $\{\Theta_i\}$ is an equilibrium allocation.

	\subsection*{Proof of Theorem~\ref{thm:me-is-pareto-ef-prop}}
	\paragraph{Pareto optimality.}
	Since $x^*$ is an equilibrium allocation, by Theorem~\ref{thm:eg-gives-me}, it is also an optimal solution of \eqref{eq:eg-primal}. If there exists $\tilde{x}\in (L^\infty(\Theta)_+)^n$, $\sum_i \tilde{x}_i \leq 1$ such that $\langle v_i, \tilde{x}_i \rangle \geq \langle v_i, x^*_i \rangle$ for all $i$ and at least one inequality is strict, then 
	\[\sum_i B_i \log \langle v_i, \tilde{x}_i \rangle > \sum_i B_i \log \langle v_i, x^*_i \rangle,\] 
	i.e., $x^*$ is not an optimal solution of \eqref{eq:eg-primal}, a contradiction. Therefore, $x^*$ is Pareto optimal.
	
	\paragraph{Envy freeness.}
	For any $j\neq i$, since $\langle p^*, x^*_i \rangle = B_i$ (Theorem~\ref{thm:eg-gives-me}) and $\langle p^*, x^*_j \rangle = B_j$, we have 
	\[ \left \langle p^*, \frac{B_i}{B_j}x^*_j \right \rangle = B_i. \]
	
	Since $x^*_i \in D_i(p^*)$ and $\frac{B_i}{B_j}x^*_j \geq 0$, we have 
	\[ \langle v_i, x^*_i \rangle \geq \left\langle v_i, \frac{B_i}{B_j} x^*_j \right \rangle \Rightarrow \frac{\langle v_i, x^*_i \rangle}{B_i} \geq \frac{\langle v_i, x^*_j \rangle}{B_j}. \]
	Therefore, $x^*$ is envy-free.
	
	\paragraph{Proportionality.}
	By the market clearance condition of ME, we have
	\[ p^*(\Theta) = \langle p^*, \ones \rangle = \sum_i \langle p^*, x^*_i\rangle = \|B\|_1. \]
	Therefore, for each buyer $i$, it holds that 
	\[ \left \langle p^*, \frac{B_i}{\|B\|_1} \ones \right \rangle = \frac{B_i}{\|B\|_1} p^*(\Theta) = \frac{B_i}{\|B\|_1} = B_i. \]
	In other words, buyer $i$ can afford the bundle $\frac{B_i}{\|B\|_1} \ones $. Hence, its equilibrium utility must be at least
	\[ \left \langle v_i,\frac{B_i}{\|B\|_1}\ones \right\rangle = \frac{B_i}{\|B\|_1}v_i(\Theta). \] 
	




	\subsection*{Proof of Lemma~\ref{lemma:equi-bounds}}
	By the characterization of $p^*$ in Corollary~\ref{cor:eg-pure-solution-gives-strong-duality}, clearly, 
	\[ p^*(\Theta) = \sum_i p^*(\Theta_i) = \sum_i B_i = \|B\|_1, \]
	where $\{ \Theta_i \}$ is a pure equilibrium allocation. 
	Clearly, we have
	\[ u^*_i = \langle v_i, x^*_i\rangle \leq v_i(\Theta) = 1. \]
	On the other hand, since $\{\Theta_i\}$ is an equilibrium allocation, it is proportional (Theorem~\ref{thm:me-is-pareto-ef-prop}), that is, $\frac{B_i}{\|B\|_1}\ones$ is a budget-feasible allocation for buyer $i$. Hence,
	\[ 
		u^*_i \geq \left \langle v_i, \frac{B_i}{ \|B\|_1 }\ones\right \rangle = \frac{B_i}{\|B\|_1} v_i(\Theta) = \frac{B_i}{\|B\|_1}. 
	\]
	The bounds on $\beta^*_i = \frac{B_i}{u^*_i}$ follow immediately.

	\subsection*{Proof of Lemma~\ref{lemma:all-linear-equilibrium-geometry}}
	Since $\beta^*_i v_i$ are linear, the equilibrium price vector $p^* = \max_i \beta^*_i v_i$ is a piecewise linear function with at most $n$ pieces. Each linear piece has a support interval corresponding to the ``winning set'' of a buyer $\{p^* = \beta^*_i v_i\}$.
	Since all $v_i$ are linear, normalized and distinct, there is no tie, i.e., no $i\neq j$ such that $\beta^*_i v_i = \beta^*_j v_i$ on a set of positive measure (otherwise we must have $v_i = v_j$ on $[0,1]$).
	Since $B_i>0$ for all $i$, each buyer must receive a positive equilibrium utility $u^*_i>0$. 
	Hence, $p^*$ consists of exactly $n$ linear pieces and each buyer must get a nonempty interval as its equilibrium allocation in order to receive a positive equilibrium utility (Lemma~\ref{lemma:equi-bounds}).
	Let the breakpoints of $p^*$ be $a^*_0 = 0 < a^*_1 < \dots < a^*_n = 1$, which are clearly unique since $\beta^*$ is unique (Lemma~\ref{lemma:beta-dual-attain}). 
	Hence, there exists a (unique) permutation $\sigma$ of $[n]$ such that $\{ p^* = \beta^*_i v_i \} = [a^*_{\sigma(i)-1}, a^*_{\sigma(i)}]$ for all $i$ (every buyer gets exactly one of the $n$ nonempty intervals). 
	
	We show that $\sigma$ must be the identity map $\sigma(i) = i$. In other words, at equilibrium, the entire interval is divided into $n$ intervals; these intervals are allocated to buyers $1, \dots, n$ from left to right, respectively. To see this, first note that any $\beta^*_i v_i$ and $\beta^*_j v_j$, $i < j$, must intersect on $[0,1]$ (otherwise one of them is completely dominated by the other, which means $p^* = \max_i \beta^*_i v_i$ cannot have $n$ pieces and one of the buyers can only receive a zero-measure set at equilibrium, a contradiction to the fact that each buyer gets a positive equilibrium utility $u^*_i > 0$). 
	Since $v_i$, $v_j$ are linear and $a^*_k$ are breakpoints of $p^* = \max_\ell \beta^*_\ell v_\ell$, we want to show that $\beta^*_i v_i(0) > \beta^*_j v_j(0)$, which will imply that, at equilibrium the interval for $i$ is on the left of the interval for $j$. 
	Suppose $\beta^*_i v_i(0) \leq \beta^*_j v_j(0)$, which implies $\beta^*_i d_i \leq \beta^*_j d_j$. By Assumption~\ref{assump:linear-distinct-decreasing-[0,1]}, $d_i > d_j$, which implies $\beta^*_i < \beta^*_j$. Furthermore, \[v_i(1) = c_i + d_i = 2- d_i < 2 - d_j = v_j(1).\] 
	Hence, 
	\[ \beta^*_i v_i(1) < \beta^*_j v_j(1). \]
	In other words, $\beta^*_i v_i < \beta^*_j v_j$ on $(0, 1]$ and buyer $i$ gets zero utility, a contradiction.
	Therefore, each interval $[a^*_{i-1}, a^*_i]$ is precisely the ``winning set'' $\{p^* = \beta^*_i v_i\}$ of buyer $i$. by Assumption~\ref{assump:linear-distinct-decreasing-[0,1]}, we have $v_i > 0$ on $(0,1)$ for all $i$, which implies $p^* > 0$ on $(0,1)$ (since $\beta^*_i\geq B_i > 0$ for all $i$, by Lemma~\ref{lemma:equi-bounds}).Therefore, by the market clearance condition of ME, every buyer must be allocated all of its winning set $[a^*_{i-1}, a^*_i]$ (except possibly the endpoints, which have measure zero). 
	Therefore, $\Theta_i = [a^*_{i-1}, a^*_i]$, $i\in [n]$ is the unique pure equilibrium allocation. 
	
	\subsection*{Proof of Lemma~\ref{lemma:pareto-opt-utility-is-equil-of-some-B}}
	Assume $u^\circ_i>0$ for all $i$. 
	If this does not hold, remove the buyers with $u^\circ_i=0$, assign them zero budgets $B_i = 0$ and consider the market without these buyers. 
	Since $u^\circ$ is on the Pareto frontier, we know that $(1+\delta)u^\circ\notin U$ for any $\delta>0$. Hence, $u^\circ$ is on the \emph{boundary} of the convex compact set $U$. By the supporting hyperplane theorem, there exists $\beta^\circ\in \RR^n$ such that 
	\begin{align}
		\langle \beta^\circ, u^\circ \rangle = 1\ \text{and}\ \langle \beta^\circ, u\rangle \leq 1,\, \forall\, u\in U.
		\label{eq:beta-supporting-hyperplane}
	\end{align}
	We can verify that $\beta^\circ_i \geq 0$ for all $i$: otherwise, if $\beta^\circ_i < 0$ for some $i$, decreasing $u^\circ_i >0$ makes $\langle \beta^\circ, u^\circ \rangle > 1$ while ensuring $u^\circ\in U$, which contradicts \eqref{eq:beta-supporting-hyperplane}.

	By Lemma~\ref{lemma:U-convex-compact}, there exists a pure allocation $\{\Theta_i\}$ such that $u^\circ_i = v_i(\Theta_i)$ for all $i$. W.l.o.g., assume that $\cup_i \Theta_i = \Theta$ a.e.
	Define $p^\circ$ as 
	\[ p^\circ(\theta) = \sum_i \beta^\circ_i v_i(\theta)\ones_{\Theta_i}(\theta). \]
	In other words, for $\theta\in \Theta_i$, $p^\circ(\theta) = \beta^\circ_i v_i(\theta)$. Clearly, $p^\circ\in L^1(\Theta)_+$ (since each $v_i\in L^1(\Theta)_+$ and $\beta^\circ \geq 0$) and, for any measurable set $A\subseteq \Theta$, 
	\[ p^\circ(A) = \sum_i \beta^\circ_i v_i(A\cap \Theta_i). \]
	
	Next, we show that $p^\circ = \max_i \beta^\circ_i v_i$ almost everywhere. It suffices to show that $\beta^\circ_i v_i \geq \beta^\circ_j v_j$ on each $\Theta_i$ for any $j\neq i$. Suppose not, i.e., there exists a measurable set $A\subseteq \Theta_j$ such that, for $\ell \neq j$,
	\[ \beta^\circ_j v_j(A) < \beta^\circ_\ell v_\ell(A). \]
	Remove the set $A$ from $\Theta_j$ and give it to buyer $\ell$ instead, i.e., $\Theta'_j = \Theta_j \setminus A$, $\Theta'_\ell = \Theta_\ell \cup A$, $\Theta'_i = \Theta_i$ for all $i \notin \{j, \ell\}$. 
	Clearly, $\{\Theta'_i\}$ is still a feasible (pure) allocation. However, its utilities $u'_i = v_i(\Theta'_i)$ satisfy
	\[ \sum_i \beta^\circ_i u'_i = \sum_i \beta^\circ_i u^\circ_i - \beta^\circ_j v_j(A) + \beta^\circ_\ell v_\ell(A) > \langle \beta^\circ, u^\circ\rangle, \]
	which contradicts \eqref{eq:beta-supporting-hyperplane}. 
	Hence,
	\[ p^\circ = \max_i \beta^\circ_i v_i \, \almeve. \]
	
	Now, we are ready to show that $(\{\Theta_i\}, p^\circ)$ is a ME for buyers with budgets $B_i = \beta^\circ_i u^\circ$. Market clearance is satisfied since we assume $\cup_i \Theta_i = \Theta$ almost everywhere. 
	To verify buyer optimality, note that for each $i$ and any $x_i \in L^\infty(\Theta)_+$ such that $\langle p^\circ, x_i \rangle \leq B_i = \beta^\circ u^\circ_i$, we have 
	\[ \beta^\circ_i v_i \leq \max_i \beta^\circ_i v_i = p^\circ \, \almeve, \]
	which implies
	\[ \beta^\circ_i \langle v_i, x_i \rangle \leq \langle p^\circ, x_i \rangle \leq B_i \ \Rightarrow \ \langle v_i, x_i \rangle \leq u^\circ_i = v_i(\Theta_i). \]
	Hence, $(\{\Theta_i\}, p^\circ)$ is a ME under budgets $B_i$ and $u^\circ_i = v_i(\Theta_i)$ are the corresponding equilibrium utilities.

	\subsection*{Proof of Lemma~\ref{lemma:all-linear-represent-any-u}}
	Denote $U = U(v, [0,1])$. 
	Assume w.l.o.g. that all $u_i > 0$: otherwise, simply remove the buyers with $u_i = 0$ and set $a_{i-1} = a_i$ in the final partition, i.e., giving an empty interval to this buyer.

	\paragraph{The case of distinct $d_i$.} 
	We prove this case and show that the general case with some $d_i$ being identical follows easily.
	Let $u^\circ \in  U$ be a Pareto optimal utility vector such that $u^\circ \geq u$ (by the definition of Pareto optimality, such $u^\circ$ exists). 
	By Lemma \ref{lemma:all-linear-represent-any-u}, there exists $B^\circ_i > 0$, $i\in [n]$ such that $u^\circ_i$ are the equilibrium utilities of buyers with budgets $B^\circ_i$ and valuations $v_i$. By Lemma~\ref{lemma:all-linear-equilibrium-geometry}, there exists 
	\[a^\circ_0 = 0 < a^\circ_1 < \dots < a^\circ_n = 1\] 
	such that $\Theta^\circ_i = [a^\circ_{i-1}, a^\circ_i]$, $i\in [n]$ is the unique equilibrium allocation under budgets $B^\circ_i$. Let $a_0 = 0$. Let $a_1 \leq a^*_1$ be such that $v_i( [a_0, a_1] ) = u_i$. Such $a_1$ exists because (i) $a_1 \mapsto v_i([0, a_1]) = \frac{c_1}{2}\cdot a_1^2 + d_1 a_1$ is continuous and strictly increasing and (ii) $ v_i([0, a^\circ_1]) = u^\circ_i$. Inductively, there exist $a_i\leq a^\circ_i$ such that $v_i([a_{i-1}, a_i]) = u_i$ for all $i\in [n]$. Here, for simplicity, always take $a_n = 1$ regardless of the value of $u_n$, which ensures $v_n([a_{n-1}, a_n]) \geq v_n([a^\circ_{n-1}, 1]) = u^\circ_n \geq u_n$ (since $a_{n-1}\leq a^\circ_{n-1}$). 
	
	\paragraph{Handling identical valuations $v_i = v_j$ ($d_i = d_j$), $i\neq j$.} 
	In fact, the above procedure easily extend to the case of some intercepts $d_i$ being equal. We can merge the buyer with the same $d_i$, where the ``aggregate buyer'' $I\subseteq [n]$ has the same valuation $v_I = v_i$, $i\in I$, budget $B_I = \sum_i B_i$ and ``target utility value'' $U_I = \sum_i u_i$. After merging all identical buyers, $(u_I)$ is still a set of feasible utilities given (distinct) valuations $v_I$, i.e., $ (u_I) \in U((v_I), [0,1])$. By the above case, we can partition $[0,1]$ into intervals, each for one aggregate buyer $I$.
	Let buyer $I$ receives interval $[l_I, h_I]\subseteq [0,1]$ such that $v_i([l_I, h_I]) = u_I$. Since $v_I$ is linear, we can easily find breakpoints $l_i, h_i$ on $[l_I, h_I]$ via ``cut'' operations such that
	\[ v_i([l_i, h_i]) = u_i,\ i\in I. \]
	This is because all buyers $i\in I$ share the same valuation $v_i = v_I$.
	
	\subsection*{Proof of Theorem~\ref{thm:U-conic-rep}}
		Define $a_0 = 0$ and $a_n = 1$. By Lemma~\ref{lemma:all-linear-represent-any-u}, for any $u\in U = U(v, [0,1])$, there exists $0\leq a_1 \leq \dots \leq a_{n-1}\leq 1$ such that 
	\begin{align}
		u_i \leq \bar{u}_i := \frac{c_i}{2}(a_i^2 - a_{i-1}^2) + d_i (a_i - a_{i-1}),\ i\in [n]. 
		\label{eq:u<=a}
	\end{align}
	For any $0\leq a_1, \dots, a_{n-1}\leq 1$, since $v_i \geq 0$ on $[0,1]$, we have 
	\[ a_1 \leq \dots \leq a_{n-1} \Leftrightarrow \bar{u}_i \geq 0, \ i\in [n]. \]
	Hence, $u\in U$ is equivalent to the following constraints involving auxiliary variables $a_i$:
	\begin{align*}
		& u\geq 0, \\
		& u_i \leq \frac{c_i}{2}(a_i^2 - a_{i-1}^2) + d_i (a_i - a_{i-1}),\, i\in [n],\\
		& 0 \leq a_i \leq 1,\, i\in [n-1]. 
	\end{align*}
	Note that \[ \frac{c_i}{2}(a_i^2 - a_{i-1}^2) + d_i(a_i - a_{i-1}) = \left(\frac{c_i}{2}a_i^2 + d_i a_i\right) - \left(\frac{c_i}{2}a_{i-1}^2 + d_i a_{i-1}\right),\, i\in [n]. \]
	Consider auxiliary variables $z_i\leq \frac{c_i}{2}a_i^2 + d_i a_i$ and $w_i \leq - (\frac{c_{i+1}}{2}a_i^2 + d_{i+1} a_i)$, $i\in [n-1]$. The above inequalities are equivalent to 
	\begin{align}
		\begin{split}
			& u\geq 0, \\
			& u_1 \leq z_1, \\
			& u_i \leq z_i + w_{i-1},\, i=2, \dots, n-1, \\
			& u_n \leq 1+w_{n-1},\\
			& z_i \leq \frac{c_i}{2} a_i^2 + d_i a_i,\ w_i \leq -\frac{c_{i+1}}{2} a_i^2 - d_{i+1}a_i, \, i \in [n-1], \\
			& 0\leq a_i \leq 1,\, i\in [n-1]. 
		\end{split}\label{eq:rep-U-u,z,w,a}
	\end{align}
	Since $a_i \in [0,1]$, $d_i \in [0,2]$, $\frac{c_i}{2} + d_i = 1$ for all $i$, we have
	\begin{align}
		\begin{split}
			& \frac{c_i}{2}a_i^2 + d_i a_i \in [0,1],\\
			& -\frac{c_{i+1}}{2}a_i^2 - d_{i+1} a_i \in [-1,0],\\
			& \left(\frac{c_i}{2}a_i^2 + d_i a_i \right) + \left(-\frac{c_{i+1}}{2}a_i^2 - d_{i+1} a_i\right) = (d_{i+1} - d_i)(a_i^2 - a_i) \geq 0, 
		\end{split} \label{eq:range-quad-expr-a}
	\end{align}
	where the last inequality is because $d_i \geq d_{i+1}$, $\frac{c_i}{2} + d_i = 1$ (for all $i\in [n]$) and $a_i \in [0,1] \Rightarrow a_i^2 - a_i \leq 0$.
	By the first two inequalities in \eqref{eq:range-quad-expr-a}, we can add additional constraints 
	\[ 0\leq z_i \leq 1, \ -1 \leq w_i\leq 0, \ z_i + w_i \geq 0,\ i\in [n-1] \]
	to the inequalities in \eqref{eq:rep-U-u,z,w,a} without affecting the feasible region of $u$. 
	Next, for each $i$, consider the set $S_i$ of $(z_i, w_i)$ satisfying the following inequalities together with some $a_i$:
	\begin{align}
		& z_i \leq \frac{c_i}{2}a_i^2 + d_i a_i,\ w_i \leq - \frac{c_{i+1}}{2} a_i^2 - d_{i+1} a_i,\ a_i \in [0,1], \nonumber \\
		& 0\leq z_i\leq 1, \, -1\leq w_i \leq 0,\, z_i + w_i \geq 0.
		\label{eq:zi-wi-ai-zi+wi>=0}
	\end{align}
	
	When $d_i = d_{i+1} \in [0,2]$, the parametric curve
	\[ \Gamma_i = \left\{ \begin{bmatrix}
		\frac{c_i}{2} \alpha^2 + d_i \alpha \\
		-\frac{c_{i+1}}{2} \alpha^2 - d_{i+1} \alpha
	\end{bmatrix}: \alpha \in [0,1] \right\}, 
	\]
	is simply the line segment $(0,0)$ to $(1,-1)$. Together with the last inequality in \eqref{eq:zi-wi-ai-zi+wi>=0}, we know that
	\[ S_i = \{ (z_i, w_i)\in [0,1]\times [-1, 0]: z_i + w_i = 0 \} \]
	is the line segment as well.

\begin{figure}
		\centering
		\includegraphics[scale=0.5]{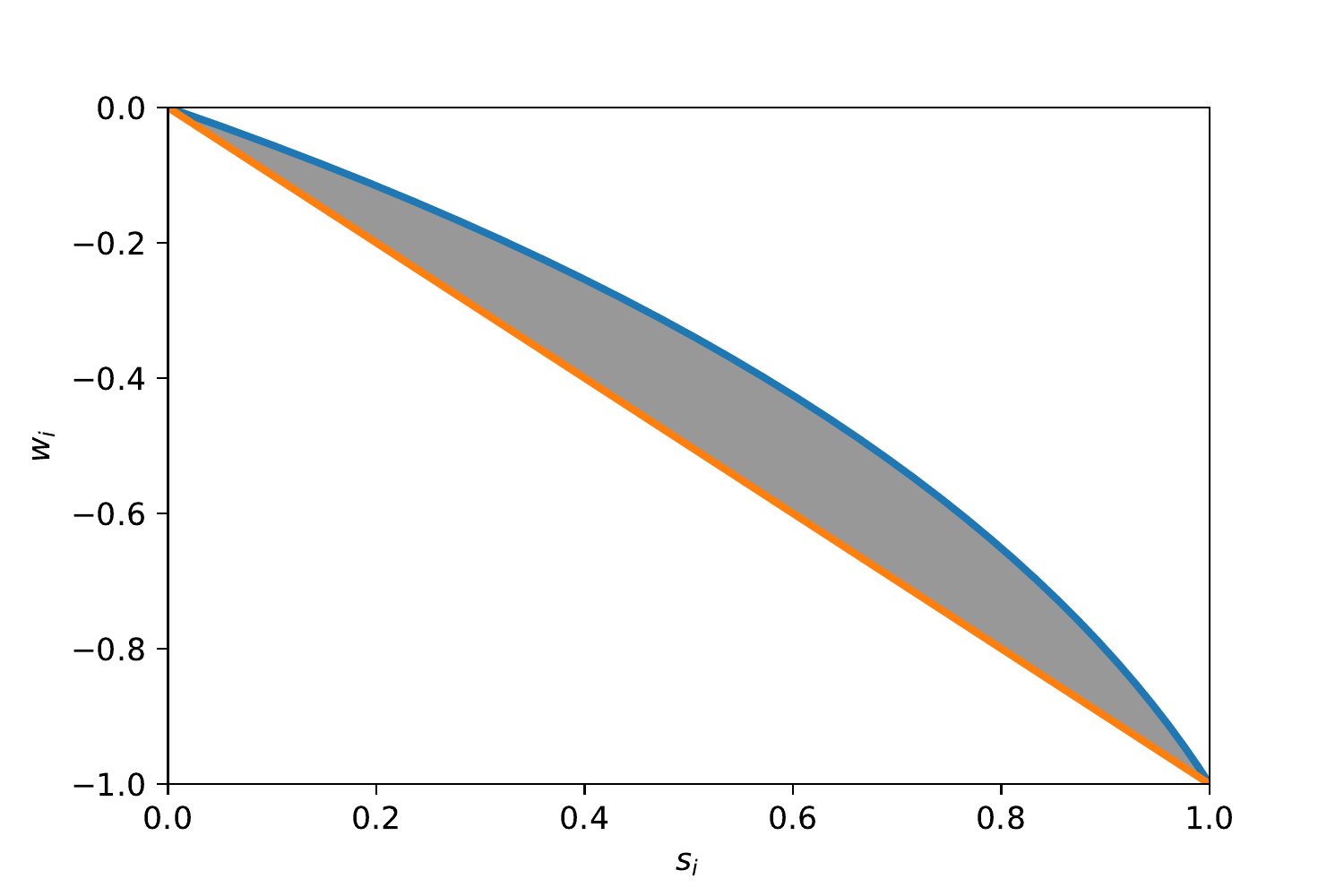}
		\caption{An illustration of the set $S_i$ for $(z_i, w_i)$, which is the region bounded by (i) the line segment between $(0,0)$ and $(1,-1)$ and (ii) the arc $\Gamma_i$ (part of a quadratic curve) on the top-right of it. In this figure, we use $d_1 = 1.5$, $d_2 = 0.8$. When $d_{i+1} = d_i$, the region becomes the line segment itself.}
		\label{fig:Si-illustr}
	\end{figure}

	When $d_i > d_{i+1}$, $\Gamma_i$ is part of a quadratic curve connecting $(0,0)$ and $(1,-1)$.
	By the last inequality in \eqref{eq:range-quad-expr-a} (with $d_{i+1} - d_i < 0$ and $0<a_i<1$), $\Gamma_i$ lies on the top-right of the line segment between the two points.
	In this case, the set $S_i$ is the region between $z_i = 0$, $w_i =-1$ and $\Gamma_i$. 
	See Figure~\ref{fig:Si-illustr} for an illustration.
	Let the \emph{entire} curve be 	
	\[ \bar{\Gamma}_i := \left\{ \begin{bmatrix}
		\frac{c_i}{2} \alpha^2 + d_i \alpha \\
		-\frac{c_{i+1}}{2} \alpha^2 - d_{i+1} \alpha
	\end{bmatrix}: \alpha \in \RR \right\}. \] 
	It is a parabola, since it is the image of the standard parabola $\{ (\alpha, \alpha^2): \alpha\in \RR \}$ under a linear transformation given by $G_i$:
	\[ G_i \begin{bmatrix}
		\alpha \\ \alpha^2
	\end{bmatrix} = \begin{bmatrix}
		\frac{c_i}{2} \alpha^2 + d_i \alpha \\
		-\frac{c_{i+1}}{2}\alpha^2 - d_{i+1} \alpha
	\end{bmatrix}. \]
	Hence, the convex hull $\conv(\bar{\Gamma}_i)$---the set of convex combinations of any finite number of points on $\bar{\Gamma}_i$---is also the image of the epigraph of the standard parabola $\mathcal{C}=\conv(\{ (\alpha, \alpha^2):\alpha\in \RR \})$ under the same linear transformation. By the convexity of $\conv(\bar{\Gamma}_i)$, we have
	\[ S_i = \conv(\bar{\Gamma}_i) \cap T_i, \]
	where 
	\[ T_i = \left\{ (z_i, w_i): z_i \in [0,1],\, w_i \in [-1,0],\, z_i + w_i \geq 0 \right\}. \]
	Therefore, the set $S_i$ can be represented by the linear constraints in $T_i$ and $ (z_i, w_i)\in G_i \mathcal{C}$. 
	The latter can be expanded with two additional variables $s_i, t_i$:
	\[ G_i \begin{bmatrix}
		s_i \\ t_i
	\end{bmatrix} = \begin{bmatrix}
		z_i \\ w_i
	\end{bmatrix},\ \begin{bmatrix}
		s_i \\ t_i
	\end{bmatrix} \in \mathcal{C}. \]
	Note that the above hold for both $d_{i+1} = d_i$ and $d_{i+1} > d_i$ (if $d_{i+1} = d_i$, then $G_i$ maps the parabola $\{(t_1, t_1^2): t_1 \in \RR \}$ into the straight line $\{ (z_i, w_i): z_i + w_i = 0 \}$ and the above characterization still works).
	Substituting this and the constraints in $T_i$ into \eqref{eq:rep-U-u,z,w,a}, we obtain the desired set of constraints that characterize $u\in U(v, [0,1])$.
	
	Note that we do not need to include $z_i + w_i \geq 0$ in the final set of constraints: this is the same as enlarging $S_i$ to contain $(z_i, w_i)$ such that $z_i + w_i < 0$, $(z_i, w_i)\in [0,1]\times [-1, 0]$, $(z_i, w_i) \in G_i \mathcal{C}$. Doing so not affect the feasible region of $u$.

	Finally, we can also easily verify that $S_i$ is the image under linear transformation $G_i$ of the convex hull of the parabola segment $\mathcal{C}^0 = \{ (s_i, s_i^2): s_i \in [0,1] \}$:
	\[ \conv(\mathcal{C}^0) = \{ 0\leq s_i \leq 1,\, 0 \leq t_i \leq 1,\, s_i^2 \leq t_i \}. \]
	Hence, the set of constraints (in particular, $z_i + w_i \geq 0$ and $(z_i,w_i)\in [0,1] \times [-1,0]$) imply
	\[0\leq s_i, t_i \leq 1. \]

\subsection*{Proof of Theorem~\ref{thm:U-conic-rep-general-vi}}

Consider $\tilde{v}_i(\theta) = (h-l)^2 c_i \theta + (h-l)(c_i l + d_i)$, $\theta\in [0,1]$ and $\varphi(\theta) = \frac{\theta-l}{h-l}$. For any $[a,b]\subseteq [l,h]$, we have
\begin{align*}
	\tilde{v}_i([\varphi(a), \varphi(b)]) & = \frac{(h-l)^2 c_i}{2}\left( \frac{(b-l)^2}{(h-l)^2} - \frac{(a-l)^2}{(h-l)^2} \right) + (h-l)(c_i l + d_i) \left( \frac{b-l}{h-l} - \frac{a-l}{h-l} \right) \\
	& = (b-a)\left( \frac{c_i}{2}(a+b) + d_i \right) = v_i([a,b]).
\end{align*}
Therefore, for any $u\in \RR_+^n$ such that $u_i = v_i([l_i, h_i])$ for a.e.-disjoint intervals $[l_i, u_i]\subseteq [l,u]$, we have 
\[ u_i = \tilde{v}_i([\varphi(l_i), \varphi(h_i)]),\, i\in [n] \]
and $[ \varphi(l_i), \varphi(h_i) ]\subseteq [0,1]$ are also a.e.-disjoint intervals. Let $\tilde{c}_i = (h-l)^2 c_i$ and $\tilde{d}_i = (h-l)(c_i l + d_i)$. By Lemma~\ref{lemma:all-linear-represent-any-u}, we have
\begin{align*}
	U(v, [l,h])
	&= \left\{ u\in \RR_+^n: u_i = \langle v_i, x_i \rangle,\, x_i \in L_{\infty}([l,h])_+,\, i\in [n],\, \sum_i x_i \leq \ones \right\} \\
	&= \left\{ u\in \RR_+^n: \text{$u_i = v_i([l_i, h_i])$ for a.e. disjoint $[l_i, h_i]\subseteq [l,h]$, $i\in [n]$} \right\} \\
	& = \left\{ u\in \RR_+^n: \text{$u_i = \tilde{v}_i([\tilde{l}_i, \tilde{h}_i])$ for a.e. disjoint $[\tilde{l}_i, \tilde{h}_i]\subseteq [0,1]$, $i\in [n]$} \right\} \\
	& = \left\{ u\in \RR_+^n: u_i = \langle \tilde{v}_i, x_i \rangle,\, x_i \in L^1([0,1])_+,\, i\in [n],\, \sum_i x_i \leq \ones \right\}\\
	& = U(\tilde{v}, [0,1]).
\end{align*}
Furthermore, let $\hat{v}_i = \tilde{v}_i/\|\tilde{v}_i\|$, where \[\|\tilde{v}_i\| = \tilde{v}_i([0,1]) =\frac{(l-h)^2c_i}{2} + (l-h)(c_i l + d_i) = \Lambda_i.\] 
The coefficients of $\hat{v}_i$ are $\hat{c}_i$ = $\tilde{c}_i / \|\tilde{v}_i\|$ and $\hat{d}_i = \tilde{d}_i / \|\tilde{v}_i \|$, which are the same as defined in the theorem statement. Then,
\[U(\tilde{v}, [0,1]) = D U(\hat{v},[0,1]) = \left\{D\hat{u}: \hat{u} \in U(\hat{v}, [0,1]) \right\}, \]
where $D \in \RR^{n\times n}$ is a diagonal matrix with $D_{ii} = \|\tilde{v}_i\| = \Lambda_i$. 
Let $P\in \{0,1\}^{n\times n}$ be the permutation matrix defined in the theorem statement. Then, 
\[ U(\hat{v}, [0,1]) = PU(\hat{v}_\sigma, [0,1]), \]
since permutation does not affect the feasibility of $x$. By Theorem~\ref{thm:U-conic-rep}, $U(\hat{v}_\sigma, [0,1])$ can be represented by $O(n)$ linear and quadratic constraints using $O(n)$ auxiliary variables.
Therefore, 
\[ U(v, [l,u]) = U(\tilde{v}, [0,1]) = D P U(\hat{v}_\sigma, [0,1]) \]
can also be represented by $O(n)$ linear and quadratic constraints using $O(n)$ auxiliary variables. 

\subsection*{Proof of Theorem~\ref{lemma:eg=>u(ik)-facts}}
First, by Lemma~\ref{lemma:U-convex-compact}, each $U_k$ is convex and compact. W.l.o.g., Assume not all $v_i$ are $0$ on $[a_{k-1}, a_k]$ (otherwise, $U_k = \{\mathbf{0}\}$ is a singleton of the $n$-dimensional zero vector and we can remove this $k$ in all summations in the following analysis).
For any $u\in U(v, [0,1])$, there exists $x\in L^\infty([0,1])_+^n$ such that $u_i = \langle v_i, x_i\rangle$.  Let $x_{ik}$ be the restriction of $x_i$ on $[a_{k-1}, a_k]$ and $u_{ik} = \langle v_i, x_{ik}\rangle$. 
Clearly, this makes the objective value of \eqref{eq:convex-prog-u(ik)} at $(u_{ik})$ equal to that of \eqref{eq:eg-primal} at $x$. Conversely, for any $(u_{ik})$ feasible to \eqref{eq:convex-prog-u(ik)}, we can also find $x$ feasible to \eqref{eq:eg-primal} that attains the same objective value. Therefore, \eqref{eq:convex-prog-u(ik)} and \eqref{eq:eg-primal} have the same optimal objective value. 
In particular, the supremum of \eqref{eq:convex-prog-u(ik)} is attained at some $(u^*_{ik})$. 
By the Pareto optimality of $(u^*_i) \in U(v, [0,1])$ (where $u^*_i$ are the unique equilibrium utilities), an optimal solution $(u^*_{ik})$ of \eqref{eq:convex-prog-u(ik)} must satisfy $u^*_i = \sum_k u^*_{ik}$ for all $i$.

By Theorems~\ref{thm:U-conic-rep} and~\ref{thm:U-conic-rep-general-vi}, each $U_k$ can be represented by $O(n)$ variables and $O(n)$ (linear and quadratic) constraints (if some $v_i$ is zero on $[a_{k-1}, a_k]$, i.e., $v_i([a_{k-1}, a_k]) = 0$, simply remove it from the set of buyers on this interval when representing the set $U_k$). 
The set $\mathcal{C}$ is the image of an transformation of the second-order cone $\mathcal{L}$:
\begin{align*}
	(t_1, t_2) \in \mathcal{C} \ \Leftrightarrow \ t_1^2 \leq t_2 \ \Leftrightarrow \ \sqrt{\left(\frac{1-t_2}{2}\right)^2 + t_1^2} \leq \frac{1+t_2}{2} \ \Leftrightarrow \ \left( \frac{1+t_2}{2}, \frac{1-t_2}{2}, t_1 \right) \in \mathcal{L}.
\end{align*}
For any $u_i>0$,
\begin{align}
	- \log u_i = \min_{q_i\geq -\log u_i} q_i = \min_{e^{q_i} \leq u_i} (-q_i) = \min_{ (u_i, 1, q_i) \in \mathcal{E}} -q_i. \label{eq:log(ui)=>min(-qi)}
\end{align}
In this way, introducing auxiliary variables $s_i, t_i, u_i$, $i\in [n]$ the objective $\max \sum_i B_i \log \left( \sum_k u_{ik} \right)$ can be written as $-\min - \sum_i B_i q_i$, with additional (linear and exponential cone) constraints
\begin{align*}
	\begin{rcases}
		u_i = \sum_k u_{ik},\\ 
		(u_i, 1, q_i)\in \mathcal{E}, \\
	\end{rcases} i\in [n].
\end{align*}
Combining the above analysis, we arrive at the overall convex conic reformulation involving only linear and (convex) conic constraints.
Recall that $i$ and $k$ are indices of buyers and linear segments of their valuations, respectively.
\begin{align}
	\begin{split}
		\min \ & - \sum_i B_i q_i \\ 
		\st \ \ 
		   & \begin{rcases}
			   & (u_i, 1, q_i) \in \mathcal{E},\ q_i'=1, \\
			   & u_i = \sum_k u_{ik}, \\
			   & u_{\sigma^k(i) k} = \Lambda_{\sigma^k(i)k} \hat{u}_{ik}, 
		   \end{rcases} \quad  \forall\, i \in [n], \\[2ex]
		   & \begin{rcases}
				& \hat{u}_{1k} \leq z_{1k}, \\
				& \hat{u}_{ik} \leq z_{ik} + w_{i-1,k}, \ \forall\, i=2,\dots, n-1,\\
				& \hat{u}_{nk} \leq 1 + w_{nk}, \\
				& G_{ik}(s_{ik}, t_{ik}) = (z_{ik}, w_{ik}), \ \left(\frac{1+t_{ik}}{2}, \frac{1-t_{ik}}{2}, s_{ik}\right) \in \mathcal{L}, \\
				& 0\leq z_{ik} \leq 1, \ -1 \leq w_{ik} \leq 0,\ z_{ik}+w_{ik}\geq 0,\ \forall\, i \in [n],
			\end{rcases} \quad\quad \forall\, k \in [K].\\[2ex]
	\end{split}
	\label{eq:overall-conic-reform} \tag{$\mathcal{CP}$}
\end{align}
In the above, the first group of constraints involving objective transformation, decomposition of buyers' utilities and scaling of them over subintervals.The second group represents $(\hat{u}_{1k}, \dots, \hat{u}_{nk}) \in U(\hat{v}_k, [0,1])$, where $U(\hat{v}_k, [0,1])$ is the ``standardized'' version of $U_k$ as described in Theorem~\ref{thm:U-conic-rep-general-vi}.
To complete the proof, we calculate the number of each type of variables in the final reformulation \eqref{eq:conic-std-form} below. 
\begin{itemize}
	\item The exponential cone variables are $(q_i, q'_i, u_i)\in \mathcal{E}$, $i\in [n]$ (with additional linear constraint $q'_i = 1$). Hence, $n_3 = O(n)$.
	\item For each $(s_{ik}, t_{ik})$, there is a second-order cone $\mathcal{L}$ (involving $3$ conic variables, $t_{ik}' = \frac{1+t_{ik}}{2}$, $t_{ik}'' = \frac{1-t_{ik}}{2}$ and $s_{ik}$). Hence, $n_2 = O(nK)$.
	\item The linear cone (nonnegative) variables are $u_{ik}$, $\hat{u}_{ik}$, $i\in [n]$, $k\in [K]$, $(z_{ik}, w'_{ik})$, $i\in [n-1]$, $k\in [K]$ ($w'_{ik} = -w'_{ik}\geq 0$) and nonnegative auxiliary variables added to transform the $O(nK)$ linear inequality constraints (in the second group of constraints in \eqref{eq:overall-conic-reform}) into equality constraints. Hence, $n_1 = O(nK)$.
	\item The linear equality constraints (which form ``$Ax=b$'' in the standard form \eqref{eq:conic-std-form}) are those above in \eqref{eq:overall-conic-reform} plus additional ones involving auxiliary conic variables: $q''_i = 1$, $t_{ik}' = \frac{1+t_{ik}}{2}$, $t_{ik}'' = \frac{1-t_{ik}}{2}$. Hence, $m = O(nK)$. 
	\item It can be easily verified that, each of the linear constraints in \eqref{eq:overall-conic-reform} (except the linear constraints $u_i = \sum_{k} u_{ik}$, $k\in[K]$), involve only a constant number of variables. 
	In particular, the linear equality constraints of $U_k = D^k P^k \hat{U}_k$ (i.e., last equality constraint in the first group of constraints in \eqref{eq:overall-conic-reform}) only consist of $(n)$ nonzeros, since $D^k$ and $P^k$ are diagonal and permutation matrices, respectively. 
	Hence, the total number of nonzeros in all linear constraints is also $O(nK)$.
\end{itemize} 
By \eqref{eq:log(ui)=>min(-qi)}, we know that the minimum $f^*$ of \eqref{eq:conic-std-form} is $-z^*$. Finally, the above reformulation does not affect feasible region of the variables $(u_i)$ and $(u_{ik})$.
Hence, in the optimal solution of the reformulation \eqref{eq:conic-std-form}, these variables correspond to an optimal solution $(u^*_{ik})$ of \eqref{eq:convex-prog-u(ik)}.

\subsection*{Proof of Theorem~\ref{thm:ellipsoid-method-for-u(ik)-convex-program}}
To make use of the above theorem, we first case our problem \eqref{eq:convex-prog-u(ik)} into \eqref{eq:generic-constrained-minimization}. Same as in the paragraph below \eqref{eq:convex-prog-u(ik)}, 
let the decision variables be $(u_i)$, $(u_{ik})$, $(\hat{u}_{ik})$, $(s_{ik})$, $(t_{ik})$, $(z_{ik})$, $(w_{ik})$, with a total number of $O(nK)$. 
Here, the variables $(u_i)$, $(u_{ik})$ correspond to those in \eqref{eq:convex-prog-u(ik)}. 
The variables $(\hat{u}_{1k}, \dots, \hat{u}_{nk})$ are used to describe each ``normalized'' set of feasible utilities $\hat{U}_k$ such that $U_k = D^k P^k \hat{U}_k$.
Denote the aggregate decision variable as $x$, which has $O(nK)$ dimensions. The (minimization) objective function is 
\[ f(x) = - \sum_i B_i \log u_i.\] 
Next, we specify the feasible region. 
Recall that $u^*_i \in [B_i, 1]$ for all $i$ at equilibrium (Lemma~\ref{lemma:equi-bounds}). 
Hence, we can add linear constraints 
\begin{align}
	\min\{B_i, \epsilon/2\} \leq u_i \leq 1
	\label{eq:ellipsoid-complexity-proof-u_i-bounds}
\end{align}
without affecting the optimal solution. 
Then, ``enlarge'' the feasible sets given by the constraints by $\epsilon$ to ensure a nonempty interior of the feasible region.
\begin{itemize}
	\item For each $i$, relax the equality constraint $u_i = \sum_k u_{ik}$ into $u_i \leq \sum_k u_{ik} + \epsilon$ for all $k$ (when there is no tolerance $\epsilon$, the equality constraint can clearly be relaxed to an inequality without affecting the optimum, since at optimality the inequality must be tight).
	\item For each $k$, $(u_{1k}, \dots, u_{nk})\in U_k + [0,\epsilon]^n$.
	\item For each $k$, for every linear constraint describing $U_k = D^k P^k \hat{U}_k$, that is, $u_{\sigma^k(j)k} = \Lambda_{\sigma(j) k} \hat{u}_{jk}$, relax it into $u_{\sigma^k(j)k} = \Lambda_{\sigma(j) k} \hat{u}_{jk} + \epsilon$.
	\item For each $k$, in the linear and quadratic constraints for $\hat{U}_k$, add $\epsilon$ to \emph{all} linear constraints involving $\hat{u}_{ik}$ and $z_{ik}$ or $w_{ik}$ (e.g., $\hat{u}_{ik} \leq z_{ik} + w_{i-1,k} + \epsilon$).
	\item For each $k$, relax the constraints $G_{ik} (s_{ik}, t_{ik}) = (z_{ik}, w_{ik})$ (Theorem~\ref{thm:U-conic-rep}, with the ``$k$th'' copies of the variables and the $G_i$ matrix) into $G_{ik} (s_{ik}, t_{ik}) = (z_{ik}, w_{ik}) + (\epsilon, \epsilon)$.
	\item For each $i$ and $k$, keep the constraints $0\leq z_{ik}\leq 1$, $-1\leq w_{ik}\leq 1$, $(s_{ik}, t_{ik})\in\mathcal{C}$ unchanged.
\end{itemize}
Constructing these constraints takes $O(nK\log n)$ time, where the $n\log n$ factor is due to sorting $\hat{d}_{1k}, \dots, \hat{d}_{nk}$ for each $k$ (see Theorem~\ref{thm:U-conic-rep-general-vi}). It is negligible compared to the running time of the ellipsoid method. 
Now, it can be easily verified that the feasible region (described by all constraints above) contains a Euclidean ball of radius $r \geq \epsilon/2$. The objective function is also convex and continuous on the feasible region. Furthermore, the total number of constraints is $O(nK)$ and each constraint only involves a constant number of variables (except $u_i = \sum_k u_{ik}$ which involves $K$ variables).

Next, we bound $R = \max_{x\in X} \|x\|$. 
\begin{itemize}
	\item Each of $z_{ik}$, $w_{ik}$, $s_{ik}$, $t_{ik}$ have absolute values $\leq 1$. There are $O(nK)$ such variables. 
	\item The variables $\hat{u}_{ik}$ have absolute values $\leq 1+\epsilon$. There are $O(nK)$ such variables. 
	\item The variables $u_{ik}$ have absolute values $\leq 1+\epsilon$, since $\Lambda_{ik} = v_i([a_{k-1}, a_k]) \leq 1$. There are $O(nK)$ such variables. 
	\item The variables $u_i$ have absolute values $\leq 1$, as we added the constraints \eqref{eq:ellipsoid-complexity-proof-u_i-bounds}. There are $n$ such variables.
\end{itemize}
Hence, $R = O(\sqrt{nK})$.

To bound $V$, first note that 
\[ f(x) \geq -\sum_i B_i \log 1 = 0. \]
Then, since $\|B\|_1 = 1$, 
\[ f(x) \leq -\sum_i B_i \log \min\{ B_i, \epsilon/2 \} \leq \sum_i B_i \log \max\{\kappa, 2/\epsilon \} \leq \log \kappa + \log \frac{2}{\epsilon}. \]
Therefore ($f^* = \min_{x\in X} f(x)$), 
\[ V = \max_{x\in X}f(x) - f^* \leq \log \kappa + \log \frac{2}{\epsilon}. \]

Hence, the overall ratio $\frac{VR}{\epsilon r}$ in the expression of the time complexity of the ellipsoid method in Theorem~\ref{thm:4.1.2-ellipsoid-bn-notes} is
\begin{align}
	\frac{VR}{\epsilon r} = O\left( \frac{\sqrt{nK} \left( \log \kappa + \log \frac{2}{\epsilon} \right) }{\epsilon^2} \right).
	\label{eq:ellipsoid-ratio-VR/eps*r}
\end{align}

\paragraph{The two oracles.}
The first-order oracle is trivial: the objective function is differentiable w.r.t. $u_i$ and a subgradient simply consists of the derivatives $\frac{B_i}{u_i}$. This oracle takes $T_\mathcal{G} = O(n)$ time. 
Next, we describe the separation oracle. Given a solution $x^0$ (consisting of $O(nK)$ variables in total), it clearly takes $O(nK)$ time to verify whether all constraints are satisfied. Suppose not all constraints are satisfied. There are two cases.
\begin{itemize}
	\item A linear constraint is violated, say, $g^\top x^0 > a$ while $g^\top x \leq a$ for all $x\in X$. Then, this constraint itself is a separating hyperplane.
	\item A quadratic constraint is violated, say, $(s_{ik}, t_{ik}) \notin \mathcal{C} = \{(t_1, t_2): t_1^2 \leq t_2\}$. By elementary calculus, the line
	\[ \{ (t_1, t_2): t_2 - s_{ik}^2 = (2s_{ik}) (t_1 - s_{ik}) \}\] is tangent to the curve $\{(t_1, t_2): t_1^2 = t_2\}$ at the point $(s_{ik}, s_{ik}^2)$ on the curve. Hence, it separates $\mathcal{C}$ and $(s_{ik}, t_{ik})$.
\end{itemize}
Since there are $O(nK)$ linear and quadratic constraints in total, the separation oracle described above takes $T_\mathcal{S} = O(nK)$ time. 

By Theorem~\ref{thm:4.1.2-ellipsoid-bn-notes}, the ellipsoid method finds a solution $x_\epsilon$ such that $f(x_\epsilon) - f^* \leq \epsilon$ in $N(\epsilon)$ number of calls of the oracles and $O(1)n^2N(\epsilon)$ additional arithmetic operations, where 
\[ N(\epsilon) = O(1) (nK)^2 \log \left( 2 + \frac{VR}{\epsilon r} \right). \]
Combining the above, \eqref{eq:ellipsoid-ratio-VR/eps*r} and the time complexity of the oracles, the overall time complexity for computing $x_\epsilon$
\begin{align}
	& N(\epsilon)(T_\mathcal{S} + T_\mathcal{G}) + O(1)(nK)^2 N(\epsilon) \nonumber \\
	&= N(\epsilon)\left(O(n) + O(nK) + O((nK)^2) \right) \nonumber \\
	& = O\left( (nK)^4 \log \frac{ \sqrt{nK}\left(\log \kappa + \log \frac{2}{\epsilon}\right) }{\epsilon} \right) \nonumber \\
	& = O\left( (nK)^4 \log \frac{ nK \log \kappa }{\epsilon} \right). 
	\label{eq:ellipsoid-comp-for-eps-solution-proof}
\end{align} 

Here, since the feasible region has been enlarged, the minimum $f^* \leq -z^*$, where $z^*$ is the true maximum of \eqref{eq:convex-prog-u(ik)}. Therefore, $x_\epsilon$ is also $\epsilon$-close to the ``true'' minimum $-z^*$. 
Furthermore, we can transform the difference in objective value to the difference in utilities using strong convexity (where $u_i$ are part of the solution $x_\epsilon$):
\begin{align*}
	\left(-\sum_i B_i \log u_i\right) - \left( -\sum_i B_i \log u^*_i \right) \geq \frac{\sigma}{2}\|u-u^*\|^2,
\end{align*}
where a strong convexity modulus is $\sigma = \min_i B_i = \frac{1}{\kappa}$, since $u_i \leq 1$.
Therefore, for each $i$,
\begin{align}
	|u_i - u^*_i| \leq \|u-u^*\| \leq 2\epsilon \kappa.
	\label{eq:ui-vs-u*-ellipsoid-solution-proof}
\end{align}

To recover a fully feasible solution $(u_{ik})$ such that $(u_{1k}, \dots, u_{nk}) \in U_k$ and $\sum_k u_{ik} \geq u^*_i - \epsilon$, it suffices to ``discount'' $x_\epsilon$ as follows.
\begin{itemize}
	\item Decrease each $\hat{u}_{ik}$ by $\epsilon$ (that is, $\hat{u}'_{ik} = \max\{ \hat{u}_{ik}-\epsilon,0 \}$), which requires each $u_{ik}$ ($i=\sigma^k(j)$) decrease by $\Lambda_{ik} \epsilon\leq \epsilon$ to ensure $u_{ik} \leq \Lambda_{ik} \hat{u}_{jk} + \epsilon$.
	\item Further decrease each $u_{ik}$ by $\epsilon$ ($u_{ik} = \max\{ u_{ik} - \epsilon, 0 \}$), which makes $u_i$ decrease by at most $(K+1) \epsilon$ to ensure $ u_i \leq \sum_k u_{ik}$.
\end{itemize}
We still use $(u_{ik})$ to denote the processed solution according to the above ``discounting'' procedure. This solution satisfies $(u_{1k}, \dots, u_{nk})\in U_k$ for all $k$ and $\sum_k u_{ik} \geq u_i - (K+1)\epsilon$ for all $i$. Combining this with \eqref{eq:ui-vs-u*-ellipsoid-solution-proof}, we know that the processed solution $(u_{ik})$ approximately attains the equilibrium utilities:
\[ \sum_i u_{ik} \geq u^*_i - 2\kappa \epsilon - (K+1)\epsilon. \]
Therefore, to ensure that $\sum_k u_{ik} \geq u^*_i - \epsilon$ for a given tolerance level $\epsilon>0$, it suffices to replace $\epsilon$ by $\frac{\epsilon}{2\kappa+ (K+1)}$, which slightly degrades the time complexity in \eqref{eq:ellipsoid-comp-for-eps-solution-proof}  (due to $\kappa$) and yields a final time complexity of 
\[ O\left( (nK)^4 \left(\log (nK) + \log \frac{\kappa}{\epsilon}  \right) \right). \]

To construct a pure equilibrium allocation given $(u_{ik})$ is easy: for each $k$, since $(u_{1k}, \dots, u_{nK})\in U_k$, simply run Algorithm~\ref{alg:interval-partition-given-feasible-u} on $[a_{k-1}, a_k]$, where no sorting is required as we reuse the sorting permutations in formulating the convex program \eqref{eq:convex-prog-u(ik)}. This produces a.e.-disjoint intervals $[l_{ik}, h_{ik}]\subseteq [a_{k-1}, a_k]$ such that $v_i([l_{ik}, h_{ik}]) \geq u_{ik}$. Therefore, the pure allocation $\Theta_i := \cup_k [l_{ik}, h_{ik}]$ satisfies
\[ v_i(\Theta_i) \geq \sum_k u_{ik} \geq u^*_i - \epsilon. \]

\subsection*{Proof of Lemma~\ref{lemma:subgrad-w.r.t.-beta-fixed-theta}}
The first half is well-known, see, e.g., \cite[Theorem 3.50]{beck2017first}. In fact, by this theorem, the subdifferential (the convex set of all subgradients) is
\[ \partial_\beta \phi(\beta, \theta) = \conv\left\{ v_{i}(\theta)\mathbf{e}^{(i)}:  i\in \argmax_i \beta_i v_i(\theta) \right\}. \]
For the second half, note that for any $\beta > 0$, since $g(\beta, \theta)\in \partial_\theta \phi(\beta, \theta)$  we have 
\[ \phi(\beta', \theta) - \phi(\beta, \theta) \geq \langle g(\beta, \theta), \beta' - \beta\rangle,\ \forall\,\beta'>0. \]
Integrate w.r.t. $\theta$ over $\Theta$ on both sides yield
\[ \phi(\beta') - \phi(\beta) \geq \int_{\Theta} \langle g(\beta, \theta), \beta'-\beta\rangle d\theta = \left \langle \int_{\Theta} g(\beta, \theta)d\theta, \beta' - \beta\right \rangle, \]
where the (component-wise) integral $\int_{\Theta} g(\beta, \theta) d\theta$ is well-defined and finite, since each component of $g(\beta, \theta)$ is uniformly bounded by the pointwise maximum $\max_i v_i$, which is integrable:
\[ 0 \leq \max_i v_i \leq \sum_i v_i \in L^1(\Theta). \]
Therefore, by the definition of subgradient, 
\[ \int_\Theta g(\beta, \theta)d\theta \in \partial \phi(\beta). \]
The integral can also be written as an expectation over $\theta\sim {\rm Unif}(\Theta)$, since the probability density of ${\rm Unif}(\Theta)$ is $\frac{1}{\mu(\Theta)}$ for all $\theta\in \Theta$.

\subsection*{Proof of Theorem~\ref{thm:sda-conv}}
The bounds on $\EE \|\beta^t - \beta^* \|^2 $ and = $\EE\|\tilde{\beta}^t - \beta^*\|^2$ are derived directly from the proof of \cite[Corollary 4]{xiao2010dual}. 
Here, the regularizer $\Psi(\beta) = -\sum_i B_i \log \beta_i$ has domain $[B, \ones] = \prod_i [B_i, 1]$. 
For $\beta_i \in [B_1, 1]$, we have
\[ \frac{\partial^2}{\partial \beta_i^2} (-B_i \log \beta_i) = \frac{B_i}{\beta_i^2} \geq B_i. \]
Hence, $\Psi$ (and the entire objective function) is strongly convex on $[B, \ones]$ with modulus $\sigma = \min_i B_i$ (c.f. Lemma \ref{lemma:equi-bounds}). 

Based on Lemma~\ref{lemma:subgrad-w.r.t.-beta-fixed-theta}, a subgradient can be computed as $g^t = g(\beta^t, \theta_t) = v_{i_t}(\theta_t)\cdot \mathbf{e}^{(i_t)}$, where $i_t \in \argmax \beta^t_i v_i(\theta_t)$ (choosing the smallest index in a tie). 
Hence, we have
\[\| g^t \|^2 = v_{i_t}(\theta_t)^2 \leq \max_i v_i(\theta_t). \]
Therefore, 
\[ \EE \|g_t\|^2 \leq \EE_\theta [\max_i v_i(\theta)^2] = G^2. \]
Here, $\max_i v_i \in L^2(\Theta)$ since
\[  0\leq \max_i v_i \leq \sum_i v_i \in L^2(\Theta). \]

The second half is derived in a straightforward manner from the discussion in \cite[pp. 2559]{xiao2010dual}. Note that
\[ \Delta_t = \frac{G^3}{2\sigma}(6 + \log t) \] 
is an upper bound on the regret in iteration $t$ in an online optimization setting \cite[\S 3.2, Eq. (20)]{xiao2010dual}. 
The constant $V$ here is an upper bound on the difference between maximum and minimum attainable objective values of \eqref{eq:sda-reg-std-form} (across all $\theta\in \Theta$). 
Since $ B_i \leq \beta_i \leq 1$, we have ($\sigma = \min_i B_i$)
\[0 \geq \sum_i B_i \log \beta_i \geq \sum_i B_i \log B_i \geq \log \sigma \]
and 
\[ 0 \leq \langle \max_i \beta_i v_i, \ones \rangle \leq \left \langle \sum_i v_i , \ones \right \rangle = n. \]
Hence,
\[ V \leq n - \log \sigma. \]

\subsection*{Proof of Theorem~\ref{thm:ql-me-duality-and-eq}}
\paragraph{Proof of Part~\ref{item:ql-primal-attainment}.}
Let $U = U(v, \Theta)$ be the set of feasible utilities defined in \eqref{eq:def-U-U(v,Theta)}.
By Lemma~\ref{lemma:U-convex-compact}, we know that $U$ is convex and compact. Hence, we can reformulate the convex program as one involving $u\in \RR_+^n$ and $\delta\in \RR_+^n$:
\begin{align}
    \begin{split}
        w^* = \sup\, & \sum_i (B_i \log u_i - \delta_i) \\
        \st & u - \delta \in U,\\ 
        & u, \delta\in \RR_+^n.
    \end{split}    
    \label{eq:ql-primal-in-u-delta}
\end{align}
Note that $w^*$ is finite: taking $u = \delta = (1, \dots, 1) \in \RR^n$ gives an objective value of $-n$, while the objective is bounded above by $\sum_i B_i \log v_i(\Theta) < \infty$. 
Similar to the proof of Part~\ref{item:eg-primal-attain} of Theorem~\ref{thm:eg-equi-opt-combined}, let $w_0 = - n \leq w^*$ and consider the set
\[ F = \left\{ (u, \delta) \in \RR_+^n \times \RR_+^n: u-\delta\in U,\, \sum_i (B_i \log u_i - \delta_i) \geq w_0 \right\}. \]
The set $F$ is convex and compact, on which the objective function is finite and continuous. Hence, 
\[ \sup \left\{\sum_i (B_i \log u_i - \delta_i): (u, \delta)\in F \right\} \]
is attained via some $(u^*, \delta^*) \in F$. The solution $(u^*, \delta^*)$ is also feasible to \eqref{eq:ql-primal-in-u-delta} and attains its supremum, since $F$ ensures feasibility and the level set constraint $\sum_i (B_i \log u_i - \delta_i) \geq w_0$ only excludes suboptimal solutions with objective value $< w_0$. Since $u^*\in U$, it can be attained by some feasible allocation $x^* \in (L^\infty(\Theta)_+)^n$, $\sum_i x^*_i \leq \ones$. Finally, Lemma~\ref{lemma:U-convex-compact} also ensures that $u^*\in U$ can be attained by a pure solution $x^*$, $x^*_i = \ones_{\Theta_i}$ for a.e.-disjoint subsets $\Theta_i \subseteq \Theta$.

\paragraph{Proof of Part~\ref{item:ql-dual-attainment}}
Note that, for any fixed $\beta\in \RR_+^n$, setting $p = \max_i \beta_i v_i$ minimizes the objective of \eqref{eq:ql-eg-dual-p-beta} subject to the constraints $p \geq \beta_i v_i$, $\forall\, i$. Hence, we can rewrite \eqref{eq:ql-eg-dual-p-beta} in terms of $\beta\in \RR_+^n$, which is a finite-dimensional convex program with a finite, strongly convex objective function (c.f. Part~\ref{item:eg-dual-p-beta-attain} of Theorem~\ref{thm:eg-equi-opt-combined}).
The proof of Lemma~\ref{lemma:beta-dual-attain} implies that
there is a unique optimal solution $\beta^*$ of \eqref{eq:ql-eg-dual-p-beta}, which means an optimal solution $(p^*, \beta^*)$ must satisfy $p^* = \max_i \beta^*_i v_i$ a.e.

Before proving the next parts, we first establish weak duality. 

\paragraph{Weak duality.}
Similar to the proof of Lemma~\ref{lemma:weak-duality}, we first establish weak duality.
For any $(x, u, \delta)$ feasible to \eqref{eq:ql-eg-primal} and any $(p, \beta)$ feasible to \eqref{eq:ql-eg-dual-p-beta}, 
\begin{align}
    \sum_i (B_i \log u_i - \delta_i) 
    & \leq \sum_i (B_i \log u_i - \delta_i) - \sum_i \beta_i (u_i - \langle v_i, x_i \rangle - \delta_i) - \left\langle p, \sum_i x_i - \ones \right\rangle \nonumber \\
    & \leq \sum_i \left( B_i \log \frac{B_i}{\beta_i} - \delta_i \right) - \sum_i \beta_i\left( \frac{B_i}{\beta_i} - \langle v_i, x_i \rangle - \delta_i \right) - \left\langle p, \sum_i x_i - \ones \right\rangle \nonumber \\ 
    & = \sum_i (\beta_i-1)\delta_i + \sum_i \langle \beta_i v_i - p, x_i \rangle + \langle p, \ones\rangle - \sum_i B_i \log \beta_i + \sum_i B_i (\log B_i - 1) \nonumber \\
    & \leq \langle p, \ones \rangle - \sum_i B_i \log \beta_i + \sum_i B_i (\log B_i - 1), \label{eq:ql-ineq-chain}
\end{align}
where the second inequality is because $u_i = \frac{B_i}{\beta_i}$ maximizes $(u_i \mapsto B_i \log u_i - \beta_i u_i)$ and the other inequalities easily follow from the feasibility assumptions on $(x, u, \delta)$ and $(p, \beta)$.
Let the supremum of \eqref{eq:ql-eg-primal} be $z^*$ and the infimum of \eqref{eq:ql-eg-dual-p-beta} be $w^*$. Then, the above inequalities imply
\[ z^* \leq w^* + \sum_i B_i (\log B_i - 1). \]

\paragraph{Strong duality and proof of Part~\ref{item:ql-kkt}.} 
We list the KKT conditions again for convenience. 
\begin{align}
	& \left \langle p^*, \ones - \sum_i x^*_i \right \rangle = 0, \label{eq:ql-market-clear} \\
	& u^*_i := \frac{B_i}{\beta^*_i}, \ \forall\, i, \label{eq:ql-u=B/beta} \\
	& \delta^*_i (1 - \beta^*_i) = 0,\ \forall\, i, \label{eq:ql-delta*(1-beta)=0} \\
	& \langle p^* - \beta^*_i v_i, x^*_i \rangle = 0, \ \forall \, i. \label{eq:ql-alloc-only-winning}
\end{align}
Both primal and dual optima ($z^*$ and $w^*$) are attained, by $(x^*, \delta^*)$ and $(p^*, \beta^*)$, respectively, if and only if the inequalities in \eqref{eq:ql-ineq-chain} are all tight. When this happens, 
the first inequality being tight implies \eqref{eq:ql-market-clear}, the second inequality being tight implies \eqref{eq:ql-u=B/beta}, the last inequality being tight implies \eqref{eq:ql-delta*(1-beta)=0} and \eqref{eq:ql-alloc-only-winning}.
Conversely, for feasible solutions $(x^*, \delta^*)$ and $(p^*, \beta^*)$, the set of conditions imply that all inequalities in \eqref{eq:ql-ineq-chain} are tight, which ensure that both optima are attained.

The following two paragraphs complete the proof of Part~\ref{item:ql-eq-iff-opt}.
\paragraph{Optimality $\Rightarrow$ QLME.} Given optimal solutions $(x^*, u^*, \delta^*)$ and $(p^*, \beta^*)$ of \eqref{eq:ql-eg-primal} and \eqref{eq:ql-eg-dual-p-beta}, respectively, by the above analysis, the KKT conditions \eqref{eq:ql-market-clear}-\eqref{eq:ql-alloc-only-winning} hold. Hence, market clearance \eqref{eq:ql-market-clear} holds.
For any $x_i\in L^\infty(\Theta)_+$ such that $\langle p^*, x_i \rangle \leq B_i$, we need to show that $\langle v_i - p^*, x^*_i \rangle \geq \langle v_i - p^*, x_i\rangle$. There are two cases.
\begin{itemize}
    \item Suppose $\beta^*_i = 1$. Then, \eqref{eq:ql-alloc-only-winning} implies 
    \[ \langle v_i - p^*, x^*_i\rangle = \langle \beta^*_i v_i - p^*, x^*_i \rangle + (1-\beta^*_i)\langle v_i, x^*_i\rangle = (1-\beta^*_i)\langle v_i, x^*_i\rangle = 0. \]
    Since $p^* \geq \beta^*_i v_i$, we have
    \[ \langle v_i - p^*, x_i \rangle \leq \langle v_i - \beta^*_i v_i, x_i\rangle = (1-\beta^*_i) \langle v_i, x_i \rangle = 0 = \langle v_i - p^*, x^*_i\rangle.\]
    \item If $\beta^*_i < 1$, then, \eqref{eq:ql-delta*(1-beta)=0} implies $\delta^*_i = 0$. By the constraint $u_i \leq \langle v_i, x_i \rangle + \delta_i$ in \eqref{eq:ql-eg-primal}, we must have
    \[ u^*_i = \langle v_i, x^*_i \rangle. \]
    Using $p^* \geq \beta^*_i v_i$ (feasibility w.r.t. \eqref{eq:ql-eg-dual-p-beta}), we have 
    \[ \beta^* \langle v_i, x_i \rangle \leq \langle p^*, x_i \rangle \leq B_i = \beta^*_i u^*_i \ \Rightarrow \ \langle v_i, x_i \rangle \leq u^*_i = \langle v_i, x^*_i \rangle.  \]
    Hence, 
    \[ \langle v_i - p^*, x_i \rangle \leq \langle v_i - \beta^*_i v_i, x_i \rangle = (1-\beta^*_i)\langle v_i, x_i \rangle \leq (1-\beta^*_i) \langle v_i, x^*_i \rangle = \langle v_i - p^*, x^*_i\rangle, \]
    where the last equality is due to \eqref{eq:ql-alloc-only-winning}.
\end{itemize}
Therefore, $(x^*, p^*)$ is a QLME.

\paragraph{QLME $\Rightarrow$ optimality.}
Let $(x^*, p^*)$ be a QLME. Then, 
\[ x^* \in \argmax \left\{ \langle v_i - p^*, x_i\rangle: \langle p^*, x_i\rangle \leq B_i,\, x_i \in L^\infty(\Theta)_+  \right\}. \] 
For each $i$, construct $\beta^*$, $u^*$ and $\delta^*$ as follows.
\begin{itemize}
    \item If $\langle p^*, x^*_i \rangle < B_i$, then, by the above analysis, $\langle v_i, x^*_i \rangle = \langle p^*, x^*_i \rangle$. Set $\beta^*_i = 1$, $u^*_i = B_i$ and $\delta^*_i = u^*_i - \langle v_i, x^*_i \rangle > 0$. 
    \item If $\langle p^*, x^*_i \rangle = B_i$, then 
	\[ \langle v_i - p^*, x^*_i \rangle \geq 0 \, \Rightarrow \, \langle v_i, x^*_i \rangle \geq \langle p^*, x^*_i \rangle = B_i > 0.\] 
	Set $u^*_i = \langle v_i, x^*_i\rangle$, $\delta^*_i = 0$, and $\beta^*_i = \frac{B_i}{u^*_i} = \frac{B_i}{\langle v_i, x^*_i \rangle} \leq 1$.
\end{itemize}
Finally, set $p^* = \max_i \beta^*_i v_i$. In this way, we have $(x^*, u^*, \delta^*)$ feasible to \eqref{eq:ql-eg-primal} and $(p^*, \beta^*)$ feasible to \eqref{eq:ql-eg-dual-p-beta} that satisfy \eqref{eq:ql-market-clear}-\eqref{eq:ql-alloc-only-winning} (where \eqref{eq:ql-market-clear} is due to the market clearance property of QLME and the other three are easily verified through our construction of $\beta^*$, $u^*$ and $\delta^*$). Hence, they make all inequalities in \eqref{eq:ql-ineq-chain} tight and are both optimal to \eqref{eq:ql-eg-primal} and \eqref{eq:ql-eg-dual-p-beta}, respectively.

\section{More details on the numerical examples} \label{app:more-on-numerical}
\paragraph{Linear $v_i$. } The buyers' budgets are $B = (B_1, B_2, B_3, B_4) = (0.1, 0.3, 0.2, 0.4)$. 
Buyers have linear valuations $v_i$ with intercepts $d = (1.2, 0.6, 0.3, 1.9)$, which give $c = (-0.4,  0.8,  1.4, -1.8)$ since $v_i$ are normalized. The descending order of buyers by $d_i$ is $\sigma = (4, 1, 2, 3)$, that is, buyer $4$ should be allocated first (from left to right), and then buyer $1$, and so on.
Solving the convex program \eqref{eq:convex-prog-u(ik)} (with only $K=1$ subinterval being the entire interval $[0,1]$) yields equilibrium utilities $u^* = (0.1241, 0.3688, 0.2834, 0.5814)$. To partition $[0,1]$ into $n$ intervals, first find $a^*_1\in [0,1]$ such that $v_3([0, a^*_1]) = \frac{c_3}{2} (a^*_1)^2 + d_3 a^*_1 = u^*_1$ (since $\sigma(1) =3$, i.e., $d_3$ is the largest and buyer $3$ should get the leftmost interval); 
solving the quadratic equation gives $a^*_1 = 0.3713$ (this is the ``cut'' operation introduced in \S\ref{subseq:charact-linear-vi-[0,1]}).
Next, find $a^*_2 \in [a^*_1, 1]$ such that $v_1([a^*_1, a^*_2]) = u^*_2\ \Rightarrow\ a^*_2 = 0.4921$ ($\sigma(2) = 1$). Similarly, $\sigma(3) = 2$, $v_2([a^*_2, a^*_3]) = u^*_2 \ \Rightarrow \ a^*_3 =  0.8199$. Figure~\ref{fig:n-linear} illustrates the equilibrium $\beta^*$ (as in $\beta^*_i v_i$) and prices $p^* = \max_i \beta^*_i v_i$, where the breakpoints of $p^*$ are precisely $a^*_i$.
The allocation is as follows: buyer $1$ gets the second interval $[l_1, h_1] = [0.3713, 0.4921]$ (since $\sigma(1) = 2$), buyer $2$ gets the third interval $[l_2, h_2] = [0.4921, 0.8199]$, buyer $3$ gets the fourth interval $[l_3, h_3] = [0.8199, 1.0]$ and buyer $4$ gets the first interval $[l_4, h_4] = [0.0, 0.3713]$. Since all $v_i$ are distinct, it is also the unique pure equilibrium allocation. As illustrated in Figure~\ref{fig:n-linear}, the interval of buyer $i$ is in fact its winning set, i.e., $ [l_i, h_i] = \{ p^* = \beta^*_i v_i \}$, where
\[\beta^*_i = B_i/u^*_i \ \Rightarrow \ \beta^* = (0.8058, 0.8135, 0.7057, 0.6880).\]
To verify that the primal solution $(x^*, p^*)$, $x^*_i = \ones_{[l_i, h_i]}$ is indeed a ME, it suffices to verify that $x^*_i = \ones_{\Theta_i}$ (the pure allocation) and $p^*, \beta^*$ satisfy the conditions in Corollary~\ref{cor:check-pure-alloc-ME}. Alternatively, by Theorem \ref{thm:eg-gives-me}, we can also verify that the duality gap is zero, i.e., the primal and dual objective values are equal, after adding back the constant $\|B\|_1 - \sum_i B_i \log B_i$ to the dual objective \eqref{eq:eg-dual-beta-p} (the constant is defined in Lemma \ref{lemma:weak-duality}). When computing the objective value of \eqref{eq:eg-dual-beta-p}, the term $\langle p^*, \ones\rangle = p^*([0,1])$ can be decomposed, according to the pure equilibrium allocation given by $\Theta_i = [l_i, h_i]$, as 
\[\langle p^*, \ones \rangle = \sum_i \beta^*_i v_i([l_i, h_i]) = \sum_i \beta^*_i u^*_i.\]

\paragraph{Piecewise linear $v_i$.}
Here, we generate random budgets 
\[ B = (B_1, B_2, B_3, B_4) = (0.2270, 0.2584, 0.2642, 0.2505) \] 
and random piecewise linear coefficients $c_{ik}$, $d_{ik}$ such that $v_i(\theta) = c_{ik}\theta +d_{ik} \geq 0$ for $\theta \in [a_{k-1}, a_k]$, $i\in [n]$, $k\in [K]$. Here, $a_k$ are the breakpoints of the predefined intervals corresponding to the linear pieces of $v_i$.
The values of $a_k$, $c_{ik}$, $d_{ik}$ are as follows:
\begin{align*}
	a &= (a_0, a_1, a_2, a_3) =  (0, 0.3741, 0.8147, 1), \\
	c &= (c_{ik}) =  \begin{bmatrix}
		1.2887 &  1.6253 &  -0.4692  \\
		-1.2494 & -0.2604 &  -0.1476  \\
		-0.4802 &  -1.7084 &  1.1019  \\
		-0.0501 &  2.5419  & 1.2096 
	\end{bmatrix}, \\
	d &= (d_{ik}) = \begin{bmatrix}
		1.9391 & -0.2972 & 1.3209 \\
		0.4674 & 0.4864 & 0.1476 \\
		0.4137 & 1.3919 & -0.0462 \\
		0.4262 & 0.6464 & 0.8471
	\end{bmatrix}.
\end{align*}

Reformulating and solving the convex program \eqref{eq:convex-prog-u(ik)} yield $u^* = (u^*_{ik})$ as follows (e.g., the amount of utility buyer $3$ receives from its allocation of interval $[a_1, a_2]$ is $u^*_{32} = 0.0191$):
\begin{align*}
 u^* = 	\begin{bmatrix}
		0.5845  & 0.0000  & 0.0000  \\
		0.0454  & 0.0579  & 0.0000  \\
		0.0000  & 0.0191  & 0.1767  \\
		0.0000  & 0.6089  & 0.0000 
	\end{bmatrix}.
\end{align*}
Then, we partition each $[a_{k-1}, a_k]$ among the buyers as follows. 
\begin{itemize}
	\item For $k=1$, since the buyers are sorted as $ \sigma^1 = (2,3,4,1)$ in descending order of $\hat{d}_{1k}$, where $\hat{d}_{:,k} = (0.8894, 2.0, 1.277 1.022)$, they also get intervals from the left $a_0 = 0$ to right $a_1 = 0.3741$ in this order. For buyer $2$, $l_{21}=0$ (the left endpoint of the subinterval of buyer $2$ in interval $[a_0, a_1]$) and $v_2([0, h_{21}]) = u^*_{21} \ \Rightarrow h_{21} = 0.1148$. For buyer $3$, since $u^*_{31} = 0$, $l_{31} = h_{31} = h_{21} = 0.1148$. The same is true for buyer $4$. Buyer $1$ gets $[l_{11}, h_{11}] = [0.1148, 1]$, which gives $v_1([l_{11}, h_{11}]) = 0.5845 = u^*_{11}$. 
	\item Similarly, for $k=2$, the order is $\sigma^k = (3,2,4,1)$; buyer $3$ gets $[l_{32}, h_{32}] = [0.3741, 0.4001]$ with utility $u^*_{32} = 0.0579$, buyer $2$ gets $[l_{22}, h_{22}] = [0.4001, 0.5604]$ with utility $u^*_{22} = 0.0191$, buyer $4$ gets $[l_{42}, h_{42}] = [0.5604, 0.8147]$ with utility $u^*_{42} = 0.6089$, buyer $1$ gets nothing. 
	\item For $k=3$, the order is $\sigma^3 = (2,1,4,3)$; only buyer $4$ gets the entire interval, $[l_{43}, h_{43}] = [a_2, 1] = [0.8147, 1]$ with utility $u^*_{43} = 0.1767$. Similarly, we can also verify that it is a pure ME by either Corollary~\ref{cor:check-pure-alloc-ME} or showing zero duality gap as in the linear example above.
\end{itemize}